\documentclass{article}

\usepackage{amsmath}
\usepackage{graphicx}
\usepackage{caption}
\usepackage[nonumberlist]{glossaries}
\usepackage[a4paper,margin=1in]{geometry}
\usepackage{microtype}
\usepackage[english]{babel}
\usepackage{accents}
\usepackage{bbm}
\usepackage{subcaption}
\usepackage{grffile}
\usepackage{url}
\usepackage{multicol}
\usepackage[normalem]{ulem}
\usepackage{bm}
\usepackage{natbib}
\usepackage{underscore}
\useunder{\uline}{\ul}{}

\usepackage{xcolor}

\ifpdf % show pdf plots
\else % DO NOT show plots
  \renewcommand{\includegraphics}[2][1]{include plot: #2}
\fi

\begin{document}

\titlepage

\large
\begin{center}

\textbf{Cause-of-death contributions to declining mortality improvements and life expectancies using cause-specific scenarios}\\
\end{center}

\begin{center}
Alexander M. T. L. Yiu, Torsten Kleinow, and George Streftaris 
\\

\medskip
School of Mathematical and Computer Sciences, Heriot-Watt University,
and
Maxwell Institute for Mathematical Sciences, UK

\end{center}

\abstract

\large

In recent years, improvements in all-cause mortality rates and life expectancies for males and females in England and Wales have slowed down. In this paper, cause-specific mortality data for England and Wales from 2001 to 2018 are used to investigate the cause-specific contributions to the slowdown in improvements.
Cause-specific death counts in England and Wales are modelled using negative binomial regression and a breakpoint in the linear temporal trend in log mortality rates is investigated. Cause-specific scenarios are generated, where the post-breakpoint temporal trends for certain causes are reverted to
pre-breakpoint rates and the effect of these changes on age-standardised
mortality rates and period life expectancies is explored. These scenarios are used to quantify cause-specific contributions to the mortality improvement slowdown. Reductions in improvements at older ages in circulatory system diseases, as well as the worsening of mortality rates due to mental and behavioural disorders and nervous system diseases, provide the greatest contributions to the reduction of improvements in age-standardised mortality rates and period life expectancies. Future period life expectancies scenarios are also generated, where cause-specific mortality rate trends are assumed to  either persist or be reverted. In the majority of scenarios, the reversion of cause-specific mortality trends in a single cause of death results in the worsening of period life expectancies at birth and age 65 for both males and females.
This work enhances the understanding of cause-specific contributions to the slowdown in all-cause mortality rate improvements from 2001 to 2018, while also providing insights into causes of death that are  drivers of life expectancy improvements. The findings can be of benefit to researchers, policy-makers and insurance professionals.\\


\newpage

\large

\section{Introduction}

\large

In recent years, England and Wales have been experiencing a reduction in the annual improvements in all-cause mortality rates. This study uses cause-specific mortality experience to investigate the observed slowdown in these annual improvements in all-cause mortality. The aim of the study is to quantify the contributions of different causes of death towards the reduction in mortality rate improvements, and also to investigate life expectancies under various scenarios for cause-specific mortality rate trends. Aiming at enhancing the understanding of the slowdown in all-cause mortality improvements, cause-specific scenarios are generated in order to explore the differences in mortality rate improvements that are observed, compared to improvements in scenarios where cause-specific trends are reverted to pre-slowdown levels. These cause-specific scenarios are useful for the purpose of investigating what would have happened if the trends for certain causes of death had remained the same throughout the observed slowdown in mortality improvements. The use of scenarios is a novel approach for quantifying cause-specific contributions. Investigating the impact of individual causes towards the slowdown in mortality improvements is important for better understanding of the drivers of increasing human longevity.
\\

Many countries in the world have experienced improvements in life expectancy throughout the 20th century into the 21st century. A number of authors have noted that reductions in infectious disease and infant mortality have led to the greatest amount of improvements in life expectancy in the first half of the 20th century for Europe and North America (\citealp*{wilmoth2000demography}; \citealp*{oeppen2002demography}; \citealp*{cutler2006determinants}; \citealp*{leon2011trends}; \citealp*{mackenbach2013life}). In the United States and Western European countries, improvements in cardiovascular disease mortality at older ages led to the continued improvement in life expectancies after 1970 (\citealp*{tuljapurkar1998mortality}; \citealp*{wilmoth2000demography}; \citealp*{levi2002trends}; \citealp*{oeppen2002demography}; \citealp*{cutler2006determinants}; \citealp*{leon2011trends};  \citealp*{mackenbach2013life}). However, Eastern European countries experienced increasing levels of cardiovascular disease mortality (\citealp*{levi2002trends}; \citealp*{leon2011trends}), which led to slower gains and even worsening of life expectancies \citep{leon2011trends}. The improvement in cardiovascular disease mortality was linked to improvements in risk factors and in treatments (\citealp*{levi2002trends}; \citealp*{o2008coronary}; \citealp*{scholes2012persistent}).  \citet{mackenbach2013life} also suggested that the improvement in cardiovascular mortality at older ages might be related to the national income of a country.\\

England and Wales have also experienced similar reductions in infectious disease and cardiovascular disease mortality during the 20th century (\citealp*{griffiths2003twentieth}; \citealp*{o2008coronary}). Along with improvements in cardiovascular disease mortality, England and Wales have shown changing trends in associated risk factors. \citet{o2008coronary} and \citet{scholes2012persistent} noted reductions in smoking prevalence, but also increases in the prevalence of obesity and diabetes. While improvements in cardiovascular disease mortality continued to occur at older ages, the rate of improvement appeared to be slowing for individuals aged 55 and younger (\citealp*{o2008coronary}; \citealp*{howse2008review}).\\

The rate at which mortality is improving may have lowered in recent years. The Office for National Statistics (ONS) identified a reduction in mortality improvement in England and Wales. In particular, using the all-cause mortality experience in England and Wales from 1990 to 2017, the ONS suggested that a change in the temporal trend of mortality rate improvements in age-standardised mortality rates occurred in the early 2010s, for both males and females \citep{office2018changing}.\\

The \citet{office2018changingb} have also explored all-cause mortality and cause-specific rate trends in the United Kingdom and have indicated that reductions in improvements in circulatory system diseases and increases in cancer mortality and mortality due to mental and behavioural disorders, to be the leading contributors to the observed increase in mortality rates at older ages across the United Kingdom. At younger ages, increases in deaths due to external causes of death, such as accidents and intentional self-harm, provided the greatest contributions to increased mortality at these ages. \cite{bennett2018contributions}, \citet{public2018review}, and \citet{steel2018changes} investigated the mortality rate trends in England and suggested that the most deprived groups in terms of socio-economic status were associated with greatest lack of improvements, in both all-cause and cause-specific mortality. This was also addressed by \cite{boumezoued2021modeling}, where they suggested that this pattern was observable across all countries.\\

The recent reduction in mortality rate improvements is not limited to England and Wales, or the rest of the United Kingdom. The \citet{office2018changingc} performed a comparison of the changes in average annual life expectancy improvements between 20 countries and the United Kingdom was identified to have experienced one of the highest reductions in improvement amongst the countries studied. The observed slowdown in improvements did not appear to be a result of the life expectancy approaching a maximum, as countries with higher life expectancy than the United Kingdom, such as Japan and Sweden, did not experience slowdowns. \cite{boumezoued2019modeling} also noted a slowdown in mortality rate improvements in the United States, with circulatory system diseases being the major cause of this slowdown. \citet{leon2019trends} discussed the lower estimated life expectancies compared to other high-income countries due to a worsening of observed mortality rates at ages 25 to 44. \citet{raleigh2019trends} provided additional discussions on the potential risk factors which contributed to the observed trends in the slowdown of mortality rate improvements. While the comparison of cause-specific mortality trends between different countries would greatly aid the understanding of observed all-cause mortality trends, differences in the recording and coding of causes of death prevent a reliable comparison between countries.\\

\citeauthor{hiam2017has} (\citeyear{hiam2017has} and \citeyear{hiam2018life}) noticed a sharp increase in mortality in England and Wales in 2015 in contrast to the trends in previous years. While reductions in mortality due to cancers and circulatory system diseases provided positive contributions to male and female life expectancies, increases in mortality due to diseases such as Alzheimer's disease and dementia resulted in a net reduction in estimated life expectancies. In addition, the oldest age group had the greatest contribution to the reduction in the estimated life expectancy \citep{hiam2017has}. Although a single year with higher than average mortality rates influences the overall temporal trend in mortality rates, the effects of the single year may be offset by the observed mortality rates in other years.\\

Modelling mortality by the cause of death introduces additional challenges that are not necessarily present when modelling all-cause mortality. Despite individuals facing a potential risk of dying due to a number of causes, normally only one cause is recorded at the actual event of death. Cause-specific mortality rates can be modelled by using competing risk methods, where an individual is exposed to multiple risks of death arising from different causes during his or her lifetime, but death can only result from a single cause of death. However, a problem arises in competing risk methods for cause-specific mortality, as the dependency structure between the different causes of death is not known \citep{tsiatis1975nonidentifiability}. \cite{dimitrova2013dependent} explored copula methods with a known dependency structure in order to model competing causes of death. \citet{alai2015modelling} investigated the dependencies between different causes of death using multinomial logistic models.\\

In addition to competing risk models, relationships between cause-specific mortality rates have also been examined. \cite{arnold2013forecasting} investigated cause-specific mortality rates using time-series models. The models included vector autoregressive models and vector error correction models in order to model the dependence of cause-specific mortality rates through time. \cite{arnold2015causes} also investigated cause-of-death mortality rates using cointegration analysis.\\

In this paper, negative binomial regression is used in order to assess the contributions of cause-specific mortality rate trends on the observed slowdown in mortality rate improvements observed in England and Wales from 2001 to 2018. The methodology being employed here differs from existing methodologies by quantifying additional rates of improvement of mortality rates and life expectancies under the assumptions of continuing cause-specific mortality rate trends. In addition to observed rates, future life expectancy scenarios are generated under differing assumptions of cause-specific mortality rate trends. These future life expectancy scenarios reveal the causes of death with the greatest influence towards life expectancy improvements in the future.\\

The paper begins with a brief outline of the dataset that is used in the analyses. An initial investigation into the log age-standardised mortality rates is performed for the purpose of determining a suitable single breakpoint location for changes in temporal trends in all-cause and cause-specific mortality rates. This breakpoint location is used with negative binomial regression in order to model cause-specific death counts and to investigate cause-specific contributions to the observed temporal trends in log age-standardised mortality rates and period life expectancies. Lastly, future life expectancy scenarios are generated for the purpose of investigating the potential effects of cause-specific trends on period life expectancies in the near future. While deaths from COVID-19 occur during the projection period, this cause is not considered for the analyses in this paper. The analyses performed in this paper and detailed discussions are elaborated in greater detail in \cite{phdthesis}.

\section{Data}\label{data}

In this paper, we use data comprising cause-specific mortality and population information in England and Wales, obtained from the Office for National Statistics (\citeyear{office2019deaths}). The dataset is chosen for the purpose of investigating the cause-specific mortality experience in England and Wales, and includes generalised mortality data about the cause-specific mortality experience of individuals from birth to age 85 and above, from 2001 to 2018. The data include death counts grouped by 5-year age bands, cause of death, gender, and the year in which the deaths are registered. Corresponding mid-year population estimates are also obtained from the same source. The causes of death were classified according the International Classification of Diseases (ICD) 10th revision \citep{world2016classification}. The time period from 2001 to 2018 is chosen in order to avoid possible problems arising from changes in cause-of-death classifications. However, the ONS documented a change in the classification from the classification system ICD-10 v2001.2 used in 2001 to ICD-10 v2010 in 2011 \citep{wells2011impact}. Another update to the classification was implemented in 2014 \citep{wells2014impact}. These changes in the classifications of deaths present a problem to cause-specific mortality analyses, as the number of deaths attributed to a specific cause may not be consistent under the different classification schemes. Some of the effects of the classification changes were discussed by the Office for National Statistics (\citeyear{wells2011impact}, \citeyear{wells2014impact}, \citeyear{lloyd2016impact}).\\

Specific causes of death are grouped into larger groups according to the chapters in the 10th revision of the International Classification of Diseases \citep{world2016classification}. A full list of the deaths by ICD-10 code and their assigned groups are provided in Table \ref{CAUSETABLE}.\\

\begin{table}[h!]
	\caption{Cause-of-death groupings for England and Wales dataset}
	\centering
	\begin{tabular}{|c|c|}
		\hline
		{\ul \textbf{Cause-of-death group}} & {\ul \textbf{ICD-10 Code}}                                                                             \\ \hline
		Cancers (CAN)                & C00-D49                                                                                                \\ \hline
		Circulatory (CIR)                        & I00-I99                                                                                                \\ \hline
		Digestive (DIG)                           & K00-K95                                                                                                \\ \hline
		Endocrine and Blood (END)                    &\begin{tabular}[c]{@{}c@{}}D50-D89,\\ E00-E89\end{tabular}                                                                                            \\ \hline
		External (EXT)                           & V00-Y99                                                                                                \\ \hline
		Genitourinary (GEN)                      & N00-N99                                                                                                \\ \hline
		Infectious (INF)                          & A00-B99                                                                                                \\ \hline
		Mental (MEN)                              & F01-F99                                                                                                \\ \hline
		\begin{tabular}[c]{@{}c@{}}Musculoskeletal and Skin\\ (MUS) \end{tabular}                    &          \begin{tabular}[c]{@{}c@{}}M00-M99,\\ L00-L99 \end{tabular}                                                                                         \\ \hline
		Nervous (NER)                             & \begin{tabular}[c]{@{}c@{}}G00-G99,\\ H00-H59,\\ H60-H95\end{tabular}                                  \\ \hline
		Respiratory (RES)                        & J00-J99                                                                                                \\ \hline
		Other (OTH)                              & \begin{tabular}[c]{@{}c@{}}O00-O9A,\\ P00-P96,\\ Q00-Q99,\\ R00-R99,\\ S00-T88,\\Z00-Z99\end{tabular} \\ \hline
	\end{tabular}
\label{CAUSETABLE}
\end{table}

Following an initial exploration into cause-specific mortality rates by age group, it appears that the patterns in the mortality rates range greatly across age groups for different causes of death. For the purpose of modelling age- and cause-specific mortality, the age groups are treated as levels of a factor covariate rather than values of a numerical variable. The age groups, $x$, are the 5-year age bands ranging from $<$1, 1-4, 5-9, ..., 80-84, and the last group of individuals aged 85 and above, giving a total of 19 age groups.
\\

\section{Cause-specific mortality rates}

Let $d_{x,t}$ denote the observed death counts in age group $x$ in calendar year $t$, where the age groups $x$ are defined in Section \ref{data}, and the calendar years $t$ range from 2001 to 2018. $E_{x,t}$ represent the corresponding mid-year population estimates for age group $x$ in calendar year $t$. Death counts and mid-year population estimates are considered here by gender separately, and for the remainder of the paper the analyses are determined using the gender-specific mortality experience in England and Wales. However, for brevity, separate notation by gender is omitted.\\

The observed all-cause age-specific mortality rate, $\check{m}_{x,t}$, is calculated as
\begin{equation*}
\check{m}_{x,t} = \frac{d_{x,t}}{E_{x,t}}.
\end{equation*}
In addition to all-cause deaths, $d_{x,t}^{[c]}$ is used to denote the observed death counts in age group $x$ in calendar year $t$ due to cause-of-death group $c$.
The observed cause-specific, age-specific mortality rate, $\check{m}_{x,t}^{[c]}$, is given as
\begin{equation*}
\check{m}_{x,t}^{[c]} = \frac{d_{x,t}^{[c]}}{E_{x,t}},
\end{equation*}
and the observed all-cause age-specific mortality rate, $\check{m}_{x,t}$, is the summation of the observed cause-specific age-specific mortality rates, such that
\begin{equation*}
\check{m}_{x,t} = \sum\limits_c \check{m}_{x,t}^{[c]} = \frac{\sum\limits_c d_{x,t}^{[c]}}{E_{x,t}}.
\label{mortsum}
\end{equation*}
Using a standardised population with exposure weights $E^S_{x}$ for each age group $x$ and the observed age-specific mortality rates $\check{m}_{x,t}$, the age-standardised mortality rate in calendar year $t$, $ASMR_t$, is calculated as
\begin{equation}
ASMR_t = \frac{\sum\limits_x \check{m}_{x,t} E^S_{x}}{\sum\limits_x E^S_{x}}.
\label{ObsASMR}
\end{equation}
In this paper, the European Standard Population in 2013 \citep{eurostat2013revision} is chosen for the standardisation above, as the population distribution is similar to that of England and Wales within the chosen period of observation.\\ 

\begin{figure}[h!]
	\begin{multicols}{2}
		\includegraphics[width=\linewidth,height=8cm]{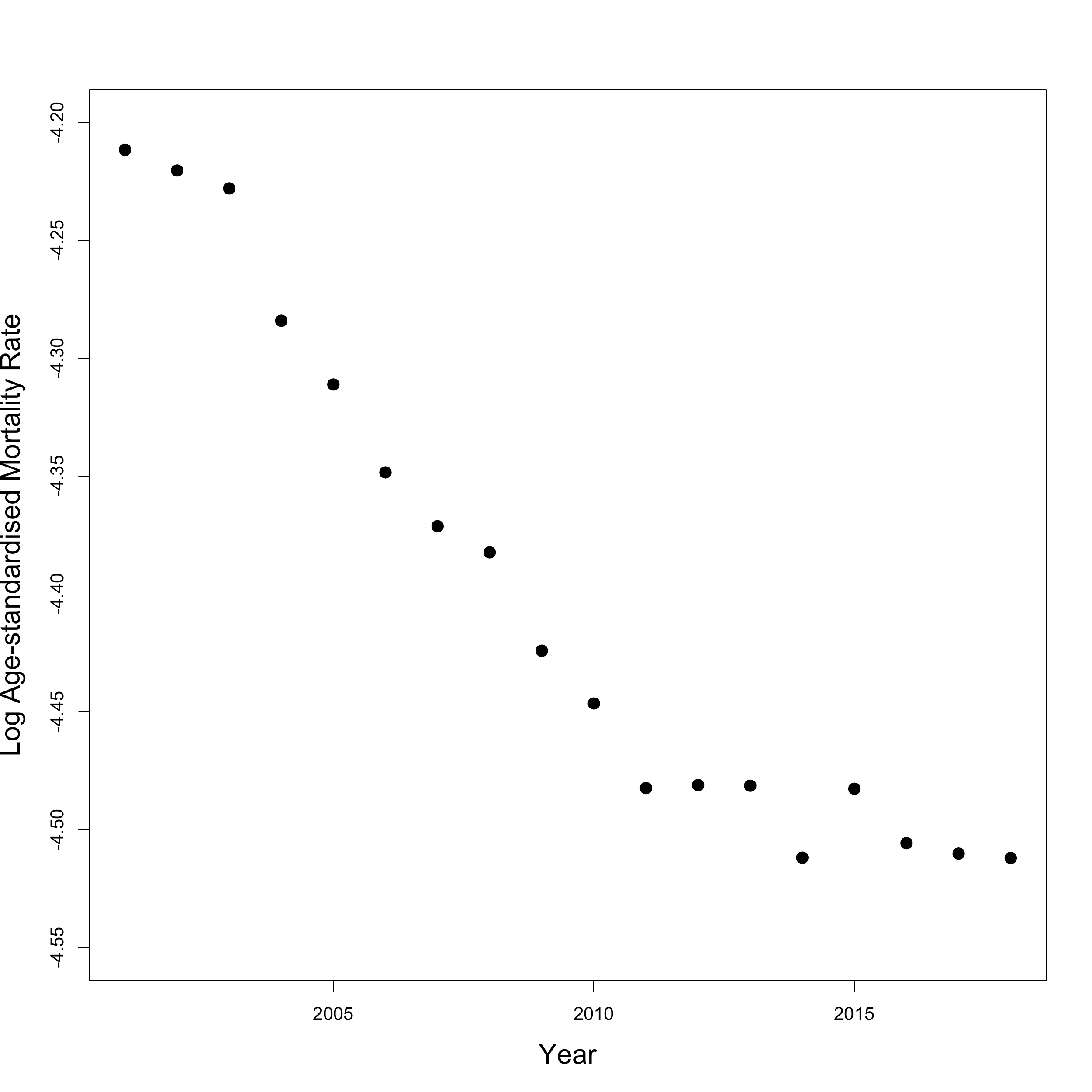}
		\subcaption{Males}
		\includegraphics[width=\linewidth,height=8cm]{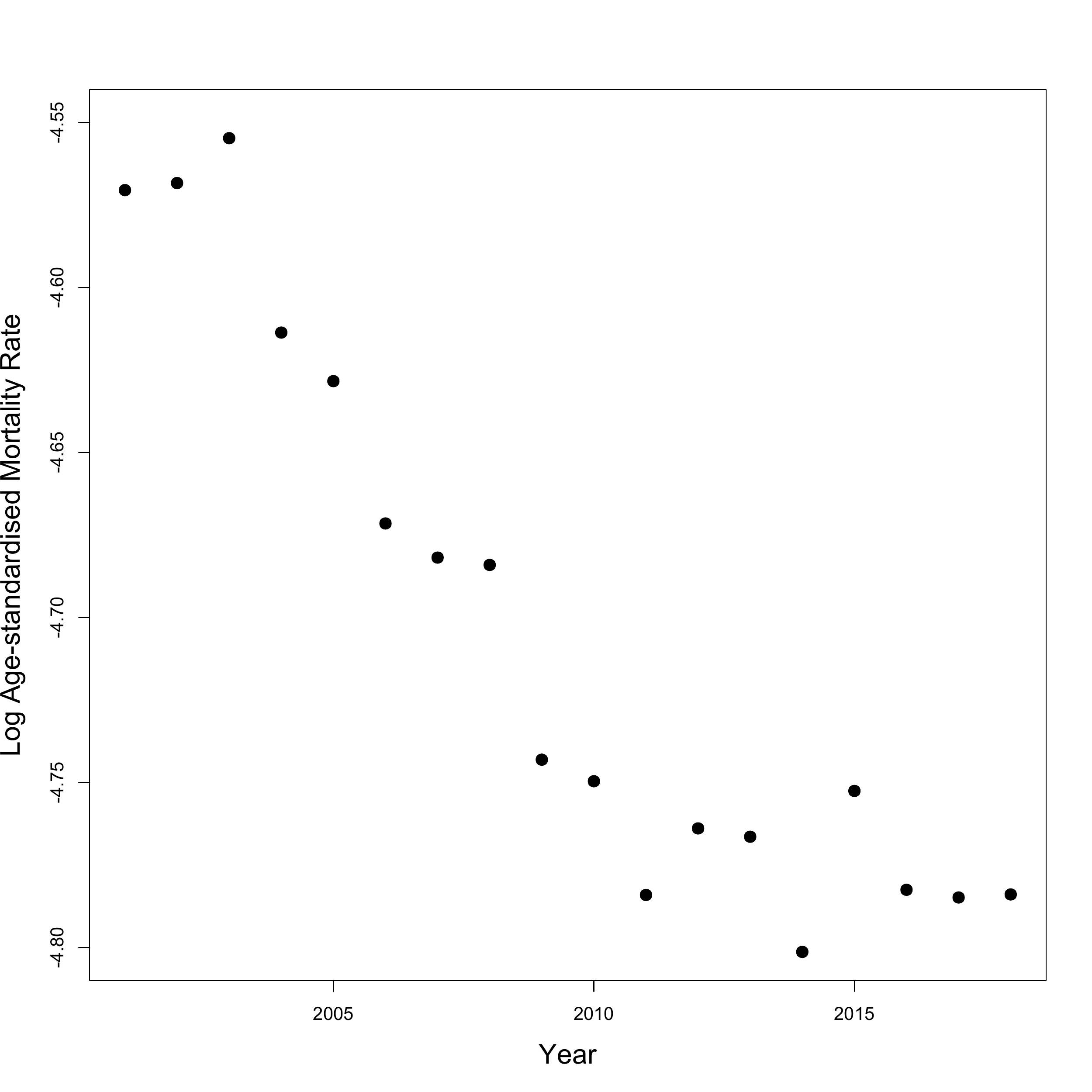}
		\subcaption{Females}
	\end{multicols}		
	\caption{Observed log age-standardised mortality rates by gender}
	\label{obsasmrboth}
\end{figure}

Figure \ref{obsasmrboth} shows age-standardised mortality rates for males and females (on the log scale), calculated using the observed age-specific mortality rates due to all causes in (\ref{ObsASMR}). The figure shows that there appears to be a possible change in a linear temporal trend which occurs around 2011 for both genders. The average rate of improvement in the log age-standardised mortality rates appears to decrease after 2011 compared to the rates of improvement pre-2011, as suggested by the slope of the plotted rates.
In addition to the reduced rate of improvement, there is a noticeable increase in the log age-standardised mortality rate in 2015 compared to that of the previous year in 2014. This increase in the mortality rate was explored by \cite{hiam2017has} and it was attributed to increases in many different causes of death, with dementia being a major contributor to the increased deaths. \cite{hiam2017has} and \cite{ho2018recent} also identified circulatory system disease and respiratory system disease as contributors to the increased deaths, especially at older ages. \cite{public2018review} and \cite{pebody2018significant} suggested that the increase in deaths was linked to an increased incidence in influenza, specifically a strain named H3N2, from 2014 to early 2015. While many of the deaths were not recorded as influenza directly, many of them concerned individuals who were admitted to hospitals with, or due to, this strain of influenza \citep{public2018review}. \cite{public2018review} and \cite{pebody2015effectiveness} commented on the lower efficacies of the available influenza vaccines against this particular strain leading to the increased incidence of influenza in the United Kingdom during this period. \cite{green2017geography} explored the increase in mortality across different regions and levels of socioeconomic deprivation in England and Wales and noted that the increase occurred across regions and different deprivation levels. Many other European countries experienced this increase in mortality \citep{molbak2015excess}, which was also observed in the United States \citep{acciai2017did}. While there is an increase in the fitted log age-standardised mortality rates in the year 2015, the log ASMRs in the following years appear to decline in a relatively linear fashion in Figure \ref{obsasmrboth}.\\

\section{Mortality rate improvement slowdown}\label{ASMRmodel}

Age-standardised mortality rates are used for all-cause mortality in England and Wales for the period from 2001 to 2018, in order to determine the location of a point at which the temporal trend in mortality rates changes. The resulting location is referred to as a breakpoint and is used in the modelling of cause-specific death counts under a negative binomial distribution in Section \ref{PoisModelling}. 
The analysis here assumes that temporal trends are linear in log age-standardised mortality rates.
Age-standardised mortality rates are calculated using observed all-cause age-specific mortality rates from 2001 to 2018, according to (\ref{ObsASMR}). The observed log age-standardised mortality rates for calendar year $t$, $\log(ASMR_t)$, are then modelled using a 
normal distribution such that
\begin{equation}
\log(ASMR_t) \sim \text{Normal}(\mu_t, \sigma^2).
\label{normal}
\end{equation}
A base model with a single linear temporal trend in the log age-standardised mortality rates is given by
\begin{equation}
\mu_t = \beta_0 + \beta_1 t ,
\label{nochange}
\end{equation}
where $\beta_1$ represents the single linear temporal trend in the log age-standardised mortality rates across the entire period of interest. The base model is compared against models where a change in the temporal trend exists at a point $t = \epsilon$. The point at which the change occurs is referred to as a breakpoint for the remainder of the paper.\\

In this research, a single breakpoint is considered following recent ONS work \citep{office2018changing}, and as it has also been suggested in the trends in
Figure \ref{obsasmrboth}. While it is possible for multiple breakpoints to exist within our chosen period of observation, the relatively small number of calendar years considered here increases the difficulty of justifying the inclusion of multiple breakpoints. 
%The analyses continue under the assumption of the existence of a single breakpoint in the linear temporal trend in log age-standardised mortality rates.\\
The models considered are parametrised such that the modelled log age-standardised mortality rates are either continuous at point $t = \epsilon$,  with
\begin{equation}
\mu_t = \beta_0 + \beta_1 t + \beta_2 (t - \epsilon) \text{I}(t \ge \epsilon) , 
\label{contASMR}
\end{equation}
where $I()$ denotes the indicator function, or that there is a discontinuity in the modelled rates at point $t = \epsilon$, with
\begin{equation}
\mu_t = \beta_0 + \beta_1 t + (\beta_2 (t - \epsilon) + \beta_3) \text{I}(t \ge \epsilon).
\label{disASMR}
\end{equation}
In both the continuous and discontinuous breakpoint models given by (\ref{contASMR}) and (\ref{disASMR}) respectively, $\beta_1$ represents the pre-$\epsilon$ temporal trend in the log age-standardised mortality rates, while $\beta_2$ represents the post-$\epsilon$ change in the linear temporal trend. The post-$\epsilon$ temporal trend is given by $\beta_1 + \beta_2$. In the case of the continuous breakpoint model, the fitted log age-standardised mortality rates are required to intersect at point $t=\epsilon$. This constraint becomes problematic if discontinuities exist in the linear temporal trends of the log age-standardised mortality rates and may result in poor model selections, depending on the magnitude of these discontinuities. With the inclusion of the $\beta_3$ term in the discontinuous case, the estimation of parameter $\beta_1$ does not depend on the log age-standardised mortality rates after the point $t = \epsilon$. \\

The location of the breakpoint is determined as follows. Model (\ref{normal}) is fitted according to (\ref{contASMR}) and (\ref{disASMR}) using integer values for $\epsilon$ between 2003 and 2016. The location of a possible breakpoint in the temporal trend, $\epsilon^*$, is then determined according to the fitted model with the lowest Bayesian Information Criterion (BIC) score. Once the location of the breakpoint is determined to occur at point $\epsilon = \epsilon^*$, the continuous and discontinuous breakpoint models, (\ref{contASMR}) and (\ref{disASMR}) are also compared against (\ref{nochange}) in terms of the BIC. The BIC values are captured in Table $\ref{obsasmrtable}$.

\begin{table}[h!]
	\centering
	\caption{Model BIC values for the determination of the location of a single breakpoint in the temporal trend in log age-standardised mortality rates by gender}
	\begin{tabular}{|c|c|c|c|c|}
		\hline
		{\ul \textbf{Breakpoint location}} & {\ul \textbf{Males Cont. (\ref{contASMR})}} & {\ul \textbf{Males Dis. (\ref{disASMR})}} & {\ul \textbf{Females Cont. (\ref{contASMR})}} & {\ul \textbf{Females Dis. (\ref{disASMR})}} \\ \hline
		\textbf{None}                      & -65.6991                   & NA          & -65.4194                     & NA            \\ \hline
		\textbf{2003}                      & -64.9625                   & -62.8782                  & -63.1398                     & -60.8058                    \\ \hline
		\textbf{2004}                      & -68.1846                   & -69.89                    & -65.1548                     & -67.6500                    \\ \hline
		\textbf{2005}                      & -71.6466                   & -72.2154                  & -67.4729                     & -68.1740                    \\ \hline
		\textbf{2006}                      & -75.6575                   & -76.3614                  & -70.4944                     & -72.2165                    \\ \hline
		\textbf{2007}                      & -79.5000                   & -78.7203                  & -73.1316                     & -72.3841                    \\ \hline
		\textbf{2008}                      & -83.4470                   & -82.0297                  & -76.0159                     & -75.0659                    \\ \hline
		\textbf{2009}                      & -89.9168                   & -90.7807                  & -81.1197                     & -84.1777                    \\ \hline
		\textbf{2010}                      & -96.8171                   & -95.6129                  & -84.7850                     & -82.9333                    \\ \hline
		\textbf{2011}                      & \textbf{-102.6420}         & -99.8451                  & \textbf{-87.0123}            & -84.1457                    \\ \hline
		\textbf{2012}                      & -97.1434                   & -99.7636                  & -83.5411                     & -84.5003                    \\ \hline
		\textbf{2013}                      & -89.2486                   & -95.8123                  & -79.8185                     & -81.4427                    \\ \hline
		\textbf{2014}                      & -83.2945                   & -87.4577                  & -76.6839                     & -77.4627                    \\ \hline
		\textbf{2015}                      & -75.2843                   & -86.7130                  & -70.9088                     & -79.7457                    \\ \hline
		\textbf{2016}                      & -70.9588                   & -73.5244                  & -68.2068                     & -68.6941                    \\ \hline
	\end{tabular}
\label{obsasmrtable}
\end{table}

Following this procedure, the location of a breakpoint in the temporal trend in the log age-standardised mortality rates is determined to occur at $\epsilon$ = 2011 for both males and females, with the continuous breakpoint models (\ref{contASMR}) having the lowest BIC values.
The results of this analysis differ slightly from the single breakpoints for the age-standardised rates that were identified by the \citet{office2018changing} with quarterly data, where the breakpoint was determined to exist between Q2 2011 to Q1 2012 for males of all ages and between Q2 2013 to Q1 2014 for females of all ages. The \citet{office2018changing} used quarterly mortality data from 1990 to 2018, while our analysis involves single year data from 2001 to 2018.
\begin{figure}[h!]
	\begin{multicols}{2}
		\includegraphics[width=\linewidth,height=8cm]{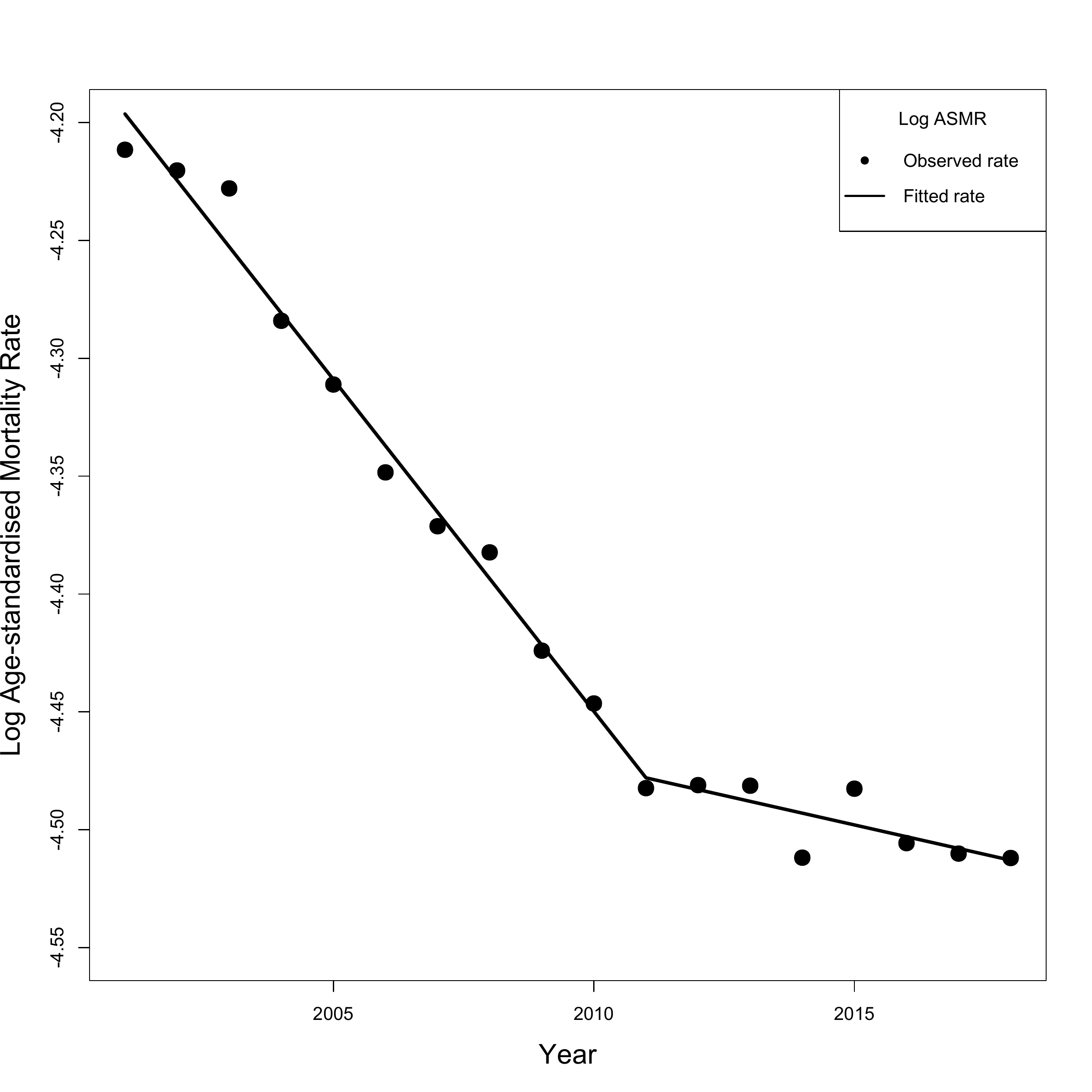}
		\subcaption{Males}
		\includegraphics[width=\linewidth,height=8cm]{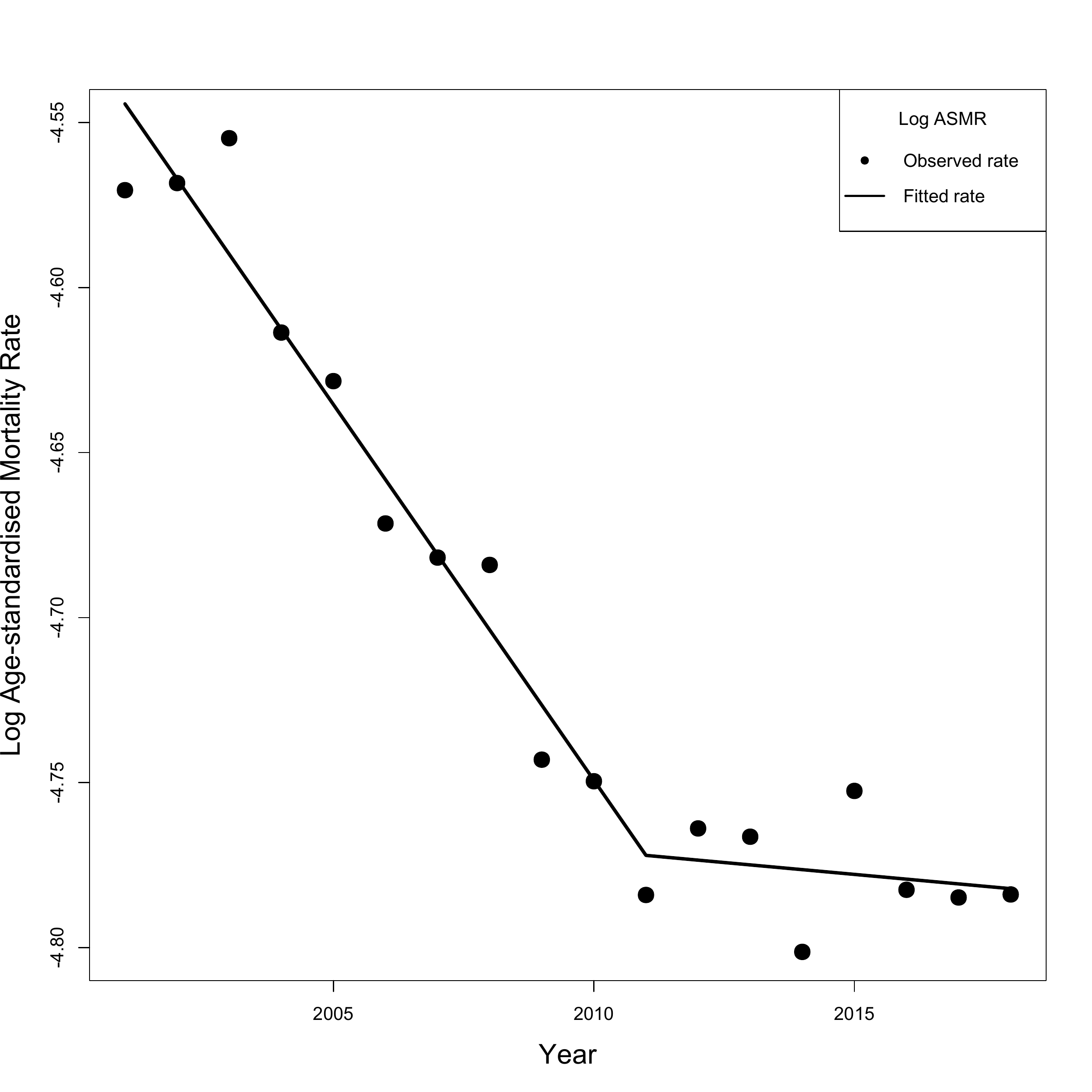}
		\subcaption{Females}
	\end{multicols}		
	\caption{Fitted log age-standardised mortality rates by gender using model (\ref{contASMR})}
	\label{obsasmrplots}
\end{figure}

Figure \ref{obsasmrplots} displays the observed log age-standardised mortality rates for males and females along with the fitted rates given a continuous breakpoint occurring in 2011. For the purpose of discussing rates of improvement in log age-standardised mortality rates, a negative change in trend (negative coefficients $\beta_1$ or $\beta_1 + \beta_2$) denotes an improvement (decrease) in mortality, while a positive change denotes an increase in mortality (worsening of rates).\\

\begin{table}[h!]
	\centering
	\caption{Annual change in mortality: pre-2011 annual improvements and post-2011 reductions in improvement in log ASMRs by gender}
	\begin{tabular}{|c|c|c|c|c|c|}
		\hline
		{\ul \textbf{Gender}}       & \textbf{\begin{tabular}{@{}c@{}}{\ul Annual}\\ {\ul change}\\ {\ul pre-2011 ($\beta_1$)} \end{tabular}} & {\ul \textbf{95\% CI}}   &\textbf{\begin{tabular}{@{}c@{}}{\ul Annual}\\{\ul reduction in} \\ {\ul improvement}\\ {\ul post-2011 ($\beta_2$)}\end{tabular}} & {\ul \textbf{95\% CI}}   & 
		\textbf{\begin{tabular}{@{}c@{}}{\ul Annual}\\ {\ul change}\\ {\ul post-2011 ($\beta_1+\beta_2$)} \end{tabular}} \\ \hline
		\textbf{Males}   & $-$0.02816                                      & $(-0.03017, -0.02617)$ & 0.02317                                                   
		& (0.01870, 0.02764) & $-0.00499$ \\ \hline
		\textbf{Females} & $-$0.02277                                      & $(-0.02585, -0.01969)$ & 0.02133                                                   & (0.01443, 0.02822) & $-0.00144$ \\ \hline
	\end{tabular}
	\label{obsAMR}
\end{table}

Table \ref{obsAMR} summarises the numerical results for the log age-standardised mortality rates, fitted using model (\ref{contASMR}). Both male and female populations in England and Wales experienced a reduction in yearly improvement in log age-standardised mortality rates post-2011, i.e., $\beta_2$ in model (\ref{contASMR}), compared to the average rate of improvement pre-2011, with males experiencing a higher reduction in the rate of decrease in mortality $(2.3\%)$ as compared to females  $(2.1\%)$. However, the overall decrease in log ASMRs experienced by males post-2011 is higher than the rate experienced by females, i.e., $-0.00499$ compared to $-0.00144$ respectively.\\

The location of the breakpoint at 2011 is suitable for further cause-specific mortality rate calculations, with the observed age- and cause-specific mortality rates having approximately linear trends pre-2011 and post-2011 for the majority of the cause-of-death groups. The breakpoint also coincides with the aforementioned change in the ICD classification at 2011 \citep{wells2011impact} and therefore the effect of the classification change in 2011 is incorporated into the models for mortality rates.
According to the bridge coding study performed by the ONS on the cause-specific mortality experience in England and Wales in 2009 \citep{wells2011impact}, the change in the classification from ICD-10 v2001.2 to ICD-10 v2010 would have resulted in a net decrease in the number of deaths being classified as deaths from diseases of the circulatory system, with these deaths being redistributed between mental and behavioural disorders, nervous system diseases, and respiratory system diseases.  However, these classification changes do not imply changes in the year to year cause-specific mortality trends. It is believed that the post-2011 changes in cause-specific mortality rate trends are the result of real trend changes rather than artificially-produced from a change in the classification.\\ 

\section{Cause-specific slowdown in mortality improvements}\label{PoisModelling}

\subsection{Fitting cause-specific mortality}\label{Fitting}

For the purpose of modelling cause-specific death counts by age through time, the data are separated into datasets for the nineteen individual age-groups and twelve causes, resulting in 228 total datasets for each sex. The random variable for cause-specific death counts, $D_{x,t}^{[c]}$, for age band $x$, cause of death $c$, and calendar year $t$ is modelled according to a negative binomial (NB) distribution,
\begin{equation}
D_{x,t}^{[c]}  \sim NB(E_{x,t} m_{x,t}^{[c]}, \theta_{x}^{[c]}),
\label{NBSpec}
\end{equation}
where $E_{x,t}$ denotes the central exposure in age group $x$ in year $t$, i.e., the estimated mid-year population size, and $m_{x,t}^{[c]}$ denotes the force of mortality, i.e., the instantaneous rate of mortality for an individual in age group $x$ and cause $c$. 
$\theta_{x}^{[c]}$ is referred to as the dispersion parameter and models the relationship between the expected number of death counts and its variance. For the purpose of discussion, the force of mortality is referred to as the mortality rate in the remainder of this paper.
Under a negative binomial distribution, the expected number of deaths is given as
\begin{equation}
	\textbf{E}(D_{x,t}^{[c]}) = E_{x,t} m_{x,t}^{[c]},
	\label{NBMean}
\end{equation}	
and the corresponding variance is
\begin{equation}
	\text{Var}(D_{x,t}^{[c]}) = E_{x,t} m_{x,t}^{[c]} + \frac{(E_{x,t} m_{x,t}^{[c]})^2}{\theta_{x}^{[c]}}.
	\label{NBVar}
\end{equation}
From (\ref{NBVar}), it is shown that the $\theta_{x}^{[c]}$ parameter models the amount of overdispersion in death counts. Overdispersion in this context refers to a greater amount of variance in the variable for cause-specific death counts compared to the expected number of deaths. This is in contrast with the commonly used Poisson mortality model (e.g., \cite{mavros2017stochastic}), under which the expected value of the variable representing the death counts is assumed to be equal to its variance. The negative binomial distribution is chosen here, due to the inherent overdispersion in the death data, which may result from the heterogeneity of the population \citep{wong2018bayesian, li2009uncertainty}. 
%The negative binomial model (\ref{NBSpec}) and variance given by (\ref{NBVar}) results from a Poisson distribution that is adjusted in order to account for this heterogeneity \citep{li2009uncertainty}. 
Further discussion regarding the suitability of the Poisson and negative binomial distributions is provided later in this section.\\

The mortality rate $m_{x,t}^{[c]}$ in (\ref{NBSpec}) is parametrised as
\begin{equation}
\log(m_{x,t}^{[c]}) = \beta_{0,x}^{[c]} + \beta_{1,x}^{[c]} t 
     + \big[\beta_{2,x}^{[c]}(t  - 2011) + \beta_{3,x}^{[c]}\big] \text{I}(t \ge 2011).
\label{disMT}
\end{equation}
Therefore, model (\ref{disMT}) implies that the age- and cause-specific death counts, $D_{x,t}^{[c]}$, are modelled using a negative binomial GLM with a linear predictor given in (\ref{disMT}) that includes a discontinuous breakpoint in 2011. The parameter $\beta_{3,x}^{[c]}$ captures the effect of the age- and cause-specific breakpoint at 2011, which represents the change in ICD-10 classification in this year, and coincides 
with the breakpoint determined in Section \ref{ASMRmodel}. While the log ASMRs in Section \ref{ASMRmodel} are modelled using model (\ref{contASMR}) and a continuous breakpoint, those log ASMRs are computed using summations of cause-specific mortality rates. In the process of summing cause-specific mortality rates in order to obtain all-cause mortality rates, the individual effects of changes in cause-of-death classifications offset one another. As model (\ref{disMT}) pertains to cause-specific mortality rather than all-cause mortality, the effect of the change in classification is kept in $\beta_{3,x}^{[c]}$.\\ 

Similar to (\ref{contASMR}) and (\ref{disASMR}), $\beta_{1,x}^{[c]}$ denotes the pre-2011 annual change in the log mortality rate for the period from 2001 to 2010. Coefficient $\beta_{2,x}^{[c]}$ represents the additional post-2011 annual change for 2011 to 2018. Together, $\beta_{1,x}^{[c]}$ + $\beta_{2,x}^{[c]}$ gives the overall annual change post 2011. Again, a negative coefficient $\beta_{1,x}^{[c]}$, or a negative sum $\beta_{1,x}^{[c]}$ + $\beta_{2,x}^{[c]}$, signifies an improvement in mortality, and a positive coefficient or sum signifies a worsening of mortality. 
Fitting (\ref{disMT}) to the data gives the 
fitted age- and cause-specific mortality rate as
\begin{equation}
\log \widehat{m}_{x,t}^{[c]} = \widehat{\beta}_{0,x}^{[c]} + \widehat{\beta}_{1,x}^{[c]} t 
+ \big[\widehat{\beta}_{2,x}^{[c]}(t  - 2011) + \widehat{\beta}_{3,x}^{[c]}\big] \text{I}(t \ge 2011).
\label{disMT2}
\end{equation}
The fitted all-cause age-specific mortality rates, $\widehat{m}_{x,t}$, are calculated as
\begin{equation}
\widehat{m}_{x,t} = \sum\limits_c \widehat{m}_{x,t}^{[c]},
\label{MTDIS}
\end{equation}
where the fitted cause-specific mortality rates are computed using (\ref{disMT2}).\\

Tables \ref{NBBeta1Male} and \ref{NBBeta1MaleB} display estimated $\widehat{\beta}_{1,x}^{[c]}$ values, while Tables \ref{NBBeta2Male} and \ref{NBBeta2MaleB} display estimated $\widehat{\beta}_{2,x}^{[c]}$ values, using (\ref{disMT}) for males. The causes of death are provided in the order given in Table \ref{CAUSETABLE}. Further estimated coefficients for females are provided in the Appendix in Tables \ref{NBBeta1FemalesA} to \ref{NBBeta2FemalesB}. Estimated coefficients corresponding to a $p$-value lower than 0.05 are shown in bold and indicate that, prior to 2011, annual changes in age specific log mortality rates are significantly different from zero. In Table \ref{NBBeta2Male}, bold values of $\widehat{\beta}_{2,x}^{[c]}$ indicate that the post-2011 changes in log mortality rates are significantly different from those for the pre-2011 period.\\

\begin{table}[h!]
	\centering
	\caption{Estimated $\widehat{\beta}_{1,x}^{[c]}$ from (\ref{disMT}) for first six cause-of-death groups - Males. 
	Coefficients corresponding to a $p$-value lower than 0.05 are shown in bold.}
	\begin{tabular}{|c|c|c|c|c|c|c|}
		\hline
		{\ul \textbf{Age Group}} & {\ul \textbf{CAN}} & {\ul \textbf{CIR}} & {\ul \textbf{DIG}} & {\ul \textbf{END}} & {\ul \textbf{EXT}} & {\ul \textbf{GEN}} \\ \hline
		\textbf{\textless{}1}    & -0.04040          & -0.03195           & \textbf{-0.06766}           & 0.00892            & \textbf{-0.08009}           & 0.09547            \\ \hline
		\textbf{1-4}             & \textbf{-0.04484}           & 0.00786            & -0.06138           & -0.00353           & \textbf{-0.07282}           & 0.02498            \\ \hline
		\textbf{5-9}             & -0.01537           & -0.02001           & 0.01162           & -0.00548           & \textbf{-0.08443}           & 0.02120           \\ \hline
		\textbf{10-14}           & \textbf{-0.05579}           & -0.03778           & -0.01826           & \textbf{-0.06334}           & \textbf{-0.07895}           & 0.00554            \\ \hline
		\textbf{15-19}           & -0.02259           & -0.01323           & -0.00460           & -0.02822           & \textbf{-0.05853}           & -0.05174           \\ \hline
		\textbf{20-24}           & -0.02651           & -0.02468           & -0.01862           & -0.01391           & \textbf{-0.04325}           & -0.02451           \\ \hline
		\textbf{25-29}           & -0.00237           & -0.01852           & -0.00250           & 0.01604            & \textbf{-0.04744}           & -0.03468           \\ \hline
		\textbf{30-34}           & \textbf{-0.01804}           & -0.01284           & 0.01462            & \textbf{-0.03659}           & \textbf{-0.03049}           & -0.04101           \\ \hline
		\textbf{35-39}           & \textbf{-0.01402}           & -0.00027           & 0.01341            & 0.00610            & \textbf{-0.01191}           & -0.01041           \\ \hline
		\textbf{40-44}           & \textbf{-0.01760}           & \textbf{-0.02794}           & \textbf{0.01106}            & 0.01108            & 0.00251            & 0.00861            \\ \hline
		\textbf{45-49}           & \textbf{-0.02937}           & \textbf{-0.04125}           & -0.00453           & -0.00158           & -0.00635           & -0.03404           \\ \hline
		\textbf{50-54}           & \textbf{-0.02733}           & \textbf{-0.03992}           & 0.00431           & \textbf{-0.01993}           & 0.00368            & 0.00053            \\ \hline
		\textbf{55-59}           & \textbf{-0.01794}           & \textbf{-0.04406}           & \textbf{0.01945}          & -0.00833          & 0.00764            & 0.00111            \\ \hline
		\textbf{60-64}           & \textbf{-0.02251}           & \textbf{-0.06339}           & \textbf{0.00779}            & \textbf{-0.06517}           & -0.00282           & -0.00926           \\ \hline
		\textbf{65-69}           & \textbf{-0.01539}           & \textbf{-0.06811}           & -0.00001           & \textbf{-0.06339}           & -0.00610           & \textbf{-0.02395}           \\ \hline
		\textbf{70-74}           & \textbf{-0.01891}           & \textbf{-0.07382}           & \textbf{-0.01966}           & \textbf{-0.05428}           & \textbf{-0.01727}           & -0.00813           \\ \hline
		\textbf{75-79}           & \textbf{-0.01970}           & \textbf{-0.06969}           & \textbf{-0.02090}           & \textbf{-0.03103}           & \textbf{-0.00911}           & 0.00350            \\ \hline
		\textbf{80-84}           & \textbf{-0.00669}           & \textbf{-0.05397}           & \textbf{-0.01011}          & \textbf{-0.02838}           & 0.01103           & \textbf{0.02510}        \\ \hline
		\textbf{85+}             & \textbf{-0.00476}           & \textbf{-0.03784}           & \textbf{-0.00567}           & \textbf{-0.01477}           & \textbf{0.00929}            & \textbf{0.03331}           \\ \hline
	\end{tabular}
	\label{NBBeta1Male}
\end{table}

\begin{table}[h!]
	\centering
	\caption{Estimated $\widehat{\beta}_{1,x}^{[c]}$ from (\ref{disMT}) for last six cause-of-death groups - Males. Coefficients corresponding to a $p$-value lower than 0.05 are shown in bold.}
	\begin{tabular}{|c|c|c|c|c|c|c|}
		\hline
		{\ul \textbf{Age Group}} & {\ul \textbf{INF}} & {\ul \textbf{MEN}} & {\ul \textbf{MUS}} & {\ul \textbf{NER}} & {\ul \textbf{RES}} & {\ul \textbf{OTH}} \\ \hline
		\textbf{\textless{}1}    & \textbf{-0.04232}           & 23.18517           & -0.00282          & \textbf{-0.04369}           & \textbf{-0.05569}           & \textbf{-0.02031}           \\ \hline
		\textbf{1-4}             & -0.03772           & -0.00032           & 0.02178            & \textbf{-0.06787}           & 0.00079            & -0.01037           \\ \hline
		\textbf{5-9}             & -0.00099          & -0.14912          & -0.20206           & -0.00260           & 0.03701            & 0.04519            \\ \hline
		\textbf{10-14}           & -0.05483           & \textbf{-0.45634}           & -0.03509           & \textbf{-0.04025}           & 0.00081           & -0.00314          \\ \hline
		\textbf{15-19}           & -0.05455           & \textbf{-0.12954}           & -0.08361           & \textbf{-0.04244}           & -0.01346           & \textbf{0.06693}            \\ \hline
		\textbf{20-24}           & \textbf{-0.08042}           & \textbf{-0.12443}           & 0.04741            & 0.00351            & -0.01945           & \textbf{0.07076}            \\ \hline
		\textbf{25-29}           & \textbf{-0.08236}           & \textbf{-0.08874}           & 0.08841            & -0.02046           & -0.03857           & \textbf{0.06587}            \\ \hline
		\textbf{30-34}           & \textbf{-0.06390}           & \textbf{-0.02924}           & -0.00384           & \textbf{-0.04793}           & -0.01570           & \textbf{0.03498}            \\ \hline
		\textbf{35-39}           & \textbf{-0.03187}           & \textbf{0.02743}            & 0.06015            & \textbf{-0.02976}           & -0.02228           & \textbf{0.04735}            \\ \hline
		\textbf{40-44}           & -0.01432           & \textbf{0.04440}            & 0.03011            & -0.01606           & 0.00965            & \textbf{0.04908}            \\ \hline
		\textbf{45-49}           & 0.00043            & 0.02052           & -0.01527           & -0.00972           & -0.01567           & \textbf{0.03879}            \\ \hline
		\textbf{50-54}           & 0.01887            & \textbf{0.02538}            & -0.01336           & -0.00223          & -0.01657           & \textbf{0.04000}            \\ \hline
		\textbf{55-59}           & \textbf{0.03516}            & \textbf{0.04440}            & 0.02362           & 0.00863          & 0.00025            & \textbf{0.04378}            \\ \hline
		\textbf{60-64}           & \textbf{0.02830}            & 0.02375            & \textbf{-0.03305}           & 0.00999            & \textbf{-0.01747}           & 0.01535            \\ \hline
		\textbf{65-69}           & 0.01149            & \textbf{0.02317}            & \textbf{-0.01790}           & \textbf{0.00396}            & -0.01417           & 0.01528            \\ \hline
		\textbf{70-74}           & 0.01055            & 0.00347            & \textbf{-0.01764}           & 0.00751           & \textbf{-0.02872}           & -0.00674           \\ \hline
		\textbf{75-79}           & 0.00413            & -0.00277           & -0.00357           & 0.00897            & \textbf{-0.03914}           & \textbf{-0.08738}           \\ \hline
		\textbf{80-84}           & 0.03075            & \textbf{0.01707}            & 0.00822            & \textbf{0.01916}            & \textbf{-0.02347}           & \textbf{-0.09152}           \\ \hline
		\textbf{85+}             & \textbf{0.04506}            & \textbf{0.01729}            & \textbf{-0.00883}           & \textbf{0.01217}            & \textbf{-0.02315}           & \textbf{-0.06958}           \\ \hline
	\end{tabular}
	\label{NBBeta1MaleB}
\end{table}

\begin{table}[h!]
	\centering
	\caption{Estimated $\widehat{\beta}_{2,x}^{[c]}$ from (\ref{disMT}) for first six cause-of-death groups - Males. Coefficients corresponding to a $p$-value lower than 0.05 are shown in bold.}
	\begin{tabular}{|c|c|c|c|c|c|c|}
		\hline
		{\ul \textbf{Age Group}} & {\ul \textbf{CAN}} & {\ul \textbf{CIR}} & {\ul \textbf{DIG}} & {\ul \textbf{END}} & {\ul \textbf{EXT}} & {\ul \textbf{GEN}} \\ \hline
		\textbf{\textless{}1}    & 0.05009            & -0.02124          & 0.05769            & -0.01388           & 0.05873            & -0.10970           \\ \hline
		\textbf{1-4}             & 0.04655            & -0.06232           & -0.11642           & 0.00606           & 0.01302            & -0.05981           \\ \hline
		\textbf{5-9}             & -0.01787           & 0.02239            & -0.10776           & -0.01449           & 0.00190            & 0.10958           \\ \hline
		\textbf{10-14}           & 0.04057            & 0.05643            & 0.08011            & 0.09166            & \textbf{0.08868}            & 0.20116            \\ \hline
		\textbf{15-19}           & \textbf{0.04615}            & 0.00659            & -0.05207           & 0.08599            & 0.06598            & 0.21546           \\ \hline
		\textbf{20-24}           & 0.01259            & 0.00924            & 0.00991            & 0.02744            & \textbf{0.05149}            & 0.04507            \\ \hline
		\textbf{25-29}           & -0.02017           & 0.00755            & -0.04515           & 0.01084            & \textbf{0.06002}            & 0.00678            \\ \hline
		\textbf{30-34}           & 0.00652            & 0.01366            & -0.02572           & 0.05556            & \textbf{0.05530}            & 0.04247            \\ \hline
		\textbf{35-39}           & \textbf{0.02313}            & -0.01589           & \textbf{-0.03161}           & \textbf{0.05062}            & \textbf{0.02427}            & 0.01300            \\ \hline
		\textbf{40-44}           & \textbf{0.01725}            & \textbf{0.02946}            & \textbf{-0.03953}           & 0.03298            & \textbf{0.02288}            & -0.06199           \\ \hline
		\textbf{45-49}           & \textbf{0.02275}            & \textbf{0.04723}            & 0.00718           & \textbf{0.07502}            & \textbf{0.04275}            & 0.01292            \\ \hline
		\textbf{50-54}           & \textbf{0.01859}            & \textbf{0.03735}            & -0.00637           & \textbf{0.09225}            & \textbf{0.01946}            & -0.02818           \\ \hline
		\textbf{55-59}           & \textbf{-0.00748}           & \textbf{0.03564}            & \textbf{-0.01867}           & \textbf{0.05754}            & 0.00799            & -0.03350           \\ \hline
		\textbf{60-64}           & 0.00284            & \textbf{0.05679}            & -0.00585          & \textbf{0.11517}            & \textbf{0.03443}            & -0.01898           \\ \hline
		\textbf{65-69}           & 0.00237            & \textbf{0.05710}            & 0.00969            & \textbf{0.10821}            & \textbf{0.05044}            & \textbf{0.00942}            \\ \hline
		\textbf{70-74}           & \textbf{-0.00542}           & \textbf{0.03910}            & \textbf{0.01467}            & \textbf{0.07891}            & \textbf{0.03672}            & -0.01303           \\ \hline
		\textbf{75-79}           & \textbf{0.00873}            & \textbf{0.03873}            & \textbf{0.01667}            & \textbf{0.04644}            & \textbf{0.05015}            & -0.02665           \\ \hline
		\textbf{80-84}           & \textbf{-0.00860}           & \textbf{0.01932}            & -0.00899           & \textbf{0.05023}            & 0.01810            & \textbf{-0.05550}           \\ \hline
		\textbf{85+}             & 0.00001            & \textbf{0.00973}            & -0.00031           & \textbf{0.05046}            & \textbf{0.02733}            & \textbf{-0.05856}           \\ \hline
	\end{tabular}
	\label{NBBeta2Male}
\end{table}

\begin{table}[h!]
	\centering
	\caption{Estimated $\widehat{\beta}_{2,x}^{[c]}$ from (\ref{disMT}) for last six cause-of-death groups - Males. Coefficients corresponding to a $p$-value lower than 0.05 are shown in bold.}
	\label{NBBeta2MalesB}
	\begin{tabular}{|c|c|c|c|c|c|c|}
		\hline
		{\ul \textbf{Age Group}} & {\ul \textbf{INF}} & {\ul \textbf{MEN}} & {\ul \textbf{MUS}} & {\ul \textbf{NER}} & {\ul \textbf{RES}} & {\ul \textbf{OTH}} \\ \hline
		\textbf{\textless{}1}    & -0.06444           & -23.17549          & -0.24279           & -0.02255           & \textbf{-0.08262}           & 0.00511            \\ \hline
		\textbf{1-4}             & -0.01096           & 0.16265            & \textbf{-0.47263}           & 0.03672            & -0.05541           & \textbf{-0.05031}           \\ \hline
		\textbf{5-9}             & -0.05689           & -0.18725           & -0.34752           & -0.08081           & -0.01595           & -0.05590           \\ \hline
		\textbf{10-14}           & 0.11564            & 0.28702            & 0.01731            & -0.01553           & 0.03530            & -0.03677           \\ \hline
		\textbf{15-19}           & 0.10938           & -0.01162           & -0.10104           & 0.00890            & 0.05167            & \textbf{-0.10147}           \\ \hline
		\textbf{20-24}           & 0.10767            & 0.18453            & \textbf{-0.22248}           & -0.04152           & \textbf{0.12321}            & \textbf{-0.07936}           \\ \hline
		\textbf{25-29}           & -0.02001           & 0.05339            & -0.18207           & 0.02802            & 0.04637            & \textbf{-0.07868}           \\ \hline
		\textbf{30-34}           & 0.01047            & \textbf{0.00260}            & -0.05604           & 0.04358            & 0.04180            & \textbf{-0.07678}           \\ \hline
		\textbf{35-39}           & -0.05846           & -0.04575           & -0.11285           & -0.01884           & 0.02798           & \textbf{-0.08496}           \\ \hline
		\textbf{40-44}           & 0.01582            & 0.01394            & -0.02331           & 0.00287            & 0.00612            & \textbf{-0.07085}           \\ \hline
		\textbf{45-49}           & -0.03006           & -0.02492           & 0.01447            & -0.01487           & \textbf{0.05339}            & -0.00318           \\ \hline
		\textbf{50-54}           & \textbf{-0.03759}           & 0.00077            & 0.01432            & -0.01481           & \textbf{0.04612}            & -0.00713           \\ \hline
		\textbf{55-59}           & -0.03309          & \textbf{-0.06366}           & -0.03311           & 0.00525           & 0.01394            & 0.01689            \\ \hline
		\textbf{60-64}           & -0.02333          & -0.00918          & 0.02946            & -0.00816          & \textbf{0.02707}            & \textbf{0.03406}            \\ \hline
		\textbf{65-69}           & 0.00447            & 0.01787            & \textbf{0.04223}            & \textbf{0.03024}            & \textbf{0.03386}            & 0.02394            \\ \hline
		\textbf{70-74}           & -0.01051           & \textbf{0.03024}            & 0.00942            & \textbf{0.03047}            & \textbf{0.02170}            & \textbf{0.06258}            \\ \hline
		\textbf{75-79}           & -0.00626           & \textbf{0.05654}            & 0.00641            & \textbf{0.04688}            & \textbf{0.04320}            & \textbf{0.13912}            \\ \hline
		\textbf{80-84}           & -0.02629           & \textbf{0.03292}            & \textbf{-0.02758}           & \textbf{0.04870}            & 0.00595            & \textbf{0.14166}            \\ \hline
		\textbf{85+}             & -0.05274           & \textbf{0.04674}            & -0.00844           & \textbf{0.08389}            & 0.00569            & \textbf{0.11150}            \\ \hline
	\end{tabular}
	\label{NBBeta2MaleB}
\end{table}

In Table \ref{NBBeta1Male}, negative estimates for $\widehat{\beta}_{1,x}^{[c]}$ indicate that  age- and cause-specific log mortality rates are improving during the period from 2001 to 2010. The estimates in Table \ref{NBBeta2Male} show the post-2011 differences in the annual changes in log mortality rates, $\widehat{\beta}_{2,x}^{[c]}$, compared to those estimated for $\widehat{\beta}_{1,x}^{[c]}$ for the pre-2011 period. While $\widehat{\beta}_{2,x}^{[c]}$ may be positive, this does not necessarily mean that the log mortality rate is worsening after 2011. For example, $\widehat{\beta}_{1,50-54}^{[CAN]}$ = $-$0.02733 and $\widehat{\beta}_{2,50-54}^{[CAN]}$ = 0.01859. Together, $\widehat{\beta}_{1,50-54}^{[CAN]}$ + $\widehat{\beta}_{2,50-54}^{[CAN]}$ = $-$0.00874 is negative, indicating a positive improvement in the log mortality rates after 2011, albeit not as strong as from 2001 to 2010. It is important to note that the improvement or worsening of rates is estimated on the logarithmic scale. As the fitted all-cause mortality rate in (\ref{MTDIS}) is a summation of fitted cause-specific mortality rates, there is a non-linear relationship between the fitted all-cause and cause-specific mortality rates. Therefore, it is not straightforward to directly compare the effects on all-cause mortality trends from these estimates across different age and cause combinations.\\ 

\begin{figure}[h!]
	\begin{multicols}{2}
		\includegraphics[width=\linewidth,height=8cm]{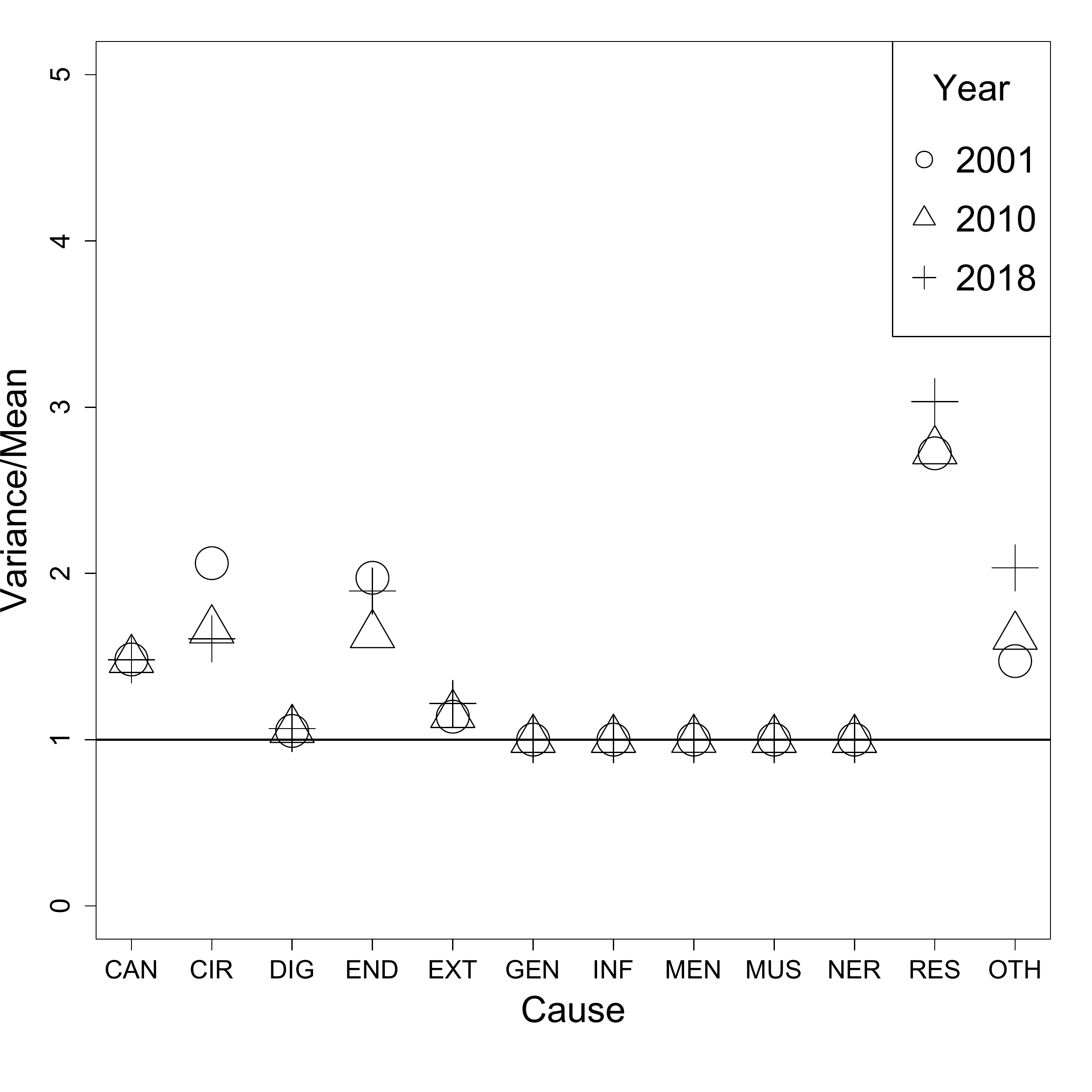}
		\subcaption{Males}
		\includegraphics[width=\linewidth,height=8cm]{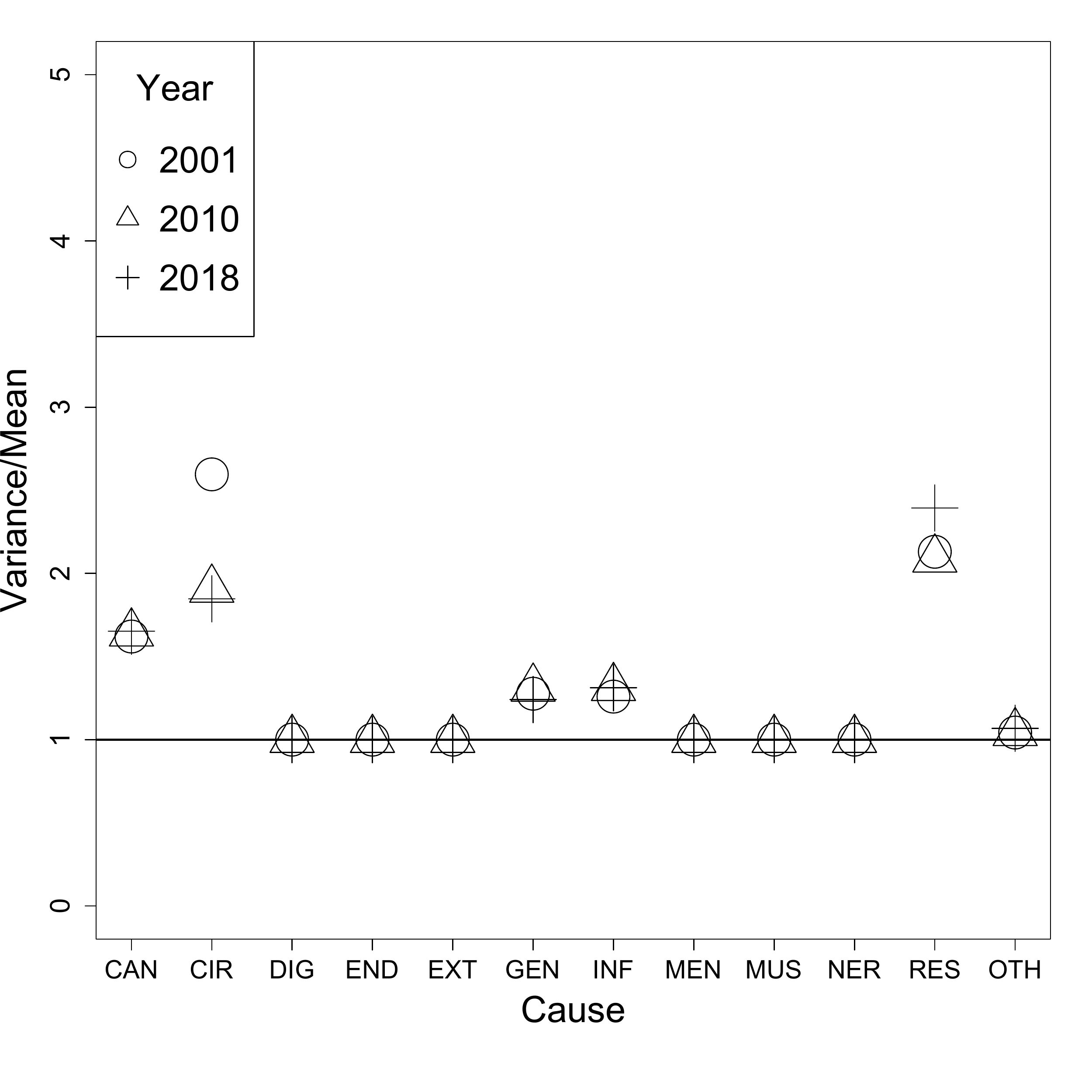}
		\subcaption{Females}
	\end{multicols}		
	\caption{Variance divided by mean for fitted death counts by gender for ages 65 - 69}
	\label{NBVarPlots65}
\end{figure}

\begin{figure}[h!]
	\begin{multicols}{2}
		\includegraphics[width=\linewidth,height=8cm]{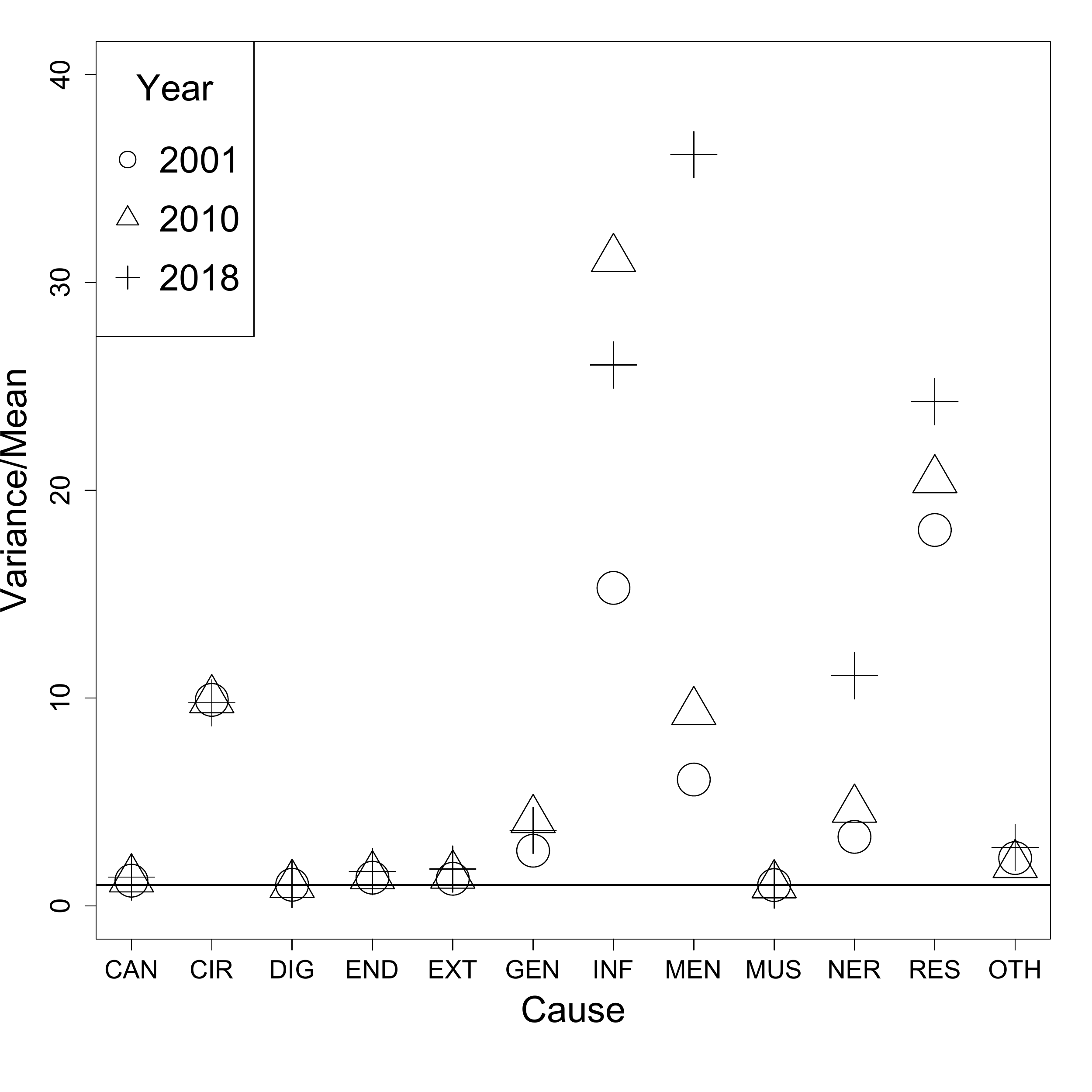}
		\subcaption{Males}
		\includegraphics[width=\linewidth,height=8cm]{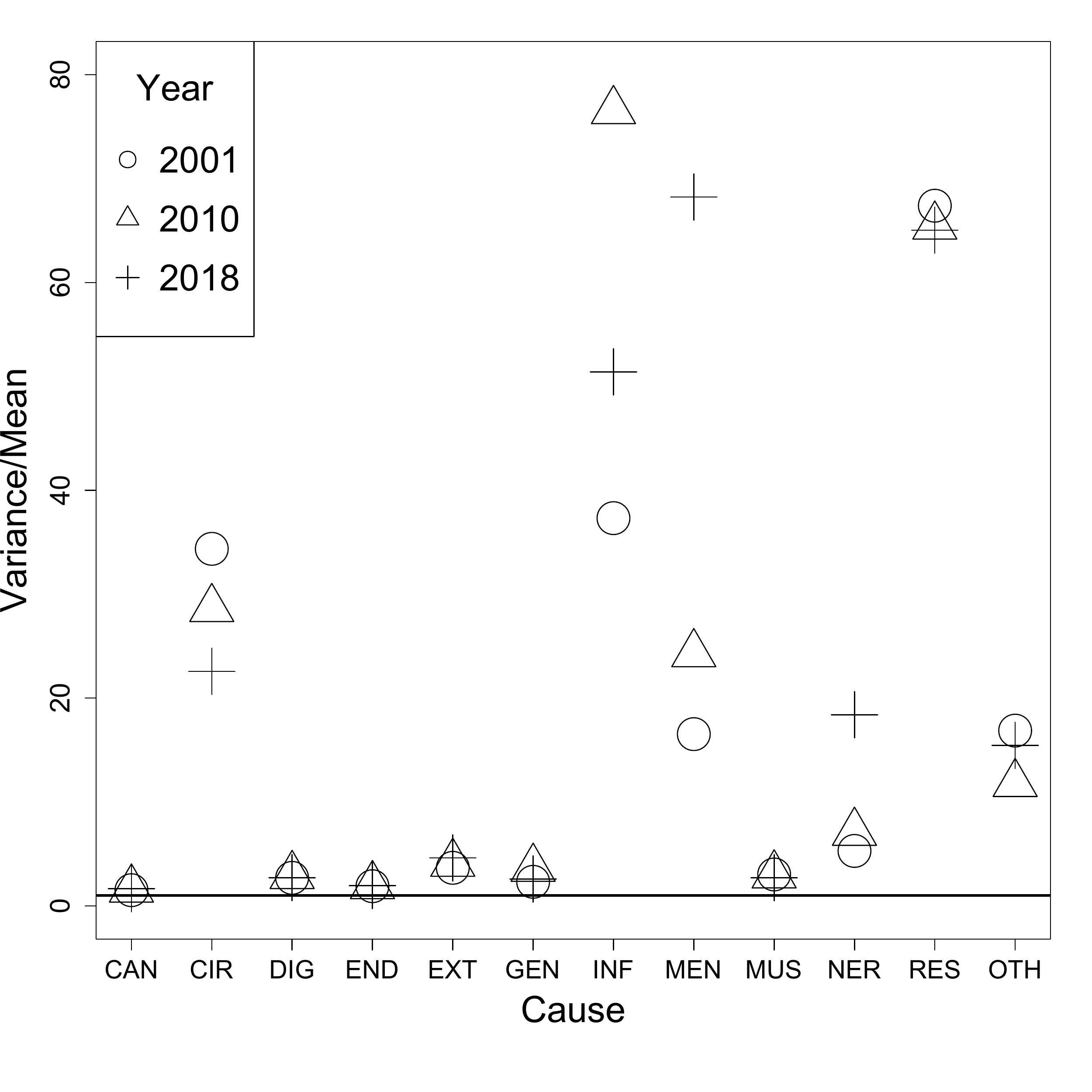}
		\subcaption{Females}
	\end{multicols}		
	\caption{Variance divided by mean for fitted death counts by gender for ages 85+}
	\label{NBVarPlots}
\end{figure}

Returning to the assumption of a negative binomial model for the numbers of deaths, it is noted here that (\ref{NBMean}) and (\ref{NBVar}) give
\begin{equation}
	\frac{\text{Var}(D_{x,t}^{[c]})}{\textbf{E}(D_{x,t}^{[c]})} = 1 + \frac{E_{x,t}m_{x,t}^{[c]}}{\theta_{x}^{[c]}},
	\label{NBvarMean}
\end{equation} 
for death counts following a negative binomial distribution.
Large values of the dispersion parameter $\theta_{x}^{[c]}$, relative to $E_{x,t}m_{x,t}^{[c]}$, result in the ratio in (\ref{NBvarMean}) approaching unity, which is the value of the ratio under the Poisson distribution.
 Using fitted death counts and estimated dispersion parameters from (\ref{disMT2}), we can compare the values of (\ref{NBvarMean}) against the baseline value of 1. Values that are close to 1 indicate lower levels of overdispersion, suggesting that the mean variance relationship of the Poisson distribution may be reasonable, whereas higher values suggest that overdispersion is present and therefore the negative binomial distribution is chosen over the Poisson distribution.
Figure \ref{NBVarPlots65} displays values from (\ref{NBvarMean}) using fitted death counts for the years 2001, 2010, and 2018 represented by circles, triangles, and crosses respectively, provided for individuals aged 65 to 69. Figure \ref{NBVarPlots} displays similar information for individuals aged 85 and above. A line is drawn at the value 1 to act as a baseline for comparison purposes.
Figure \ref{NBVarPlots65} demonstrates that causes such as digestive system diseases (DIG) and genitourinary system diseases (GEN) have values close to 1 for each of the three years across both genders. This is a result of the estimated dispersion parameter being relatively high compared to the fitted death counts and implies that the level of overdispersion is not high for these causes of death and age group. For causes such as circulatory system diseases (CIR) and respiratory system diseases (INF), the estimated dispersion parameter is relatively low compared to the fitted death counts and this is shown by the higher values of the variance divided by the fitted counts. In the case of circulatory system diseases, the estimated variance in death counts for the three years varies from 1.5 to 2 multiplied by the expected number of deaths.\\

Similar to Figure \ref{NBVarPlots65}, Figure \ref{NBVarPlots} displays high values in the estimated variance divided by the expected death counts. In particular, for causes such as infectious diseases (INF) and mental and behavioural illnesses (MEN), the estimated number of deaths is considerably higher than the estimated dispersion parameter. This suggests that negative binomial distribution is more suited to model these death counts given the parametrisation in (\ref{disMT}). 
%Ultimately, the negative binomial is chosen to model each of the age and cause combinations due to the presence of overdispersion in death counts.

\subsection{Quantification of the contribution of a single cause}\label{quant}

Aiming to quantify the contribution of each cause of death towards the slowdown in mortality improvements, we introduce a general method for generating hypothetical mortality scenarios for years 2011 to 2018. In these scenarios, two related quantitative measures of mortality improvements are studied: the log age-standardised mortality rate (ASMR) in Section \ref{sec_ASMRscenarios};  and life expectancy in Section \ref{sec_life_expect}.\\

A mortality scenario is generated for each cause of death by replacing the post-2011 trend of an age-specific mortality rate, $m_{x,t}^{[c=k]}$ for a single cause of death $k$, with its trend observed before 2011. Rates for all other causes remain unadjusted. More specifically, for a considered cause $k$, we set the parameter $\beta_{2,x}^{[k]} = 0$ for all ages $x$, while all other parameters are set to their estimated values, $\widehat{\beta}_{i,x}^{[c]}$, $i$ = 0, 1, and 3.
Therefore, for a scenario specific to cause $k$, mortality rates are determined as
\begin{equation} \label{mortproj_1}
\log \tilde{m}_{x,t}^{[c,k]}  =
\left\{ \begin{array}{ll}
\widehat{\beta}_{0,x}^{[k]} + \widehat{\beta}_{1,x}^{[k]} t 
+ \widehat{\beta}_{3,x}^{[k]} \text{I}(t \ge 2011) &, c = k,\\
\log \widehat{m}_{x,t}^{[c]} &, c \neq k,
\end{array} \right.
\end{equation}
where the estimated values $\widehat{\beta}$ are obtained by fitting the model in (\ref{disMT}). The $\widehat{\beta}_{3,x}^{[k]}$ term is kept in (\ref{mortproj_1}) in order to preserve the effect of the change in cause-of-death classification which occurs in 2011.
The all-cause mortality rates with adjustments for cause $k$ are then calculated as usual, compare with (\ref{MTDIS}),
\begin{equation}
\tilde{m}_{x,t}^{[\cdot,k]} = \sum\limits_{c} \tilde{m}_{x,t}^{[c,k]} =
 \tilde{m}_{x,t}^{[k,k]} + \sum\limits_{c \ne k} \widehat{m}_{x,t}^{[c]},
\label{causeScen}
\end{equation}
where the adjusted cause $k$ is explicitly included in the notation.
This gives mortality rates relating to scenarios where a trend of an individual cause $k$ had not changed after 2011. Comparing all-cause mortality rates, $\tilde{m}_{x,t}^{[\cdot,k]}$, in each of those hypothetical mortality scenarios with the unadjusted all-cause mortality, $\widehat{m}_{x,t}$, enables the computation of the contribution of the changing trend in a single cause to the overall slowdown in mortality improvement rates. \\

When fitting these mortality scenarios, certain adjustments are made aiming to avoid unrealistic post-2011 mortality trends in cases where there are zero, or very low, numbers of deaths as a result of the mortality risk being practically zero for certain combinations of age and cause of death. 
%In particular, for certain causes of death, there are zero or very low numbers of deaths observed for some age groups during the entire period of study. 
An extreme example is the case of zero deaths below the age of 1 attributed to mental and behavioural illnesses.
%, for the majority of the years from 2001 to 2018 in England and Wales. 
This results in extreme estimated values $\widehat{\beta}_{1,x}^{[c]}$, which are problematic for later analyses, where age-specific mortality rates for a certain cause of death are reverted by setting $\widehat{\beta}_{2,x}^{[c]} = 0$, such that pre-2011 trends are set to persist past 2011. For this reason, we adopt a pragmatic assumption, where for cause-age combinations with average annual deaths count during 2001-2018 being less than one, the fitted number of deaths  are set to 0 for the entire period, i.e., $\widehat{m}_{x,t}^{[c]} = 0$ for all $t$.
While this assumption may result in the underprediction of mortality from a certain cause $c$, the observed low number of deaths from that cause means that the fitted mortality rate $\widehat{m}_{x,t}^{[c]}$ is very close to zero. For the calculation of $\widehat{m}_{x,t}$ in (\ref{MTDIS}), those rates are therefore of minor importance. 
This assumption is particularly relevant when applying bootstrapping for simulating death counts in Sections \ref{sec_ASMRscenarios} and \ref{sec_life_expect}, where it is important to avoid simulating data for age-cause combinations with negligible mortality risk, thus also avoiding associated uncertainty.

\subsection{Scenario specific ASMRs} \label{sec_ASMRscenarios}

In this section, the following scenarios and age-standardised mortality rates are considered:

\begin{itemize}
	\item \textbf{Unadjusted rates scenario}
	
	In this scenario we do not consider any adjustments to rates and the ASMRs are given as
	\begin{equation}
		\widehat{ASMR}_t^{[Obs]} = \frac{\sum\limits_x \widehat{m}_{x,t} E^S_{x}}{\sum\limits_x E^S_{x}},
		\label{ASMRfit}
	\end{equation}
	where $\widehat{m}_{x,t}$ are the fitted rates obtained by (\ref{disMT}) and (\ref{MTDIS}), and $E^S_x$ refers to the European Standard Population in 2013 \citep{eurostat2013revision}. Also, we let $w^{[Obs]}$ be the average rate of improvement in $\log \widehat{ASMR}_t^{[Obs]}$, from 2011 to 2018 under the Unadjusted rates scenario. We obtain 
	$w^{[Obs]}$ by fitting a linear regression on $\log \widehat{ASMR}_t^{[Obs]}$ for $t = 2011, 2012, ..., 2018$ such that 
	\begin{equation*}
		\log \widehat{ASMR}_t^{[Obs]} = w_0 + w^{[Obs]} t + \epsilon_{t}.
	\end{equation*} 
    For interpretation purposes, we present $w^{[Obs]}$ multiplied by a factor of $-$1, so that a negative change in log $ASMR^{[Obs]}$ is interpreted as improvement in mortality.
	%$w^{[Obs]}$ can also be defined as the slope of the fitted regression line on $\log \widehat{ASMR}_t^{[Obs]}$ for the years 2011 through 2018 multiplied by a factor of $-$1. A negative change in the log ASMRs is considered a positive improvement. 
	
	\newpage
	
	\item \textbf{Cause-$k$-adjusted scenario}
	
	Under this scenario, rates for a single cause $k$ are adjusted to their pre-2011 levels, with rates for all remaining causes being unadjusted. The ASMRs are given as
	\begin{equation}
		\widehat{ASMR}_t^{[k]} = \frac{\sum\limits_x \tilde{m}_{x,t}^{[\cdot,k]} E^S_{x}}{\sum\limits_x E^S_{x}},
		\label{ASMRscenario}
	\end{equation}
	where $\tilde{m}_{x,t}^{[\cdot,k]}$ are scenario-specific rates for a single adjusted cause $k$, obtained by (\ref{mortproj_1}) and (\ref{causeScen}). $w^{[k]}$ represents the average rate of improvement in cause-$k$ adjusted scenario rates $\log \widehat{ASMR}_t^{[k]}$, and is obtained in a similar manner to $w^{[Obs]}$ in the Unadjusted rates scenario.
	
	\item \textbf{All-causes-adjusted scenario}
	
	We now assume that post-2011 trends are reverted to pre-2011 trends for all causes of death.
	The ASMRs are given as
	\begin{equation}
		\widehat{ASMR}_t^{[\cdot]} = \frac{\sum\limits_x \tilde{m}_{x,t}^{[\cdot]} E^S_{x}}{\sum\limits_x E^S_{x}},
		\label{trendscen}
	\end{equation}
	where the mortality rates $\tilde{m}_{x,t}^{[\cdot]}$ are a sum of scenario cause-specific mortality rates $\tilde{m}^{[c]}_{x,t}$, i.e., $\tilde{m}_{x,t}^{[\cdot]} = \sum\limits_c \tilde{m}^{[c]}_{x,t}$, where 
	\begin{equation}
		\log \tilde{m}^{[c]}_{x,t}  = \widehat{\beta}_{0,x}^{[c]} + \widehat{\beta}_{1,x}^{[c]} t + \widehat{\beta}_{3,x}^{[c]} \text{I}(t \ge 2011),
		\label{Unchanged}	
	\end{equation}
	for all causes $c$. Under this scenario,  $w^{[\cdot]}$ represents the average rate of improvement from 2011 to 2018 in $\log \widehat{ASMR}_t^{[\cdot]}$ and is determined similarly to $w^{[Obs]}$ and $w^{[k]}$ in the previous two scenarios.
\end{itemize}

\begin{figure}[h!]
	\centering
	\includegraphics[width=10cm,height=10cm]{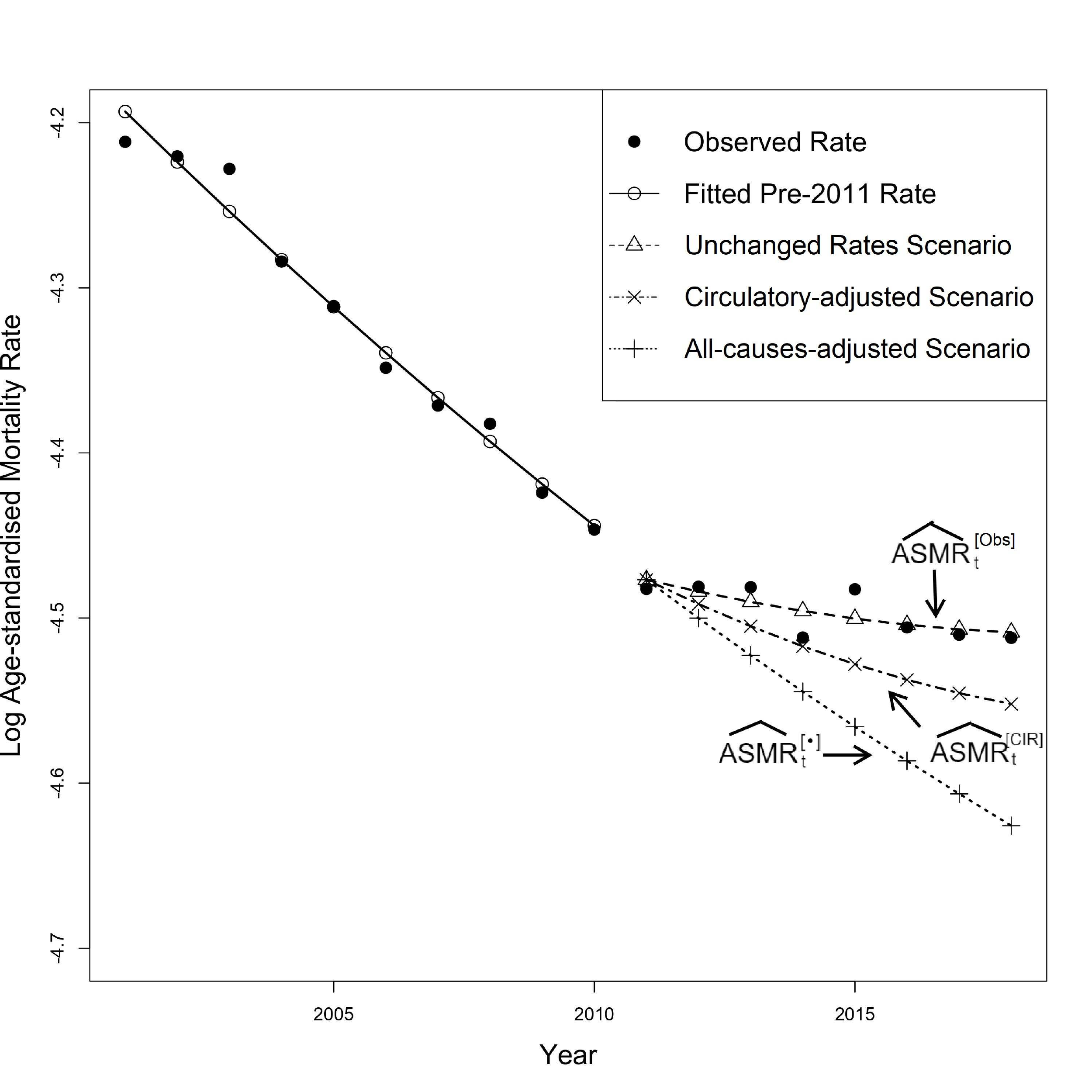}
	\caption{Fitted log age-standardised mortality rates under various scenarios - Males}
	\label{CIRscen}
\end{figure}

Figure \ref{CIRscen} shows the ASMRs for the cause-$k$-adjusted scenario, with cause $k$ being circulatory disease. For comparison, the rates corresponding to observed mortality after 2011 is also included, as well as the rates corresponding to the all-causes-adjusted scenario(\ref{Unchanged}). Figure \ref{CIRscen} illustrates that the changing trend in mortality rates from circulatory disease had a significant impact on the overall slowdown in mortality, as shown by $w^{[CIR]}$ being considerably greater than $w^{[Obs]}$.
\\

To quantify the uncertainty around the values of annual improvement rates, $w^{[k]}$, we have estimated their distributions through bootstrap methodology. The bootstrap procedure is presented in Section \ref{boot} in the Appendix. Figure \ref{ASMRComp} displays the distributions of $w^{[k]}$, obtained under Cause-$k$-adjusted scenarios for the following causes: cancer, circulatory system diseases, external causes of mortality, mental and behavioural illnesses, nervous system diseases, and respiratory system diseases. For comparison, the annual improvement rate for the Unadjusted rates scenario, $w^{[Obs]}$, is also shown as a solid line. 
Numerical results supporting Figure \ref{ASMRComp} and the remaining causes of death, can be found in Tables \ref{MaleNBASMR} to \ref{FemaleNBASMR} in the Appendix.\\

\begin{figure}[h!]
	\begin{multicols}{2}
		\includegraphics[width=\linewidth,height=8cm]{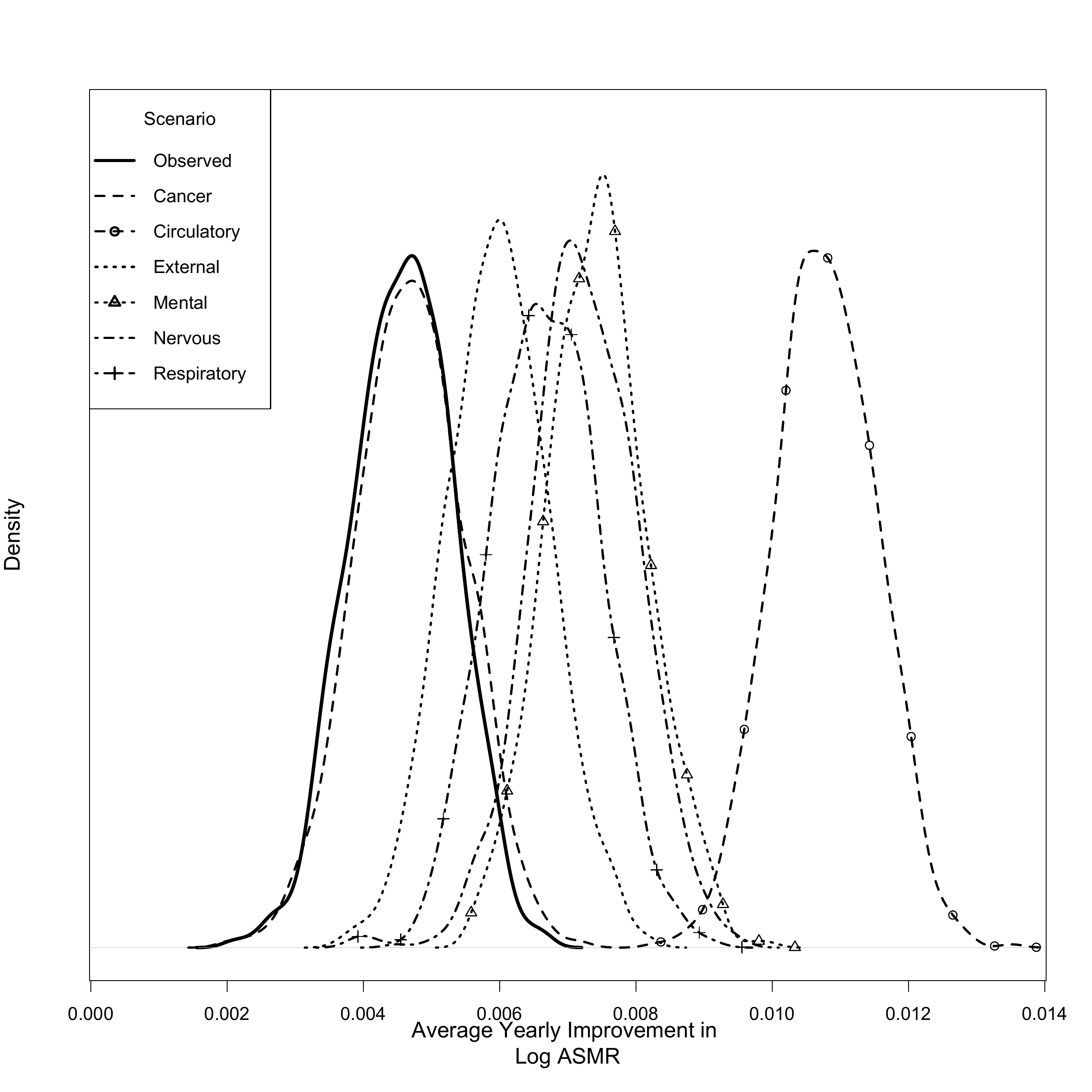}
		\subcaption{Males}
		\includegraphics[width=\linewidth,height=8cm]{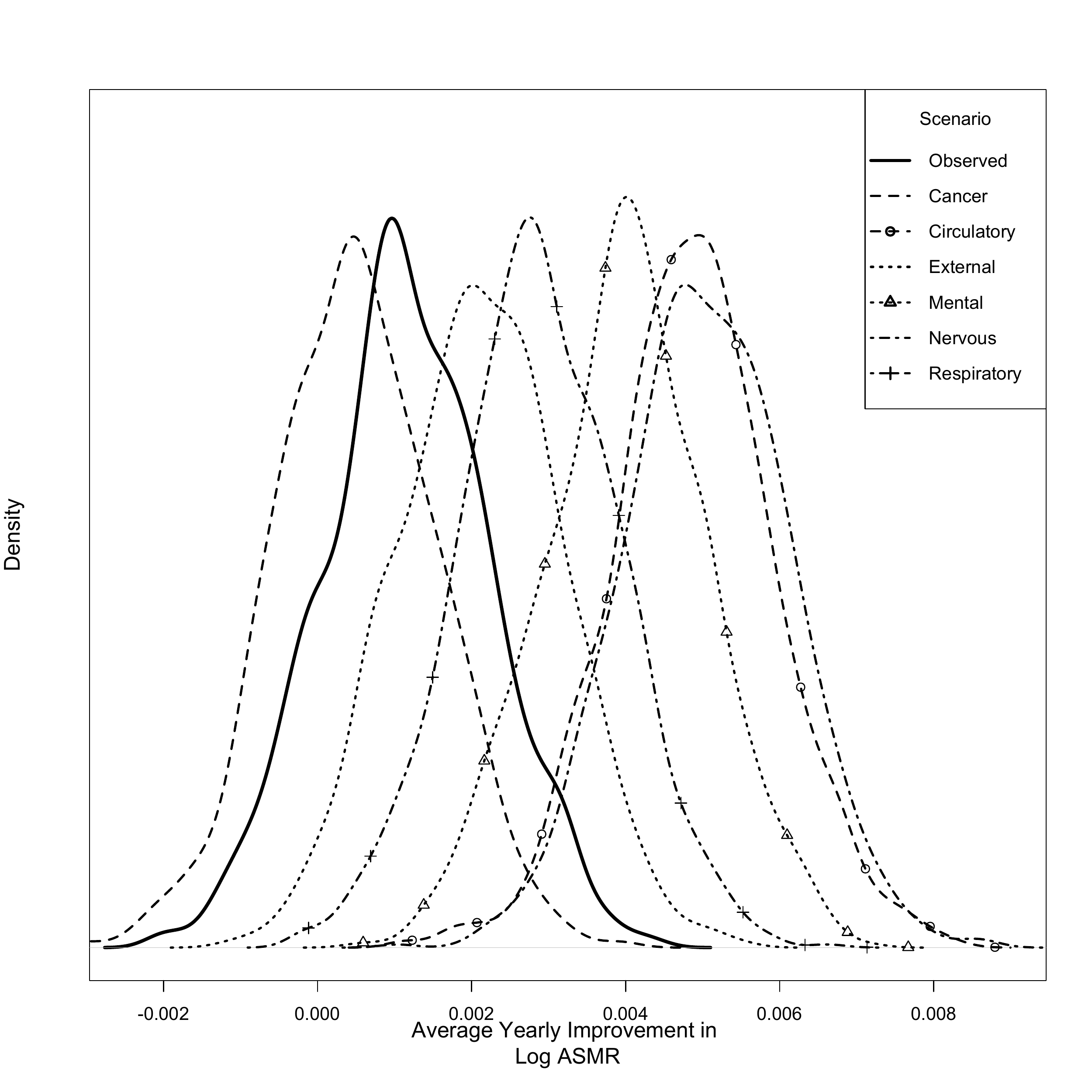}
		\subcaption{Females}
	\end{multicols}		
	\caption{Average yearly improvement in log age-standardised mortality rates post-2011, $w^{[k]}$, in cause reversion scenarios by gender; $w^{[Obs]}$ also shown.}
	\label{ASMRComp}
\end{figure} 

Figure \ref{ASMRComp} shows that the changing trend in the mortality from circulatory system diseases has a very significant impact on the slowdown of overall rates for both sexes: while the actual annual improvement in mortality rates for men is $w^{[Obs]}$ = 0.00457 (95$\%$ CI: 0.00316, 0.00596), it would have been $w^{[CIR]}$ = 0.01078 (0.00937, 0.01219) if the temporal trends in the circulatory system deaths did not change after 2011. For females, the annual improvement would have been  $w^{[CIR]}$ = 0.00485 (0.00284, 0.00707), compared to the actual improvement of $w^{[Obs]}$ = $0.00107$  $(-0.00100, 0.00325)$.\\

The contribution of an individual cause is measured by comparing the post-2011 improvement rate of the three scenarios outlined previously. This is similar to the methodology employed by \cite{victora2017contribution}, where the contribution of a single source to the total change was computed as the difference between the total change and the change once the source was removed. We determine the cause-specific contribution to the slowdown in mortality rate improvements for a cause $k$, $\rho^{[k]}$, as

\begin{equation}
	\rho^{[k]} = \frac{w^{[k]} - w^{[Obs]}}{w^{[\cdot]} - w^{[Obs]}}.
	\label{contributionratio}
\end{equation}   

Given that $w^{[\cdot]} > w^{[Obs]}$, a positive value of $ \rho^{[k]}$ implies that the rate of improvement for a cause-$k$-adjusted scenario is greater than that under the Unadjusted rates scenario, i.e., $w^{[k]}$ $>$ $w^{[Obs]}$, and this denotes a contribution towards the slowdown in mortality rate improvements. Conversely, $\rho^{[k]} < 0 $ denotes a net contribution for cause $k$ towards the opposite direction, i.e., towards mortality improvement.
\\ 

Table \ref{ASMRContProp} summarises the cause-specific contributions, $\rho^{[k]}$,  to the slowdown in mortality rate improvements. The average rates of improvement, $w^{[k]}, w^{[Obs]},$ and $w^{[\cdot]}$, used in (\ref{contributionratio}) are given in Tables \ref{MaleNBASMR} and \ref{FemaleNBASMR} in the Appendix. Table \ref{ASMRContProp} shows that the post-2011 reduction of improvements in mortality from circulatory system diseases, with a 37.17$\%$ contribution, is  one of the major contributors to the overall slowdown in mortality for males. For females, the 28.76$\%$ contribution from circulatory system diseases is second only to nervous system diseases, which shows a 30.72$\%$ contribution. Table \ref{ASMRContProp} also reveals that the other major contributors to the overall slowdown in mortality improvements are deaths from mental diseases and diseases of the nervous system. \\

\begin{table}[h!]
	\centering
	\caption{Cause-specific contributions to slowdown in ASMR improvements, $\rho^{[k]}$, calculated using (\ref{contributionratio}).
	Note that the contributions do not add up to $100\%$, as a consequence of using the logarithmic scale for the analysis of the age-standardised mortality rates.}
	\begin{tabular}{|c|c|c|}
		\hline
		{\ul \textbf{Cause}} & {\ul \textbf{Male}} & {\ul \textbf{Female}} \\ \hline
		\textbf{Cancer}         & 0.68\%              & -4.43\%               \\ \hline
		\textbf{Circulatory}         & 37.17\%             & 28.76\%               \\ \hline
		\textbf{Digestive}         & -0.32\%             & -2.00\%               \\ \hline
		\textbf{Endocrine}         & 5.35\%              & 4.53\%                \\ \hline
		\textbf{External}         & 8.50\%              & 7.92\%                \\ \hline
		\textbf{Genitourinary}         & -5.49\%             & -13.99\%              \\ \hline
		\textbf{Infectious}         & -1.94\%             & -5.50\%               \\ \hline
		\textbf{Mental}         & 17.30\%             & 22.47\%               \\ \hline
		\textbf{Musculoskeletal}         & -0.27\%             & -0.31\%               \\ \hline
		\textbf{Nervous}         & 18.91\%             & 30.72\%               \\ \hline
		\textbf{Respiratory}         & 12.72\%             & 13.90\%               \\ \hline
		\textbf{Other}         & 5.87\%              & 14.91\%               \\ \hline
	\end{tabular}
	\label{ASMRContProp}
\end{table}

%A consequence of the choice of analysis on the logarithmic scale for age-standardised mortality rates, is that the exact cause-specific contributions to the overall difference between average rates of improvement of the All causes adjusted scenario and Unadjusted rates scenario cannot be directly obtained as the sum of the differences between the average rates of improvement of the Unadjusted rates scenario and the cause-specific scenarios. Therefore, the cause-specific contributions do not sum to 1.

\subsection{Cause-specific mortality rate trends and life expectancies} \label{sec_life_expect}

The impact of changing cause-specific mortality rates on period life expectancies is also investigated. In this study, life expectancy refers to the period complete expectation of life. The computations for period life expectancy are provided in Section \ref{lifedefinition} in the Appendix. We denote the period life expectancy, or expected remaining lifetime, for an individual aged $h$ in calendar year $t$ by $\overset{\circ}e_{h,t}$. The analysis of the contribution to the slowdown in improvements in life expectancies is similar to the analysis involving ASMRs. The average improvements in life expectancies under cause-specific scenarios for the mortality rates in (\ref{causeScen}) are compared to the baseline scenario where mortality rates are computed from (\ref{MTDIS}).\\

Scenario-specific period life expectancies, $\overset{\circ}e_{h,t}^{[Scen]}$, are generated using scenarios similar to those given in Section \ref{sec_ASMRscenarios} and denoted as $\overset{\circ}e_{h,t}^{[Obs]}$, $\overset{\circ}e_{h,t}^{[k]}$ or $\overset{\circ}e_{h,t}^{[\cdot]}$ for the Unadjusted rates scenario, Cause-$k$-adjusted scenario, and All-causes-adjusted scenario respectively. Similar to the analysis on log ASMRs, average rates of improvement in the period life expectancies $\overset{\circ}e_{h,t}^{[Scen]}$, as denoted by $v^{[Scen]}$ for each corresponding scenario, are obtained by fitting a linear regression model of the form,
\begin{equation*}
	\overset{\circ}e_{h,t}^{[Scen]} = v_0 + v^{[Scen]} t + \epsilon_{t},
\end{equation*}
where $t = 2011, 2012, ..., 2018$. $v^{[Obs]}$, $v^{[k]}$ and $v^{[\cdot]}$ are the average rates of improvement in period life expectancies for the Unadjusted rates scenario, Cause-$k$-adjusted scenario, and All-causes-adjusted scenario respectively. For period life expectancies, $v^{[Scen]} > 0$ denotes a positive rate of improvement. Let $\phi^{[k]}$ denote the cause-$k$ contribution to the reduction in improvements in period life expectancy. Similar to the analysis of ASMRs, 
\begin{equation}
	\phi^{[k]} = \frac{v^{[k]} - v^{[Obs]}}{v^{[\cdot]} - v^{[Obs]}}.
	\label{contributionratioLE} 
\end{equation}	

Firstly, life expectancy at birth is considered for males and females. These results are based on period life expectancies calculated using fitted mortality rates according to (\ref{disMT}) and (\ref{MTDIS}) under the negative binomial distribution. Numerical results are provided in Tables \ref{NBLE0Male} and \ref{NBLE0Female} in the Appendix. For males, observed average life expectancy increases by 3.872 (3.773, 3.971) months per year until 2011, while post-2011 improvement is $v^{[Obs]}$ = 0.656 (0.544, 0.813) months per year on average. For females, the pre-2011 annual improvement is 3.023 (2.928, 3.111) months per year, while the post-2011 improvement is $v^{[Obs]}$ = 0.234 (0.133, 0.387) months. If the pre-2011 trends had persisted past 2011, the average annual improvement from 2011 to 2018 would have been $v^{[\cdot]}$ = 2.986 (2.885, 3.088) months per year for males and $v^{[\cdot]}$ = 1.992 (1.896, 2.008) months per year for females. Similar to the results for age-standardised mortality rates, the analysis shows very pronounced reductions in life expectancy improvement for both genders. \\

\begin{figure}[h!]
	\begin{multicols}{2}
		\includegraphics[width=\linewidth,height=8cm]{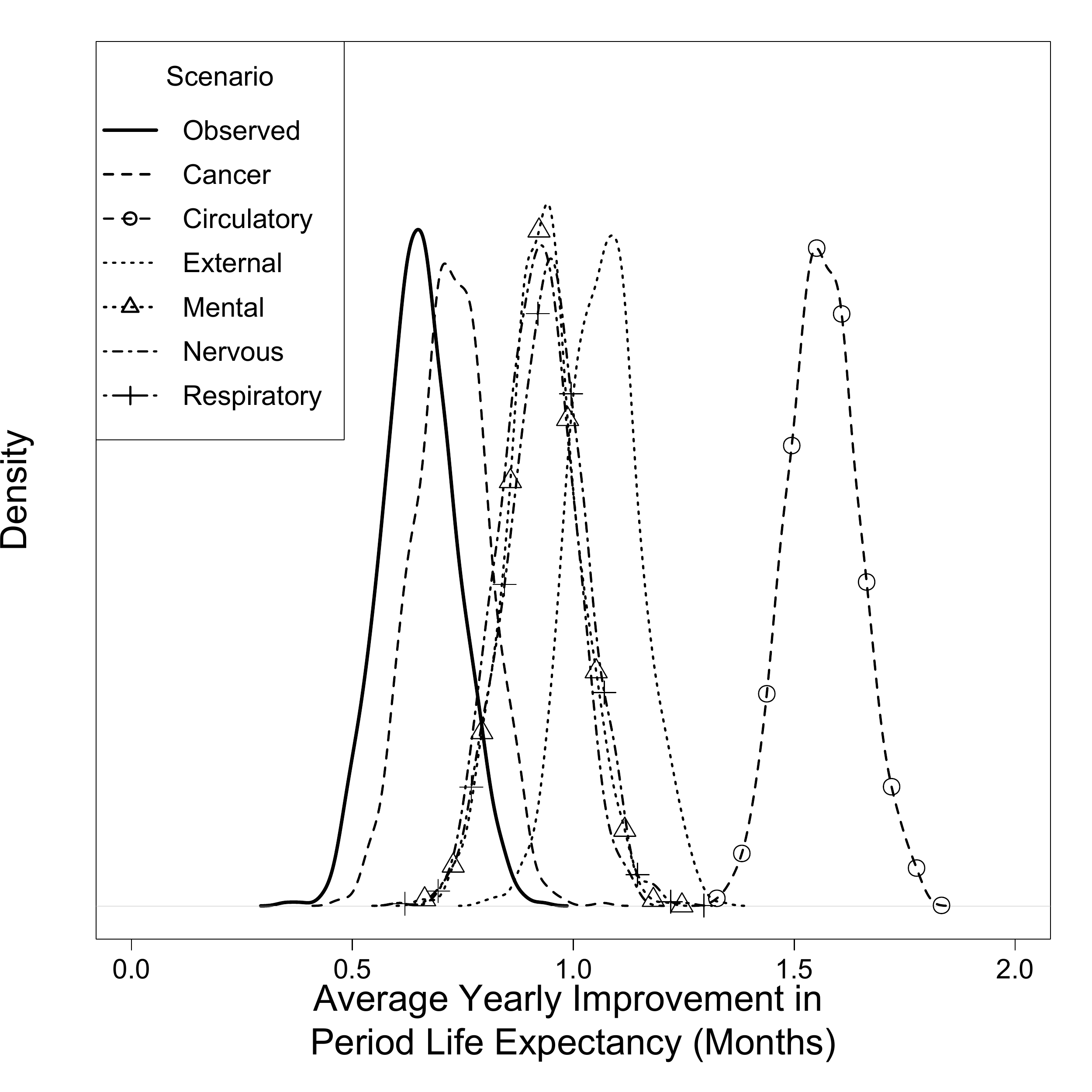}
		\subcaption{Males}
		\includegraphics[width=\linewidth,height=8cm]{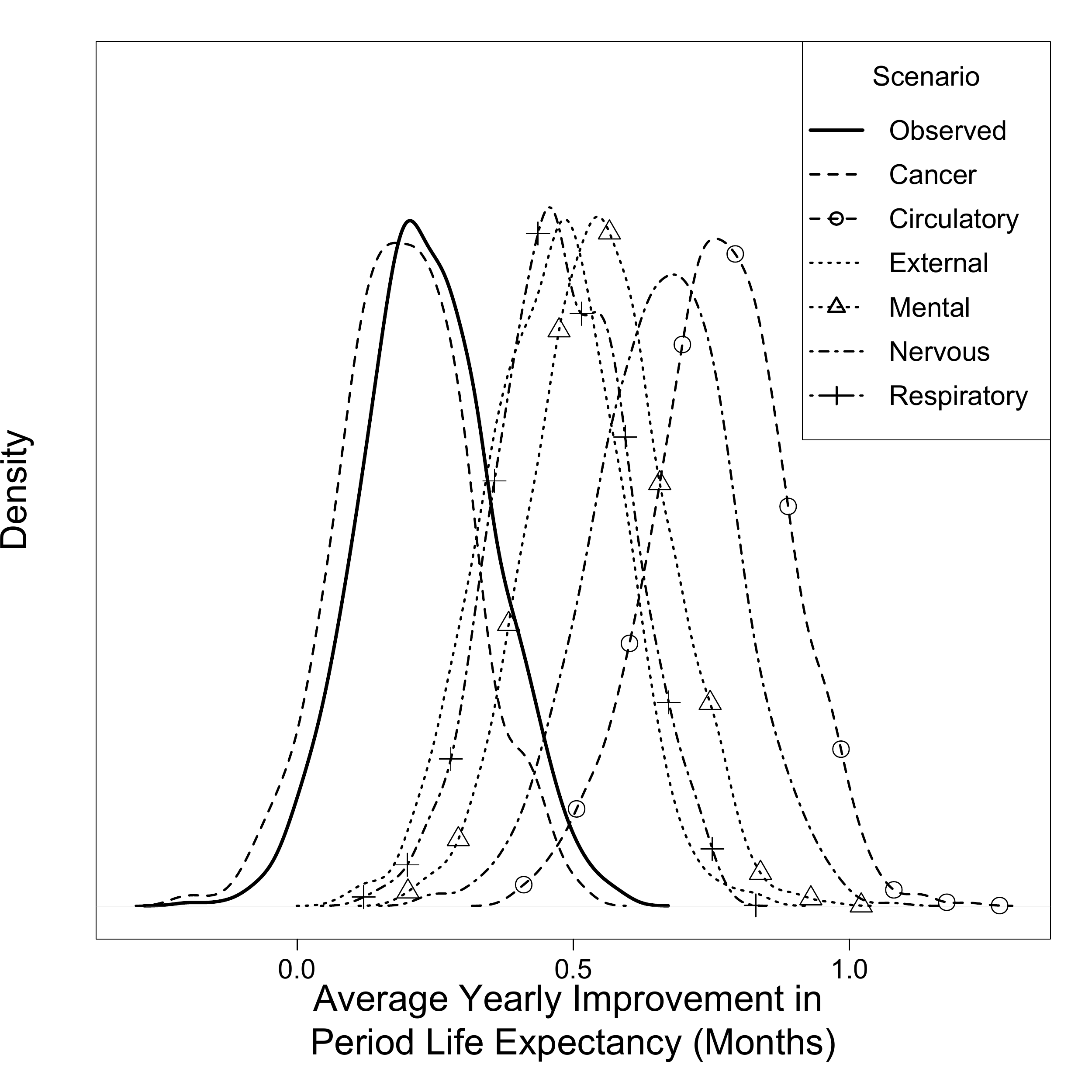}
		\subcaption{Females}
	\end{multicols}		
	\caption{Average improvements in period life expectancies at birth post-2011, $v^{[k]}$, in cause reversion scenarios by gender; $v^{[Obs]}$ also shown.}
	\label{0LEComp}
\end{figure}

Turning to the cause-specific scenarios for mortality rates after 2011 as outlined in Section \ref{sec_ASMRscenarios}, Figure \ref{0LEComp} displays {the distribution of average improvement in life expectancy under cause-$k$-adjusted scenarios, $v^{[k]}$, obtained using bootstrap methodology.} Each distribution corresponds to the average annual improvement in life expectancy that would have been observed if pre-2011 trends were restored for a specific cause, i.e.,  
with life expectancy being calculated using the mortality rates $\tilde{m}^{[\cdot,k]}_{x,t}$ computed in (\ref{causeScen}) rather than $\widehat{m}_{x,t}$ given in (\ref{MTDIS}). 
%For example, if the trend for caner mortality is restored to the trend estimated before 2011, it is found that overall period life expectancy would increase by approximately 0.069 months. %In this example, $k$ is set to cancer in (\ref{mortproj_1}) and (\ref{causeScen}).\\
As expected, the causes of death associated with high improvement are the same as those in Figure \ref{ASMRComp} where the impact on ASMR was considered.
Deaths from diseases of the circulatory system are associated with the highest improvement for life expectancy at birth.
If the post-2011 mortality rate trends for circulatory system diseases were reverted to those observed from 2001 to 2010, this would have resulted in an additional $v^{[CIR]} - v^{[Obs]}$ = 0.911 expected months of life for males per year, and approximately $v^{[CIR]} - v^{[Obs]}$ = 0.533 months for females. Interestingly, the reversion of trends in external causes of mortality for males, results in the second highest gain of an additional $v^{[EXT]} - v^{[Obs]}$ = 0.413 months per year. Referring back to Table \ref{NBBeta2Male}, the estimates for $\widehat{\beta}_{1,x}^{[EXT]}$ are all positive. Thus, setting these equal to 0 in the corresponding scenario, results in additional improvements in period life expectancies at birth for males. Note that, as a result of the non-linear relationship between mortality rates and life expectancy, as well as changes being on the scale of the log mortality rates, cause-specific reductions in life expectancy improvements do not add up to the overall observed post-2011 reduction, $v^{[\cdot]} - v^{[Obs]}$.\\

In addition to life expectancies at birth, expected remaining lifetimes at age 65 are also investigated. At age 65, males have an estimated average yearly improvement in the remaining lifetime of 3.049 (2.953, 3.141) months per year pre-2011, and an average improvement of $v^{[Obs]}$ = 0.591 (0.451, 0.735) months per year post-2011. For females, the estimated average yearly improvement is 2.501 (2.332, 2.667) months per year pre-2011 and $v^{[Obs]}$ = 0.102 ($-$0.138, 0.337) months per year post-2011. If pre-2011 trends persisted from 2011 to 2018, the average improvements would have been $v^{[\cdot]}$ = 2.405 (2.292, 2.511) months per year for males, and $v^{[\cdot]}$ = 1.559 (1.417, 1.713) months per year for females. Tables \ref{NBLE65Male} and \ref{NBLE65Female} in the Appendix display improvement rates in expected remaining lifetimes at age 65 under various cause scenarios for males and females respectively.\\

\begin{figure}[h!]
	\begin{multicols}{2}
		\includegraphics[width=\linewidth,height=8cm]{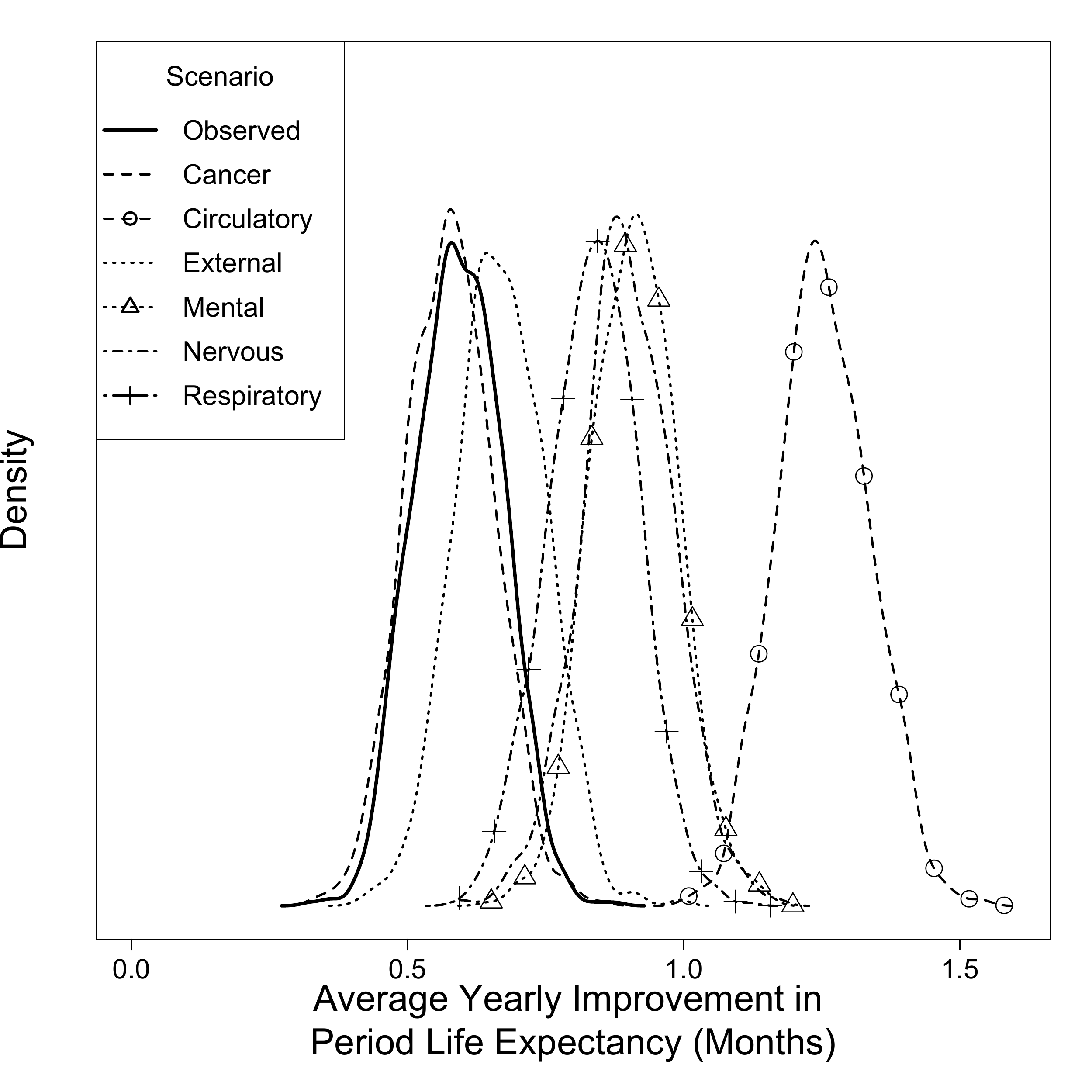}
		\subcaption{Males}
		\includegraphics[width=\linewidth,height=8cm]{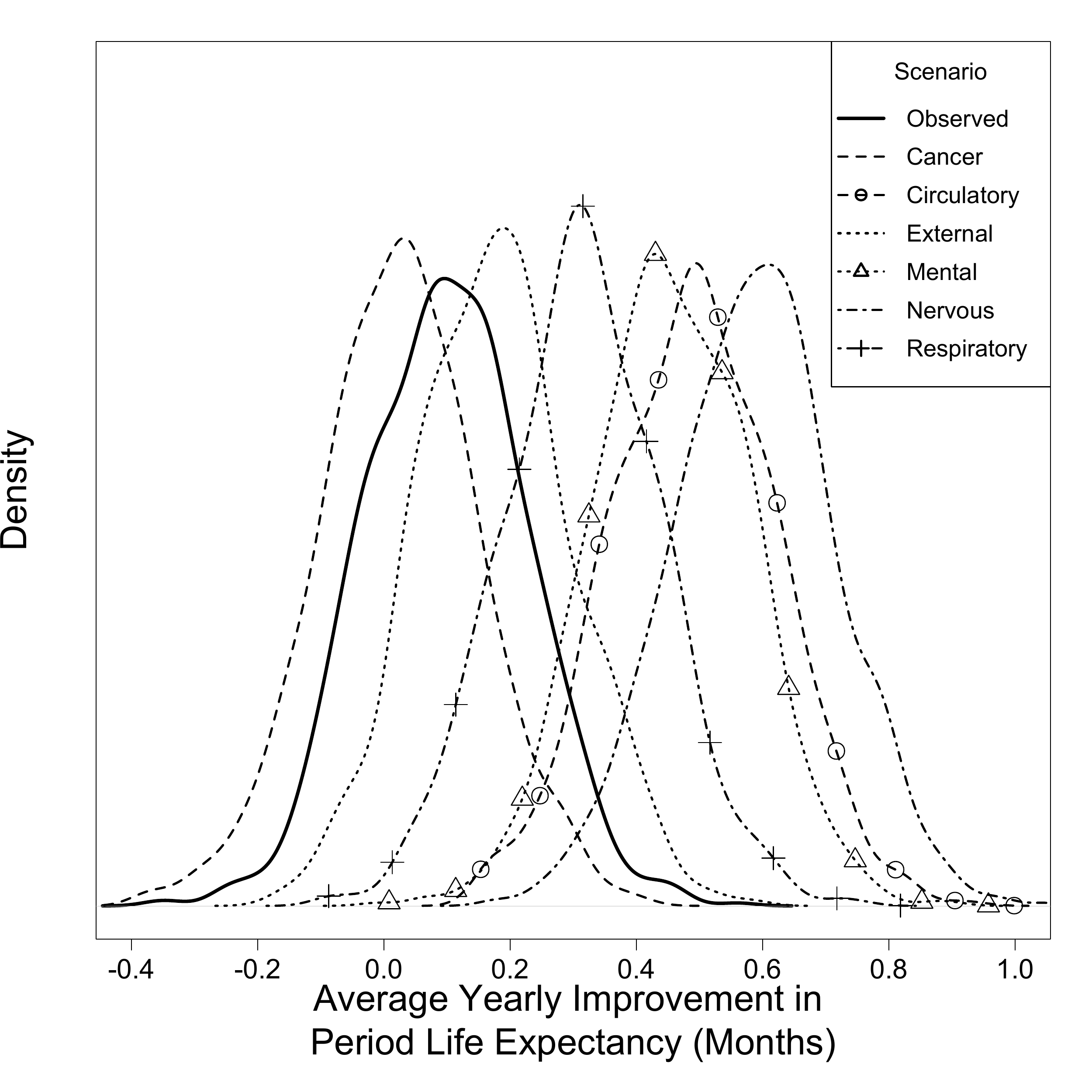}
		\subcaption{Females}
	\end{multicols}		
	\caption{Average improvements in period expected remaining lifetimes at age 65 post-2011, $v^{[k]}$, in cause reversion scenarios by gender; $v^{[Obs]}$ also shown.}
	\label{65LEComp}
\end{figure}

For comparison, the equivalent improvements in life expectancy from age 65 are also plotted in Figure \ref{65LEComp} and a very similar pattern to that observed for life expectancy at birth is found. However, the causes of death that have significant contributions towards the reduction in improvements in life expectancy from age 65 are more similar to those observed for the log ASMRs, compared to those for life expectancy at birth. The similar pattern in significant causes of death for log ASMRs and life expectancies from age 65 is the result of higher mortality rates occurring at the oldest ages.\\

Table \ref{LEContProp} summarises the cause-specific contributions to the slowdown in period life expectancy, $\phi^{[k]}$, as computed using (\ref{contributionratioLE}). \\

\begin{table}[h!]
	\centering
	\caption{Cause-specific contributions to slowdown in period life expectancy improvements, $\phi^{[k]}$, calculated using (\ref{contributionratioLE})}
	\begin{tabular}{|c|cc|cc|}
		\hline
		& \multicolumn{2}{c|}{{\ul \textbf{Males}}}                                & \multicolumn{2}{c|}{{\ul \textbf{Females}}}                              \\ \hline
		{\ul \textbf{Cause}} & \multicolumn{1}{c|}{{\ul \textbf{At Birth}}} & {\ul \textbf{At Age 65}} & \multicolumn{1}{c|}{{\ul \textbf{At Birth}}} & {\ul \textbf{At Age 65}} \\ \hline
		\textbf{Cancer}         & \multicolumn{1}{c|}{2.96\%}                  & -0.71\%                  & \multicolumn{1}{c|}{-2.43\%}                 & -5.43\%                  \\ \hline
		\textbf{Circulatory}         & \multicolumn{1}{c|}{39.10\%}                 & 36.70\%                  & \multicolumn{1}{c|}{30.30\%}                 & 26.44\%                  \\ \hline
		\textbf{Digestive}         & \multicolumn{1}{c|}{-1.69\%}                 & 1.26\%                   & \multicolumn{1}{c|}{-1.88\%}                 & -1.53\%                  \\ \hline
		\textbf{Endocrine}         & \multicolumn{1}{c|}{5.93\%}                  & 4.76\%                   & \multicolumn{1}{c|}{4.82\%}                  & 4.08\%                   \\ \hline
		\textbf{External}         & \multicolumn{1}{c|}{17.72\%}                 & 4.42\%                   & \multicolumn{1}{c|}{12.68\%}                 & 4.96\%                   \\ \hline
		\textbf{Genitourinary}         & \multicolumn{1}{c|}{-3.65\%}                 & -5.11\%                  & \multicolumn{1}{c|}{-10.83\%}                & -14.59\%                 \\ \hline
		\textbf{Infectious}         & \multicolumn{1}{c|}{-1.69\%}                 & -1.53\%                  & \multicolumn{1}{c|}{-4.62\%}                 & -5.59\%                  \\ \hline
		\textbf{Mental}         & \multicolumn{1}{c|}{11.64\%}                 & 17.54\%                  & \multicolumn{1}{c|}{17.94\%}                 & 24.40\%                  \\ \hline
		\textbf{Musculoskeletal}         & \multicolumn{1}{c|}{-0.38\%}                 & -0.08\%                  & \multicolumn{1}{c|}{-0.27\%}                 & -0.27\%                  \\ \hline
		\textbf{Nervous}         & \multicolumn{1}{c|}{11.08\%}                 & 16.93\%                  & \multicolumn{1}{c|}{24.31\%}                 & 33.46\%                  \\ \hline
		\textbf{Respiratory}         & \multicolumn{1}{c|}{12.29\%}                 & 13.78\%                  & \multicolumn{1}{c|}{13.71\%}                 & 14.10\%                  \\ \hline
		\textbf{Other}         & \multicolumn{1}{c|}{1.89\%}                  & 6.27\%                   & \multicolumn{1}{c|}{12.60\%}                 & 15.75\%                  \\ \hline
	\end{tabular}
	\label{LEContProp}
\end{table}

For causes of death such as circulatory system diseases, mental and behavioural illnesses, and nervous system diseases, the greatest changes in the mortality rate trends are observed at older ages. Differences in the trends for these causes at younger ages result in differing contributions to the slowdown in period life expectancy improvements at birth compared to age 65. This is shown for females in Table \ref{LEContProp}, where the contribution to the slowdown by circulatory system diseases, with the greatest contribution of 30.30$\%$ at birth, is surpassed by the contribution of nervous system diseases at age 65 (33.46$\%$).

\section{Life Expectancy in Future Mortality Scenarios}

In this section we apply scenario-based analysis to outline possible mortality scenarios in future years, in particular, the ten years from 2019 to 2028. This short time horizon for projections reflects the relatively short period (18 years) for which observed cause-specific rates are analysed in this study. The effect of trend changes in specific causes on life expectancy is considered.\\

For the purpose of projecting life expectancies past 2018, fitted cause-specific mortality rates, obtained using (\ref{disMT}), are used for the years 2001 to 2018. 
%The observed cause-specific mortality trends are kept due to the fact that they have already occurred. 
The focus of this analysis is to investigate the potential effects of changes in mortality trends on period life expectancies after 2018. In order to define cause-specific mortality rates in years $t>2018$, the model in (\ref{disMT}) is employed with the addition of an extra time trend component which permits the modification of the trend in future years. More specifically, the model becomes:
\begin{equation} \label{future}
	\begin{aligned}
		\log m_{x,t}^{[c]} & = && \widehat{\beta}_{0,x}^{[c]} + \widehat{\beta}_{1,x}^{[c]} t + \Big(\widehat{\beta}_{2,x}^{[c]}(t  - 2011) + \widehat{\beta}_{3,x}^{[c]} \Big) \text{I}(t \ge 2011)\\
		&  && + \beta_{4,x}^{[c]}(t - 2018)\text{I}(t \ge 2018),\\
	\end{aligned}
\end{equation}
where $\widehat{\beta}_{0,x}^{[c]}$, $\widehat{\beta}_{1,x}^{[c]}$, $\widehat{\beta}_{2,x}^{[c]}$, and $\widehat{\beta}_{3,x}^{[c]}$ are obtained from fitting the age- and cause-specific death counts according to (\ref{disMT}).
Parameter $\beta_{4,x}^{[c]}$ permits the investigation of scenarios for cause-specific trends. For example, $\beta_{4,x}^{[c]}=0$ for all causes $c$ corresponds to a scenario in which pre-2018 trends continue unchanged. This will be the baseline scenario. By changing this parameter for a specific cause $k$, the impact of changes in the improvement rate of $m_{x,t}^{[k]}$ on all-cause life expectancy can be investigated. Again, the all-cause mortality rates in each scenario are calculated as the summation of scenario cause-specific mortality rates as in (\ref{mortsum}).\\

\subsection{Scenarios for life expectancies}\label{seclife}

Scenarios are generated for the purpose of investigating the effect of changing or continuing mortality rate trends in cause-specific mortality, on the period life expectancies past 2018. In each of the scenarios, the all-cause mortality rates at 2018 are equal to the fitted rates in 2018 according to the Unadjusted rates scenario outlined in Section \ref{quant}.\\ 

Scenarios are listed as follows: 

\begin{enumerate}
	\item Future Scenario 1 (FS1): No changes in cause-specific mortality trends. We set $\beta_{4,x}^{[c]} = 0$ for all causes $c$ and age groups $x$. This is the baseline scenario.
	\item Future Scenario 2 (FS2): Reversion of mortality trends for a single cause of death $k$ to pre-2011 rates. For a cause $k$, we assume $\beta_{4,x}^{[k]} = - \widehat{\beta}_{2,x}^{[k]}$ from (\ref{disMT}) for all age groups $x$. We set $\beta_{4,x}^{[c]} = 0$ for all other causes of death $c \ne k$. In this scenario, it is assumed that the change in trend for a single cause of death from 2011 to 2018 is temporary, and the trend returns to its pre-2011 trends from 2019 onwards. 
	\item Future Scenario 3 (FS3): All cause-specific mortality trends are reverted to pre-2011 rates, i.e. $\beta_{4,x}^{[c]} = - \widehat{\beta}_{2,x}^{[c]}$ for all causes $c$ and age group $x$. In this scenario, it is assumed that all cause-specific mortality trends from 2011 to 2018 are temporary, and return to their pre-2011 levels after 2018. 
	\item Future Scenario 4 (FS4): All-cause mortality rates at each age are reduced by an additional rate of $z$ per year, with $ 0 \leq z \leq 1$. We set $\beta_{4,x}^{[c]} = \log(1 - z)$ for all causes $c$ and age groups $x$.
	\item Future Scenario 5 (FS5): A World Health Organization (WHO) scenario is developed matching projections from the WHO, that has projected cause-specific mortality rates for high income countries for the years 2016, 2030, 2045, and 2060 (World Health Organization, \citeyear{world2019projections}). 
	Causes of deaths are grouped herein according to the cause-of-death groups outlined in Table \ref{CAUSETABLE}. We performed linear interpolation on log cause-specific mortality rates between the years 2016 and 2030 in order to obtain the projected average annual change in log mortality rates during this time period.
	Table \ref{WHOMalesA} displays the WHO projected average annual changes in log mortality rates for six causes of death. Additional rates are provided in Tables \ref{WHOMalesB} to \ref{WHOFemalesB} in the Appendix. In Scenario 5,  $\beta_{4,x}^{[c]} = \beta_{WHO,x}^{[c]} - (\widehat{\beta}_{1,x}^{[c]} + \widehat{\beta}_{2,x}^{[c]})$ for each cause $c$ and age group $x$, where $\beta_{WHO,x}^{[c]}$ are the WHO projected average annual changes in log mortality rates. The age groups used by the WHO are 0-4, 5-14, 15-29, 30-49, 50-59, and 70+. 
	
	\begin{table}[h!]
		\centering
		\caption{WHO projected annual change in log mortality rates by cause - male - Part 1}
		\begin{tabular}{|c|c|c|c|c|c|c|}
			\hline
			{\ul \textbf{Age Group}} & {\ul \textbf{CAN}} & {\ul \textbf{CIR}} & {\ul \textbf{DIG}} & {\ul \textbf{END}} & {\ul \textbf{EXT}} & {\ul \textbf{GEN}} \\ \hline
			\textbf{\textless{}1}    & -0.02419           & -0.01383           & -0.01634           & -0.01736           & 0.00070            & -0.01727           \\ \hline
			\textbf{1-4}             & -0.02419           & -0.01383           & -0.01634           & -0.01736           & 0.00070            & -0.01727           \\ \hline
			\textbf{5-9}             & -0.02175           & -0.01978           & -0.02630           & -0.01100           & 0.00447            & -0.01062           \\ \hline
			\textbf{10-14}           & -0.02175           & -0.01978           & -0.02630           & -0.01100           & 0.00447            & -0.01062           \\ \hline
			\textbf{15-19}           & -0.01773           & -0.01257           & -0.01199           & -0.00509           & -0.00332           & -0.01749           \\ \hline
			\textbf{20-24}           & -0.01773           & -0.01257           & -0.01199           & -0.00509           & -0.00332           & -0.01749           \\ \hline
			\textbf{25-29}           & -0.01773           & -0.01257           & -0.01199           & -0.00509           & -0.00332           & -0.01749           \\ \hline
			\textbf{30-34}           & -0.02034           & -0.01720           & -0.01587           & 0.00181            & -0.00593           & -0.00846           \\ \hline
			\textbf{35-39}           & -0.02034           & -0.01720           & -0.01587           & 0.00181            & -0.00593           & -0.00846           \\ \hline
			\textbf{40-44}           & -0.02034           & -0.01720           & -0.01587           & 0.00181            & -0.00593           & -0.00846           \\ \hline
			\textbf{45-49}           & -0.02034           & -0.01720           & -0.01587           & 0.00181            & -0.00593           & -0.00846           \\ \hline
			\textbf{50-54}           & -0.01867           & -0.02167           & -0.01253           & -0.00575           & -0.01044           & -0.00406           \\ \hline
			\textbf{55-59}           & -0.01867           & -0.02167           & -0.01253           & -0.00575           & -0.01044           & -0.00406           \\ \hline
			\textbf{60-64}           & -0.01867           & -0.02167           & -0.01253           & -0.00575           & -0.01044           & -0.00406           \\ \hline
			\textbf{65-69}           & -0.01867           & -0.02167           & -0.01253           & -0.00575           & -0.01044           & -0.00406           \\ \hline
			\textbf{70-74}           & -0.00820           & -0.01633           & -0.00716           & 0.00085            & -0.00174           & 0.00297            \\ \hline
			\textbf{75-79}           & -0.00820           & -0.01633           & -0.00716           & 0.00085            & -0.00174           & 0.00297            \\ \hline
			\textbf{80-84}           & -0.00820           & -0.01633           & -0.00716           & 0.00085            & -0.00174           & 0.00297            \\ \hline
			\textbf{85+}             & -0.00820           & -0.01633           & -0.00716           & 0.00085            & -0.00174           & 0.00297            \\ \hline
		\end{tabular}
		\label{WHOMalesA}
	\end{table}
\end{enumerate}

\begin{figure}[h!]
	\centering
	\includegraphics[width=\linewidth,height=10cm]{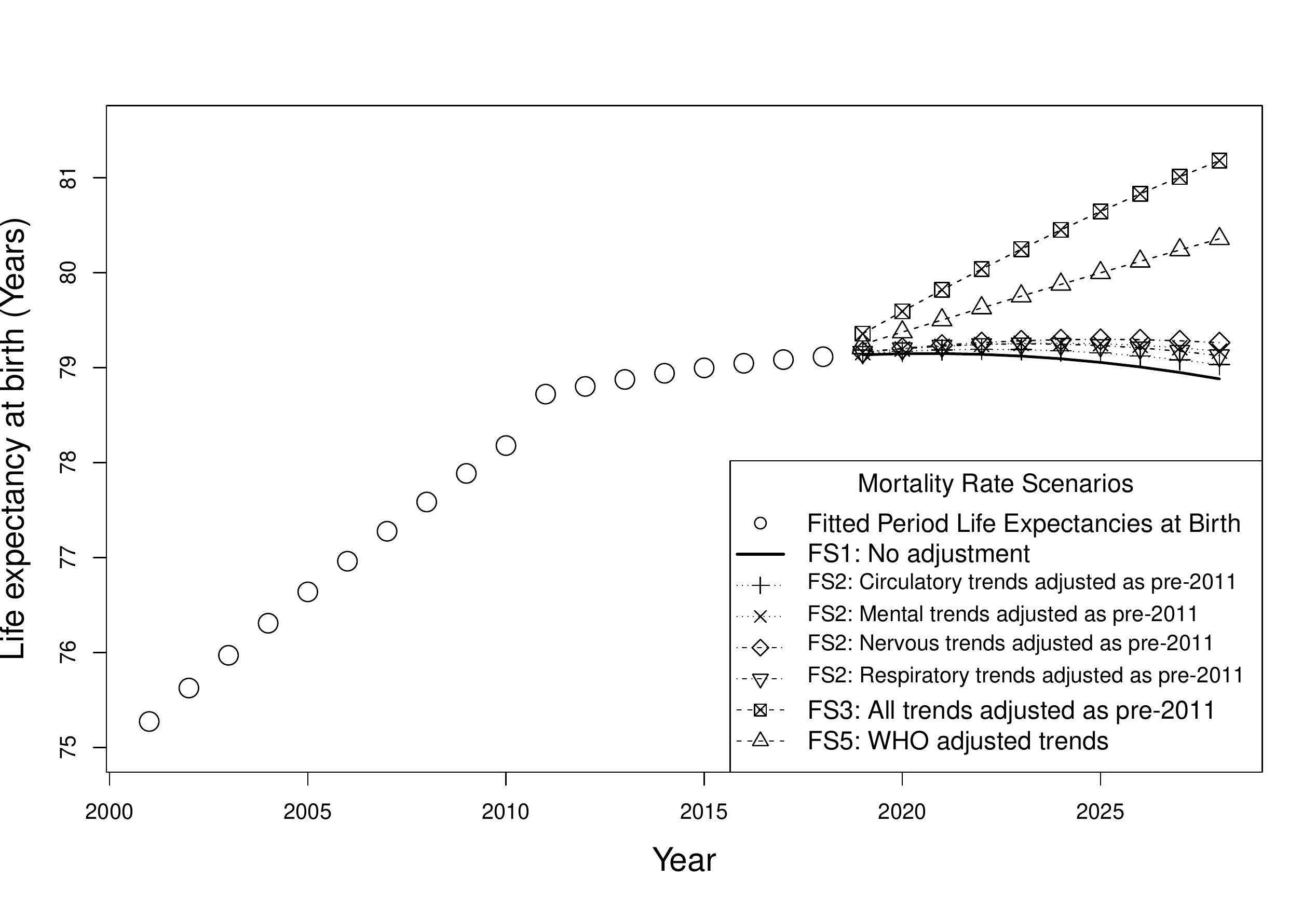}
	\caption{Future period life expectancies at birth under various scenarios - Males}
	\label{Male1LE}
\end{figure}

\begin{figure}[h!]
	\centering
	\includegraphics[width=\linewidth,height=10cm]{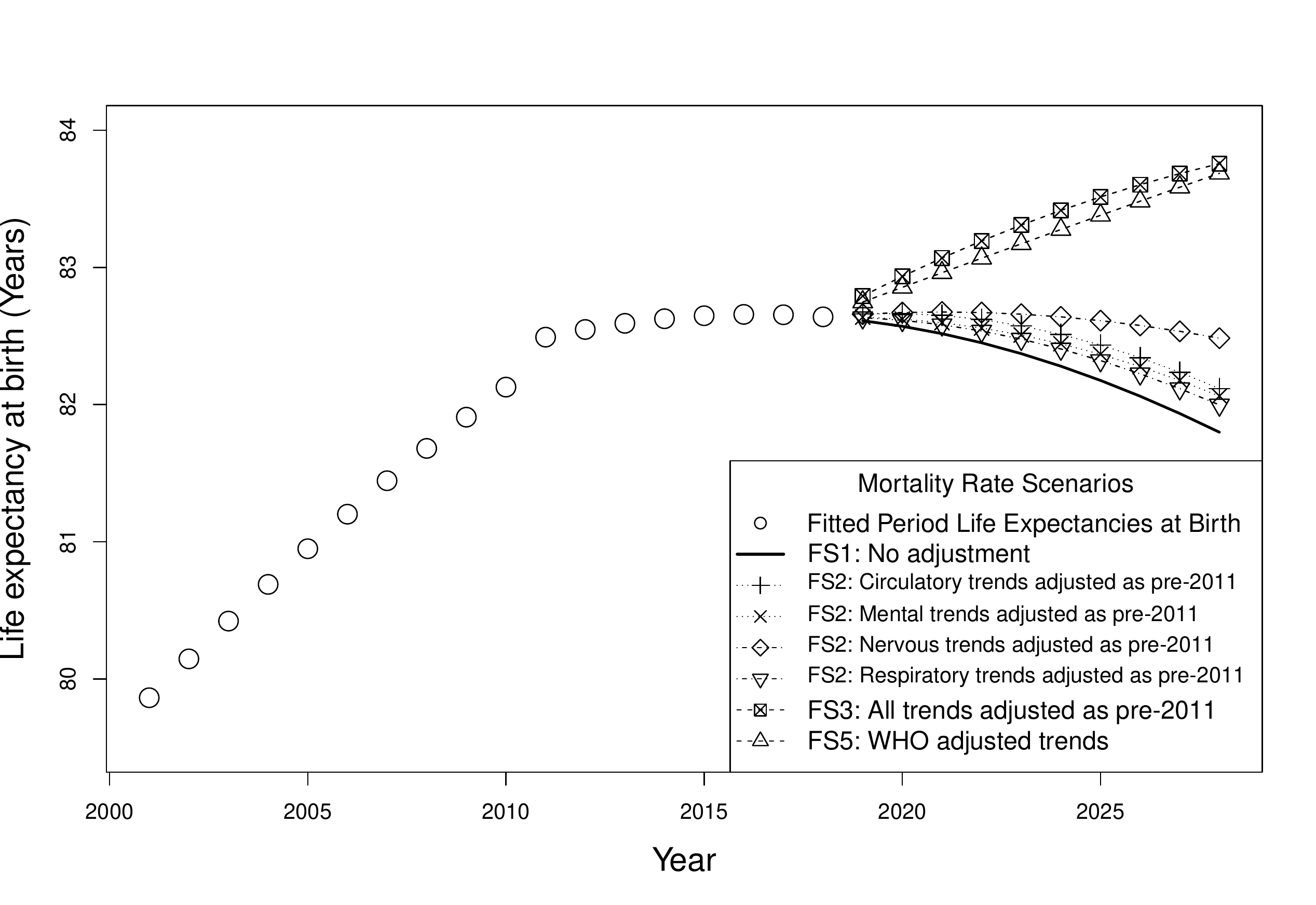}
	\caption{Future period life expectancies at birth under various scenarios - Females}
	\label{Female1LE}
\end{figure}

\begin{figure}[h!]
	\centering
	\includegraphics[width=\linewidth,height=10cm]{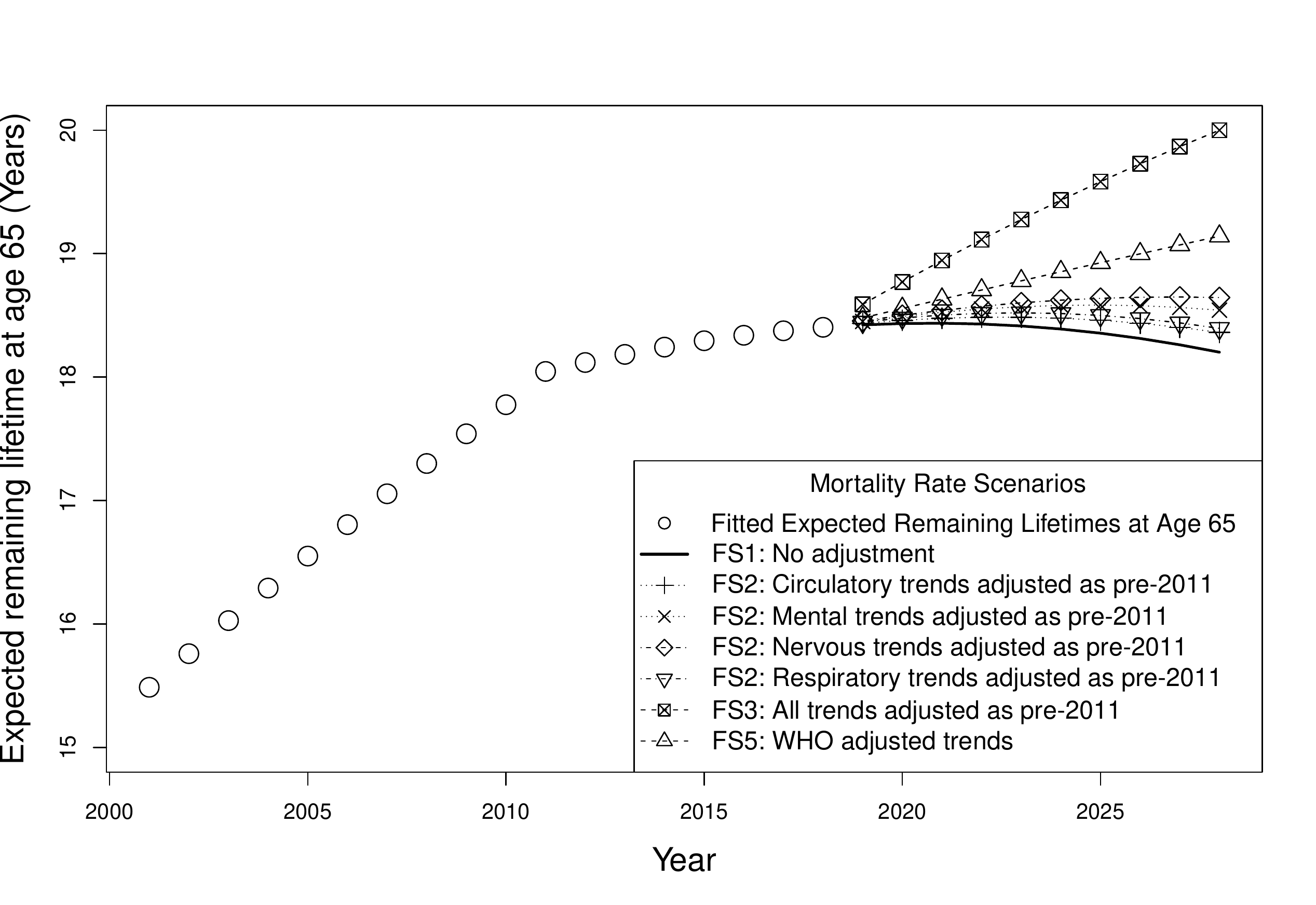}
	\caption{Future period expected remaining lifetimes at age 65 under various scenarios - Males}
	\label{Male65LE}
\end{figure}

\begin{figure}[h!]
	\centering
	\includegraphics[width=\linewidth,height=10cm]{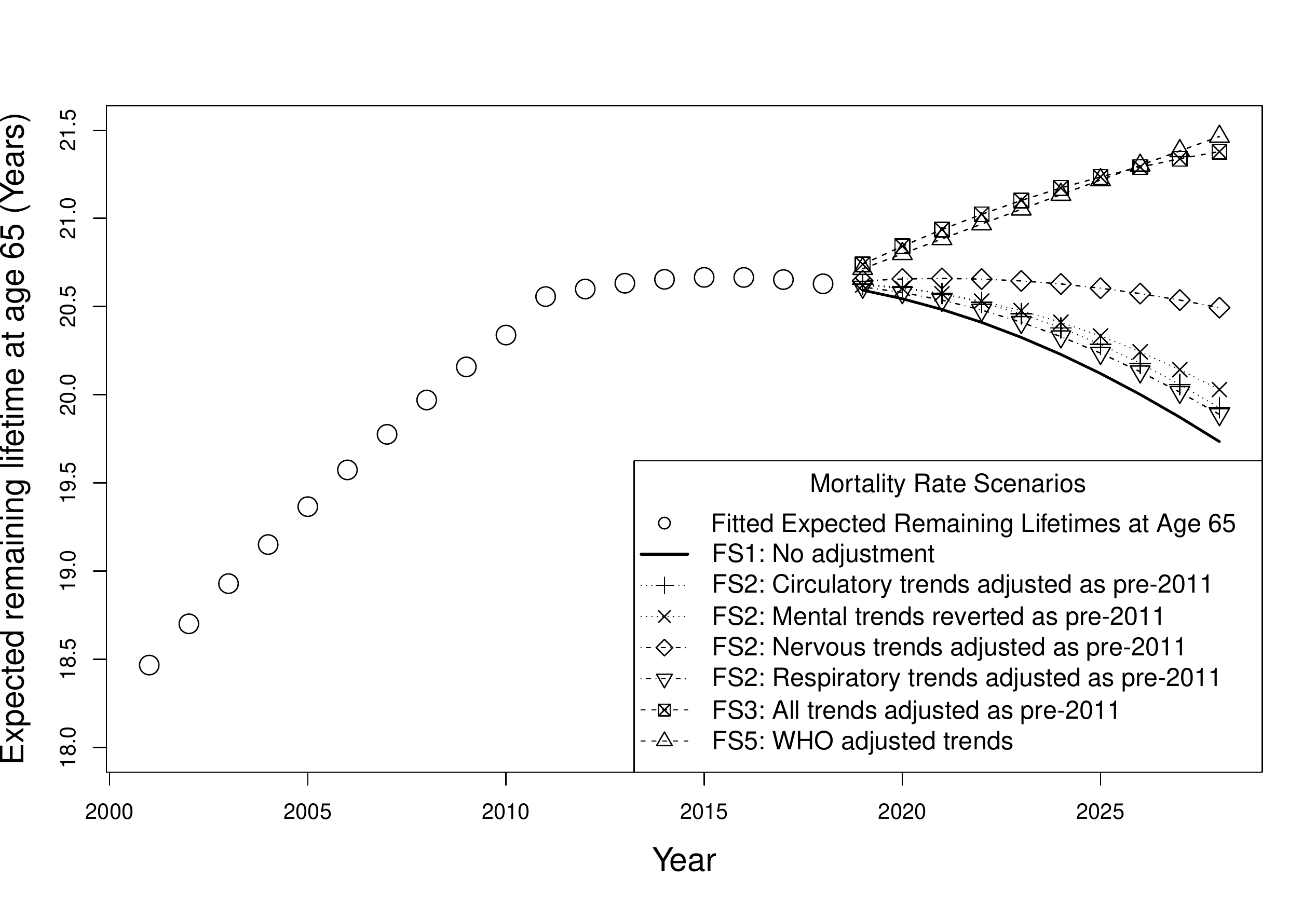}
	\caption{Future period expected remaining lifetimes at age 65 under various scenarios - Females}
	\label{Female65LE}
\end{figure}

Figures \ref{Male1LE} and \ref{Female1LE} show results for all-cause life expectancy at birth, where the period life expectancies from 2001 through 2018 are computed using fitted deaths according to (\ref{disMT2}). Figures \ref{Male65LE} and \ref{Female65LE} show corresponding results for expected remaining lifetimes at age 65. For the purpose of projecting the period life expectancies for 2019 and beyond, mortality rates are projected according to (\ref{future}). In each figure, the solid line represents the projected period life expectancies, under the baseline Future Scenario 1, where the trends in the mortality rates post 2011 are not adjusted for the years after 2018, i.e., $\beta_{4,x}^{[c]} = 0$ for all age groups $x$ and cause-of-death groups $c$. \\

In Future Scenario 1, the life expectancies are projected to worsen after 2020 for males and continue to worsen for females, as shown in Figures \ref{Male1LE} to \ref{Female65LE}. However, for life expectancy at age 65, Figures \ref{Male65LE} and \ref{Female65LE} additionally show that  the expected remaining lifetime for males increases until 2022 and begins to worsen past 2022, while the expected remaining lifetime for females continues its decreasing trend from 2016 onwards. In addition, Future Scenario 2 is shown in Figures \ref{Male1LE} to \ref{Female65LE}, where the mortality rate trend in one cause of death is adjusted. In all considered cases, life expectancy is projected to be higher compared to the baseline Future Scenario 1.\\

The comparison between Future Scenario 3 and Future Scenario 5 is also of interest. These two scenarios contrast life expectancy under the assumption of all trends being reverted to pre-2011 levels, to life expectancy under the WHO projections. 
In both these scenarios, the projected remaining lifetime increases from 2019 through 2028. However, restoring all cause-specific mortality rate trends to their pre-2011 levels in Future Scenario 3 results in a greater level of improvement for males compared to adjusting the trends according to the WHO projections. In the case of life expectancy for females at age 65, the restoration of mortality rate trends to pre-2011 levels has an overall effect similar to that under the WHO projections (\ref{Female1LE}).\\

\begin{figure}[h!]
	\centering
	\includegraphics[width=\linewidth,height=10cm]{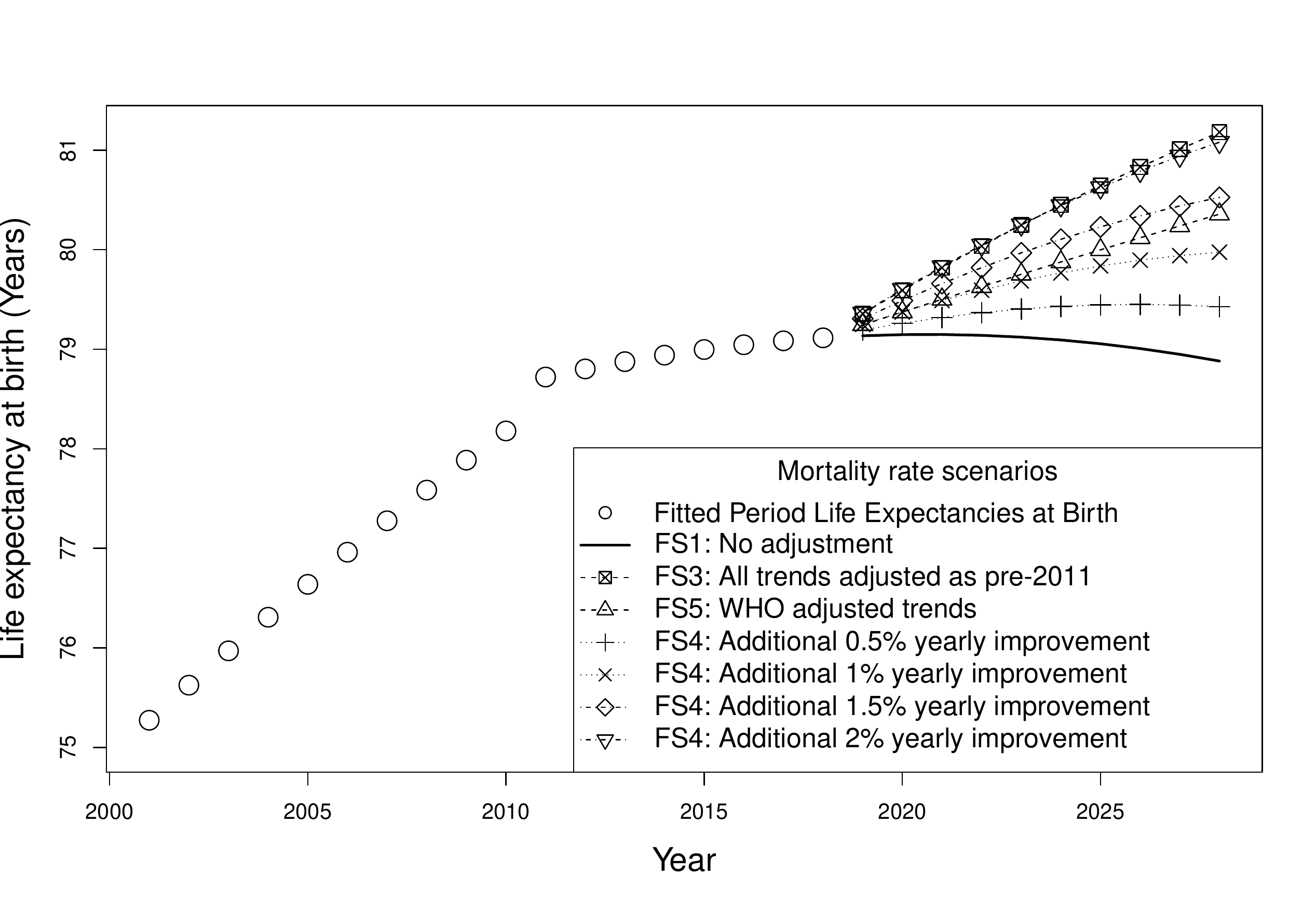}
	\caption{Further future period life expectancies at birth under various scenarios - males}
	\label{Male1LE2}
\end{figure}

\begin{figure}[h!]
	\centering
	\includegraphics[width=\linewidth,height=10cm]{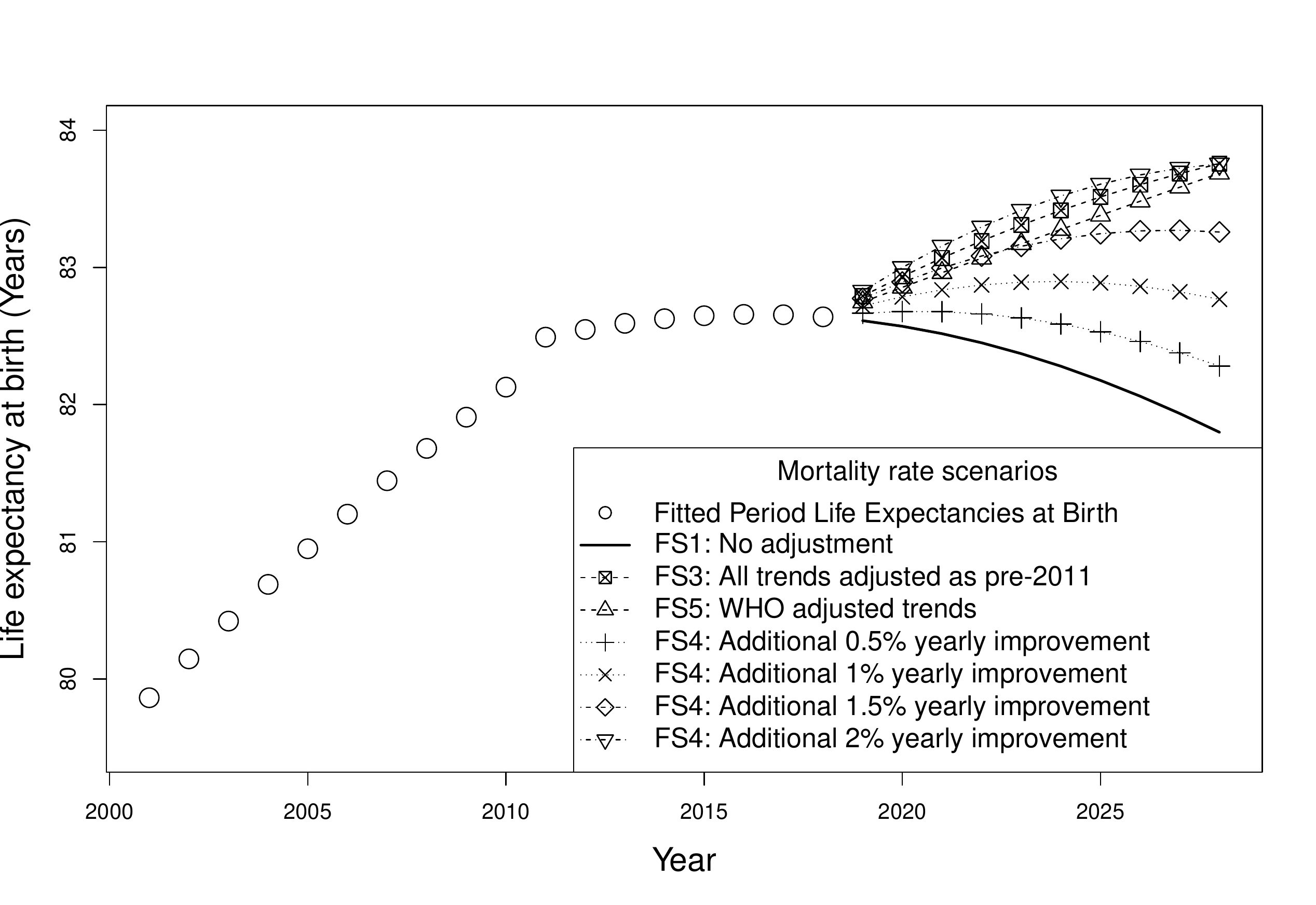}
	\caption{Further future period life expectancies at birth under various scenarios - females}
	\label{Female1LE2}
\end{figure}

Figures \ref{Male1LE2} and \ref{Female1LE2} display projected life expectancies for Future Scenario 4 compared to Future Scenarios 1, 3 and 5. In Future Scenario 4, all age- and cause-specific mortality rates are assumed to experience a constant adjustment, $z$, from 2019 to 2028. For example, in a scenario where there is an additional 1$\%$ yearly improvement over the 10 years of projections, all age- and cause-specific mortality rate experience an additional 1$\%$ improvement over the fitted post-2011 mortality rate trend.
In the case of life expectancy for males at birth, Figure \ref{Male1LE2} suggests that each of the age- and cause-specific mortality rates would need to experience an additional yearly improvement of between 1$\%$ and 1.5$\%$ to result in the same overall improvement as using the WHO trend projections. 
However, in order to experience the same rate of improvement in remaining lifetime as that in Future Scenario 3 (where all trends are restored to pre-2011 levels), each of the age- and cause-specific mortality rates would need to experience a  additional yearly improvement of more than 2$\%$. 
For female life expectancy at birth (Figure \ref{Female1LE2}), each of the age- and cause-specific mortality rates would need to experience an additional yearly improvement of between 1.5 $\%$ and 2$\%$ in order to obtain the improvements in remaining lifetimes projected in the pre-2011 mortality rate trends and the WHO-based trend projections (Future Scenarios 3 and 5 respectively).

\newpage

\section{Discussion}

\subsection{Modelling assumptions}

%In Section \ref{Fitting}, fitting the NB model (\ref{NBSpec}) with the linear predictor in (\ref{disMT}) to each group individually, resulted in increased parameter uncertainty. In particular, the parameters $\beta_{2,x}^{[c]}$ and $\beta_{3,x}^{[c]}$ may not always be significantly different from zero. However, this is not an issue for the analyses in this paper, as the setting of $\beta_{2,x}^{[c]} = 0$ in the different cause scenarios has minimal effect on the results.\\
	
The choice of the cause-of-death groupings is important as it may potentially affect the trends that are observed. For example, in cancer mortality, although the log age-standardised mortality rates are not observed to change considerably overall, the grouping of all cancers into a single group results in some loss of information about the trends for single types of cancer. However, the results indicate that the standardised mortality rate due to all cancers is improving.\\

The use a single breakpoint across all  cause-specific and age-specific mortality trends,  resulted in discontinuities in the calculated log age-standardised mortality rates, period life expectancies at birth, and period expected remaining lifetimes at different ages at the year 2011. However, the location of this breakpoint allows the effect of the change in the classification of the causes of death in 2011 to be captured. It is not straightforward to quantify the exact contribution of the change in classification to the observed and fitted trends in the cause-specific mortality rates using the available data, and we assume that the post-2011 trends are mostly a result of a real change in the temporal trends for each cause of death.

\subsection{Analysis and results}

The reduction in the improvements in all-cause mortality rates follows the reduction in the improvements in circulatory system disease mortality after 2011. Circulatory system disease mortality is one of the leading causes of death at older ages, and therefore any changes in the trends for this cause will have a greater impact on the overall mortality compared to other causes with fewer deaths.
While the age-standardised mortality rates are a weighted average of the age-specific mortality rates, the higher number of deaths occurring at the highest ages results in the trends in age-standardised mortality rates following more closely the mortality rate trends of these higher age groups. At the same time, while the period life expectancies are affected by the higher mortality rates at the older ages, these higher mortality rates do not influence period life expectancies for younger individuals to the same extent as in the age-standardised mortality rates.\\

 The results in Section \ref{sec_ASMRscenarios} suggest that mortality from circulatory system diseases is the leading contributor to the slowdown in log ASMR improvements for males. This may be explained as a combination of two reasons: (a) deaths from diseases of the circulatory system are very common, so any reduced improvements will have a significant effect on all-cause rates, and (b) improvements in the mortality from this cause have slowed down substantially after 2011. The trends in age-specific mortality rates due to circulatory system diseases, as given in Tables \ref{NBBeta1Male} and \ref{NBBeta2Male}, reveal that  mortality rates are still decreasing after 2011, i.e., $\widehat{\beta}_{1,x}^{[CIR]} + \widehat{\beta}_{2,x}^{[CIR]} < 0$, but at a lower rate than before 2011, i.e., $\widehat{\beta}_{2,x}^{[c]} > 0$. In contrast, the other very common cause of death, cancer, has no significant change in improvement rates after 2011. Therefore, assuming a continuation of the pre-2011 trend for cancer does not materially change the observed trend in all-cause mortality.\\
	
 Furthermore, the results in Section \ref{sec_ASMRscenarios} showed that mental and behavioural illnesses, and nervous system diseases were also major contributors to the slowdown in improvements. Again, there may be a number of reasons for this. Improvement rates for those causes are lower after 2011, compared to earlier years. More importantly, the substantial reduction in deaths from other causes that are observed in the last few decades, e.g., circulatory system diseases and cancer, means that deaths from mental disease and diseases of the nervous system have become relatively more common in recent years. In other words, many of those who would have died of heart attack are now living long enough to develop other diseases, such as diseases of the nervous system. Therefore, the increased number of deaths from mental diseases and diseases of the nervous system are testimony of the successful reduction of death rates from diseases of the circulatory system. A consequence of this increased importance of such illnesses is that any slowdown of improvements in deaths caused by mental and behavioural illnesses and nervous system diseases can potentially have rather significantly negative effects on all-cause improvements.\\

Our findings also show that while age-standardised mortality rates are estimated to improve at a lower rate for females compared to those for males, mortality for males is still higher than mortality for females, as shown in the comparison of ASMRs in Figure \ref{obsasmrplots}. The discrepancy between the male and female mortality rates results in the higher projected period life expectancies for females compared to males. \\

%It is important to note that in Section \ref{sec_life_expect}, the sum of the possible additional improvements in life expectancies is not equal to the total change in life expectancy. This is a consequence of the choice of analysis, where the trends for a single cause of death are reverted to pre-breakpoint levels, therefore ignoring inter-dependencies among causes of death.
%As the dependencies between cause-specific mortalities at different ages are unknown, e.g., the effect of reducing cancer mortality at age 25 on the mortality rate due to circulatory system diseases at age 60, the study 
%{\color{red} [Which study? Our study, or \citet{ponnapalli2005comparison}? If this is 'our study', this statement is a bit awkward, as ealrier sections highlight the contributions (even if not 'exact')]}
%focuses on the differences in the rate of improvement, rather than the exact contributions of different causes of death. For example, for causes of death such as circulatory system diseases and nervous system diseases, the greatest changes in mortality trends occur at older ages. As a consequence, changes in mortality rates from other causes at younger ages, will also influence the impact of the changes in mortality rate trends at older ages.\\

In the future period life expectancy scenarios in Section \ref{seclife}, the eventual decline in life expectancies is driven by reductions in improvements in mortality rates due to circulatory system diseases and the increase in mortality due to mental and behavioural illnesses and nervous system diseases. However, this depends on a continued increase in mortality in these cause-of-death groups. For dementia, the incidence rate has been projected to decrease in the United Kingdom \citep{prince2016recent, ahmadi2017temporal}.
\subsection{Conclusion}

This research highlights the impact of changes in mortality rates for different causes of death on overall mortality trends and period life expectancies.
While reduction in improvements in mortality from circulatory system diseases is the greatest contributor to the observed slowdown in improvements for log age-standardised mortality rates and life expectancies, increasing mortality due mental and behavioural illnesses and nervous system diseases also provide significant contributions. 
Using data from 2001 to 2018, our findings suggest that 
if the mortality rate trends in England and Wales persist past 2018, life expectancies for males and females will show a consistent decrease. This work and the corresponding findings can enhance the understanding of cause-specific contributions to mortality rate trends, which is of value to researchers, policymakers, and insurance professionals.\\

\textbf{Acknowledgements}

The first author would like to acknowledge the financial support by the James Watt Scholarship at Heriot-Watt University. 

\newpage

\bibliography{referencesPaper}
\bibliographystyle{chicago}

\newpage

\appendix
\setcounter{table}{0}
\renewcommand{\thetable}{A\arabic{table}}
%\chapter{Appendix}
%\label{ch:Appendix}

\section{Appendix}
\label{ch:Appendix}

\subsection{Parametric bootstrap}\label{boot}

The parametric bootstrap is employed for the purpose of construction of confidence intervals for the age-standardised mortality rates and period life expectancies.

\begin{enumerate}
	\item Fit the negative binomial GLM according to (\ref{disMT}) for each age group $x$ and cause of death $c$ combination to obtain fitted death counts $E_{x,t}\widehat{m}_{x,t}^{[c]}$ and estimated dispersion parameters $\widehat{\theta}_{x}^{[c]}$.
	\item For each bootstrap iteration:
	\begin{enumerate}
		\item New cause-specific death counts, $\tilde{d}_{x,t}^{[c]}$, are generated such that 
		\begin{equation*}
			\tilde{d}_{x,t}^{[c]} \sim NB(E_{x,t} \widehat{m}_{x,t}^{[c]}, \widehat{\theta}_{x}^{[c]}).
		\end{equation*}
		\item The total number of deaths for an age $x$ and calendar year $t$, $\tilde{d}_{x,t}$, is calculated as 
		\begin{equation*}
			\tilde{d}_{x,t} = \sum\limits_{c} \tilde{d}_{x,t}^{[c]}.
		\end{equation*}
		\item The new age-specific mortality rates, $\tilde{m}_{x,t}$, are calculated as
		\begin{equation*}
			\tilde{m}_{x,t} = \frac{\tilde{d}_{x,t}}{E_{x,t}}.
		\end{equation*}
		\item Use these new age-specific mortality rates, $\tilde{m}_{x,t}$, to calculate the ASMRs or life expectancies.
	\end{enumerate}
\end{enumerate}

In the case of a cause-specific scenario where the trends in the age- and cause-specific mortality rates are restored to pre-breakpoint levels for a single cause of death $c=k$, i.e., in the cause-$k$ adjusted scenario, set $\widetilde{\beta}_{2,x}^{k} = 0$ in each bootstrap iteration.\\

 In the scenarios and bootstrap iterations of the death counts in Sections \ref{sec_ASMRscenarios} and \ref{sec_life_expect}, exposures are kept at the original values from the mid-year population estimates. 
%However, in the case where death counts deviate from those observed, the rate of exposures is no longer the same as those from the mid-year population estimates. 
However, exposures in subsequent years for a single age group can increase by the number of individuals who enter the group from the previous age group, or as an effect of immigration, as well as from births in the case of the youngest age group. On the other hand, exposures may decrease as individuals exit the age group through the advancement into the next age group, death, and emigration. As the differences between the fitted and cause-scenario mortality rates and resulting death counts are not large, the given mid-year population estimates are taken to be satisfactory for the purpose of estimating the exposures in each year.\\

\subsection{Period life expectancies}\label{lifedefinition}

 Period life expectancies are used in this paper. In contrast to cohort life expectancy, period life expectancy relies on the age structure of mortality in a particular year to calculate survival probabilities. This is equivalent to calculating cohort life expectancy under the assumption that mortality will not change in future years, which is calculated as follows: given all-cause mortality rates $m_{x,t}$ for age group $x$ and calendar year $t$, the probability for a person who is $h$ years old in year $t$ to survive for a further $n$ years, ${}_n p_{h,t}$, is obtained by
\begin{equation*}
	{}_n p_{h,t} = \exp\left(- \sum\limits_{i=0}^{n-1} m_{h+i,t}\right),
\end{equation*}
where $m_{h,t}$ is the mortality rate for an individual aged $h$ and then the period complete expectation of life at age $h$ in calendar year $t$, $\overset{\circ}e_{h,t}$, is calculated as
\begin{equation*}
	\overset{\circ}e_{h,t} = \sum\limits_{n=1}^{\infty} {}_n  p_{h,t} + \frac{1}{2}.
\end{equation*}

As the observations are grouped in five year age groups, it is assumed that the mortality rates in each group are constant across all ages in that group, i.e., \begin{equation*}
	m_{h,t} = m_{x,t},
\end{equation*}
for all ages $h$ in age group $x$. It is also assumed that the rates $m_{x,t}$ for all ages $h$ greater than the maximum age of 85 in the study are constant and equal to the rate at 85, i.e., $m_{x,t} = m_{85,t}$ for all $x \ge 85$.\\

The period curtate expectation of life at the oldest age group for individuals aged 85 and above in calendar year $t$, $e_{85+,t}$, is computed as
\begin{equation*}
	e_{85+,t} = \frac{{}_1 p_{85+,t}}{1 - {}_1 p_{85+,t}}.
\end{equation*}

\subsection{Tabular results}

Tables \ref{NBBeta1FemalesA} through \ref{NBBeta2FemalesB} contains the estimated parameters for the negative binomial GLMs fitted according to (\ref{disMT}). $\widehat{\beta}_{1}$ is analogous to the pre-2011 trend in log age-specific mortality rates, $\widehat{\beta}_{2}$ is analogous to the post-2011 change in the temporal trend in log age-specific mortality rates. Bold values correspond to $p$-values lower than 0.05.\\  

\begin{table}[h!]
	\centering
	\caption{Estimated $\widehat{\beta}_{1,x}^{[c]}$ from (\ref{disMT}) for the first six cause-of-death groups - Females}
	\begin{tabular}{|c|c|c|c|c|c|c|}
		\hline
		{\ul \textbf{Age Group}} & {\ul \textbf{CAN}} & {\ul \textbf{CIR}} & {\ul \textbf{DIG}} & {\ul \textbf{END}} & {\ul \textbf{EXT}} & {\ul \textbf{GEN}} \\ \hline
		\textbf{\textless{}1}    & -0.01025           & 0.02699            & -0.03069           & -0.04713           & \textbf{-0.10561}           & -0.07106           \\ \hline
		\textbf{1-4}             & -0.01544           & \textbf{-0.06408}           & \textbf{-0.08336}           & -0.04349           & \textbf{-0.04930}           & 0.08830           \\ \hline
		\textbf{5-9}             & \textbf{-0.03822}           & 0.03708            & \textbf{-0.25335}           & 0.00341            & \textbf{-0.07863}           & -0.06593           \\ \hline
		\textbf{10-14}           & -0.02764           & -0.04570           & -0.02750           & 0.02450            & \textbf{-0.06288}           & 0.06689            \\ \hline
		\textbf{15-19}           & \textbf{-0.04194}           & \textbf{-0.05956}           & 0.02339            & -0.01062           & \textbf{-0.03951}           & -0.09769           \\ \hline
		\textbf{20-24}           & \textbf{-0.02936}           & \textbf{-0.04458}           & \textbf{-0.07405}           & -0.02647           & \textbf{-0.04397}           & \textbf{-0.12101}           \\ \hline
		\textbf{25-29}           & \textbf{-0.02694}           & -0.01347           & 0.01558            & 0.00795            & \textbf{-0.03284}           & -0.06831           \\ \hline
		\textbf{30-34}           & \textbf{-0.03259}           & -0.00936           & 0.01696            & 0.00872            & \textbf{-0.01570}           & -0.06061           \\ \hline
		\textbf{35-39}           & \textbf{-0.02178}           & \textbf{-0.02261}           & 0.01423            & -0.00409           & \textbf{-0.02110}           & -0.00001           \\ \hline
		\textbf{40-44}           & \textbf{-0.01640}           & \textbf{-0.02550}           & 0.00022            & -0.00562           & -0.01169           & -0.02843          \\ \hline
		\textbf{45-49}           & \textbf{-0.02713}           & \textbf{-0.04206}           & \textbf{-0.01751}           & 0.00737            & -0.00882           & -0.01772           \\ \hline
		\textbf{50-54}           & \textbf{-0.02430}           & \textbf{-0.03018}           & 0.00961            & 0.01229            & 0.00251           & -0.00319           \\ \hline
		\textbf{55-59}           & \textbf{-0.01481}           & \textbf{-0.04701}           & \textbf{0.01456}            & -0.00066           & 0.00071            & -0.00114           \\ \hline
		\textbf{60-64}           & \textbf{-0.01435}           & \textbf{-0.07203}           & -0.00629           & \textbf{-0.05577}           & -0.00797           & -0.00722           \\ \hline
		\textbf{65-69}           & \textbf{-0.00941}           & \textbf{-0.07611}           & \textbf{-0.01821}           & \textbf{-0.05631}           & -0.00077           & -0.00079           \\ \hline
		\textbf{70-74}           & \textbf{-0.01298}           & \textbf{-0.07577}           & \textbf{-0.01468}           & \textbf{-0.04365}           & \textbf{-0.01921}           & 0.01325            \\ \hline
		\textbf{75-79}           & \textbf{-0.01103}           & \textbf{-0.06975}           & \textbf{-0.01713}           & \textbf{-0.02679}           & \textbf{-0.01687}           & \textbf{0.03020}            \\ \hline
		\textbf{80-84}           & 0.00205            & \textbf{-0.05184}           & \textbf{-0.00136}           & \textbf{-0.01994}           & 0.00000            & \textbf{0.04363}            \\ \hline
		\textbf{85+}             & 0.00094            & \textbf{-0.03619}           & -0.00282           & \textbf{-0.01183}           & 0.00083            & \textbf{0.06024}            \\ \hline
	\end{tabular}
	\label{NBBeta1FemalesA}
\end{table}

\begin{table}[h!]
	\centering
	\caption{Estimated $\widehat{\beta}_{1,x}^{[c]}$ from (\ref{disMT}) for the last six cause-of-death groups - Females}
	\begin{tabular}{|c|c|c|c|c|c|c|}
		\hline
		{\ul \textbf{Age Group}} & {\ul \textbf{INF}} & {\ul \textbf{MEN}} & {\ul \textbf{MUS}} & {\ul \textbf{NER}} & {\ul \textbf{RES}} & {\ul \textbf{OTH}} \\ \hline
		\textbf{\textless{}1}    & \textbf{-0.05069}           & -0.02314           & 0.15471            & \textbf{-0.03773}           & -0.03004           & \textbf{-0.01836}           \\ \hline
		\textbf{1-4}             & -0.00128           & -0.27680           & -0.17763           & -0.00230           & 0.01252            & -0.01307           \\ \hline
		\textbf{5-9}             & -0.01094           & \textbf{0.44313}            & -0.14887           & -0.03617           & -0.00355           & 0.00236            \\ \hline
		\textbf{10-14}           & -0.01329           & -0.13107          & 0.08407            & 0.00339            & 0.02335            & -0.00987           \\ \hline
		\textbf{15-19}           & \textbf{-0.06130}           & \textbf{-0.12688}           & 0.04579           & -0.01723           & 0.0060            & \textbf{0.06186}            \\ \hline
		\textbf{20-24}           & -0.05026          & \textbf{-0.12750}           & -0.02527           & 0.00406            & -0.04089           & 0.02811            \\ \hline
		\textbf{25-29}           & \textbf{-0.05740}           & \textbf{-0.07525}           & 0.02656            & -0.01164           & -0.00523           & 0.00834            \\ \hline
		\textbf{30-34}           & \textbf{-0.04175}           & -0.01619           & \textbf{-0.07299}           & \textbf{-0.04508}           & 0.00451            & 0.00145            \\ \hline
		\textbf{35-39}           & -0.00048           & \textbf{0.05173}            & -0.05258           & \textbf{-0.02951}           & 0.00700            & \textbf{0.05528}            \\ \hline
		\textbf{40-44}           & 0.01601           & \textbf{0.05638}            & 0.03195            & -0.01461           & 0.01291            & \textbf{0.04897}            \\ \hline
		\textbf{45-49}           & 0.00726            & 0.02729           & -0.00329           & \textbf{-0.02416}           & -0.00354           & \textbf{0.02843}            \\ \hline
		\textbf{50-54}           & 0.01115            & 0.01818            & -0.00315           & -0.00435           & -0.00877           & \textbf{0.02848}            \\ \hline
		\textbf{55-59}           & 0.00802            & \textbf{0.08236}            & 0.00410           & 0.01349            & -0.00427           & -0.00213           \\ \hline
		\textbf{60-64}           & -0.01237           & \textbf{0.04627}            & \textbf{-0.02149}           & \textbf{0.01686}            & \textbf{-0.01238}           & 0.02434            \\ \hline
		\textbf{65-69}           & 0.00785            & \textbf{0.02399}            & \textbf{-0.01949}           & \textbf{0.01182}            & \textbf{-0.01743}           & -0.01719           \\ \hline
		\textbf{70-74}           & 0.02403            & \textbf{0.01707}            & \textbf{-0.02295}           & \textbf{0.00135}            & -0.02447           & \textbf{-0.03853}           \\ \hline
		\textbf{75-79}           & 0.02384            & 0.00476            & \textbf{-0.02050}           & \textbf{0.01117}            & \textbf{-0.02355}           & \textbf{-0.11014}           \\ \hline
		\textbf{80-84}           & \textbf{0.05745}            & \textbf{0.01899}            & \textbf{-0.01311}           & \textbf{0.02141}            & -0.00843           & \textbf{-0.09503}           \\ \hline
		\textbf{85+}             & \textbf{0.06626}            & \textbf{0.02973}            & \textbf{-0.01901}           & \textbf{0.02322}            & \textbf{-0.01845}           & \textbf{-0.05830}           \\ \hline
	\end{tabular}
	\label{NBBeta1FemalesB}
\end{table}

\clearpage
\newpage

\begin{table}[h!]
	\centering
	\caption{Estimated $\widehat{\beta}_{2,x}^{[c]}$ from (\ref{disMT}) for the first six cause-of-death groups - Females}
	\begin{tabular}{|c|c|c|c|c|c|c|}
		\hline
		{\ul \textbf{Age Group}} & {\ul \textbf{CAN}} & {\ul \textbf{CIR}} & {\ul \textbf{DIG}} & {\ul \textbf{END}} & {\ul \textbf{EXT}} & {\ul \textbf{GEN}} \\ \hline
		\textbf{\textless{}1}    & 0.01834            & -0.02561           & -0.10749           & 0.02327            & 0.12182            & -0.07024           \\ \hline
		\textbf{1-4}             & -0.04718           & 0.09278            & -0.05623           & -0.00034           & 0.03770            & -0.04908           \\ \hline
		\textbf{5-9}             & 0.03693            & -0.03174           & 0.16632            & -0.00028           & 0.05856            & 0.11940            \\ \hline
		\textbf{10-14}           & 0.02235            & \textbf{0.15764}            & 0.08888          & \textbf{-0.12127}           & 0.05594            & -0.05726          \\ \hline
		\textbf{15-19}           & \textbf{0.06093}            & 0.05725            & -0.05928           & 0.00048           & \textbf{0.08736}            & 0.24749           \\ \hline
		\textbf{20-24}           & -0.00715          & 0.04628            & 0.05500            & 0.00871            & \textbf{0.05850}            & 0.04851           \\ \hline
		\textbf{25-29}           & 0.01051            & 0.01573            & -0.06189           & 0.04191            & \textbf{0.06500}            & 0.11712            \\ \hline
		\textbf{30-34}           & \textbf{0.02654}            & 0.00767            & -0.03522           & 0.04150            & \textbf{0.03827}            & -0.00845           \\ \hline
		\textbf{35-39}           & \textbf{0.03781}            & 0.00972            & -0.01837           & 0.02134            & \textbf{0.04580}            & 0.01176            \\ \hline
		\textbf{40-44}           & -0.00310           & \textbf{0.03763}            & -0.01008           & 0.04886            & \textbf{0.05895}            & 0.02697            \\ \hline
		\textbf{45-49}           & \textbf{0.01521}            & \textbf{0.05255}            & \textbf{0.03404}            & \textbf{0.04315}            & \textbf{0.05423}            & 0.02910            \\ \hline
		\textbf{50-54}           & 0.00327            & \textbf{0.03171}            & -0.01215           & 0.03002            & \textbf{0.02540}            & 0.01968            \\ \hline
		\textbf{55-59}           & -0.00423           & \textbf{0.04019}            & \textbf{-0.02244}           & \textbf{0.04798}            & 0.01584            & 0.00410            \\ \hline
		\textbf{60-64}           & \textbf{-0.00624}           & \textbf{0.07572}            & 0.01287            & \textbf{0.09450}            & \textbf{0.05486}            & 0.01005            \\ \hline
		\textbf{65-69}           & -0.00567           & \textbf{0.06366}            & \textbf{0.02317}            & \textbf{0.11576}            & \textbf{0.05429}            & -0.02373           \\ \hline
		\textbf{70-74}           & -0.00244           & \textbf{0.03687}            & -0.00506          & \textbf{0.04719}            & \textbf{0.05410}            & \textbf{-0.05008}           \\ \hline
		\textbf{75-79}           & \textbf{0.00712}            & \textbf{0.02995}            & -0.00463           & \textbf{0.03518}            & \textbf{0.05006}            & \textbf{-0.06600}           \\ \hline
		\textbf{80-84}           & \textbf{-0.00958}           & \textbf{0.00798}            & \textbf{-0.01288}           & \textbf{0.02359}            & \textbf{0.01969}            & \textbf{-0.08196}           \\ \hline
		\textbf{85+}             & \textbf{-0.00470}           & 0.00652            & -0.00957           & \textbf{0.02718}            & \textbf{0.02624}            & \textbf{-0.09638}           \\ \hline
	\end{tabular}
	\label{NBBeta2FemalesA}
\end{table}

\begin{table}[h!]
	\centering
	\caption{Estimated $\widehat{\beta}_{2,x}^{[c]}$ from (\ref{disMT}) for the last six cause-of-death groups - Females}
	\begin{tabular}{|c|c|c|c|c|c|c|}
		\hline
		{\ul \textbf{Age Group}} & {\ul \textbf{INF}} & {\ul \textbf{MEN}} & {\ul \textbf{MUS}} & {\ul \textbf{NER}} & {\ul \textbf{RES}} & {\ul \textbf{OTH}} \\ \hline
		\textbf{\textless{}1}    & -0.07159           & 23.82964           & -0.07615           & -0.03858           & -0.04859           & \textbf{0.01425}            \\ \hline
		\textbf{1-4}             & -0.03768           & 0.32293            & 0.25301            & \textbf{-0.08825}           & -0.04501           & -0.03432           \\ \hline
		\textbf{5-9}             & 0.01783            & -0.21160           & 0.18935            & -0.03911           & 0.02587            & -0.01581           \\ \hline
		\textbf{10-14}           & 0.06677            & 0.05947            & -0.16357           & -0.06121           & -0.03342           & -0.08521           \\ \hline
		\textbf{15-19}           & 0.07691            & 0.01657            & -0.10083           & -0.02213           & 0.02558            & -0.07223           \\ \hline
		\textbf{20-24}           & 0.04862            & \textbf{0.21665}            & 0.04357            & -0.05516           & 0.04992            & \textbf{-0.07436}           \\ \hline
		\textbf{25-29}           & -0.01421           & 0.06460            & -0.12182           & -0.02722           & -0.05241           & -0.03184           \\ \hline
		\textbf{30-34}           & 0.00676           & 0.00464           & 0.10626            & 0.04054            & 0.04061            & \textbf{-0.05569}           \\ \hline
		\textbf{35-39}           & -0.03510           & -0.03192           & -0.03455           & 0.01683            & -0.00428           & \textbf{-0.09326}           \\ \hline
		\textbf{40-44}           & -0.02824           & -0.05831           & -0.01695           & -0.02188           & -0.00320           & -0.02590           \\ \hline
		\textbf{45-49}           & -0.05322           & 0.04058            & 0.01886            & 0.01360            & 0.04827            & -0.03802           \\ \hline
		\textbf{50-54}           & 0.00129            & -0.00711           & -0.00447           & 0.01070            & \textbf{0.04303}            & -0.01097           \\ \hline
		\textbf{55-59}           & -0.01229           & \textbf{-0.10568}           & -0.01607           & -0.00518           & \textbf{0.02175}            & \textbf{0.04564}            \\ \hline
		\textbf{60-64}           & 0.01058            & \textbf{-0.05589}           & 0.01950           & -0.01356           & \textbf{0.03018}            & \textbf{0.04750}            \\ \hline
		\textbf{65-69}           & 0.00006            & 0.01249            & 0.00956            & 0.01953            & \textbf{0.03979}            & \textbf{0.05433}            \\ \hline
		\textbf{70-74}           & -0.02546           & \textbf{0.03003}            & 0.00984            & \textbf{0.04135}            & \textbf{0.02697}            & \textbf{0.05628}            \\ \hline
		\textbf{75-79}           & -0.03955           & \textbf{0.03776}            & 0.01027            & \textbf{0.05651}            & \textbf{0.03427}            & \textbf{0.11740}            \\ \hline
		\textbf{80-84}           & \textbf{-0.08158}           & \textbf{0.01796}            & \textbf{-0.02701}           & \textbf{0.06372}            & -0.00248           & \textbf{0.11481}            \\ \hline
		\textbf{85+}             & \textbf{-0.07657}           & \textbf{0.02918}            & -0.00232           & \textbf{0.09715}            & 0.00579            & \textbf{0.08222}            \\ \hline
	\end{tabular}
	\label{NBBeta2FemalesB}
\end{table}

\clearpage
\newpage

Tables \ref{MaleNBASMR} and \ref{FemaleNBASMR} display the effects of the restoration of mortality rate trends for specific causes of death on the log age-standardised mortality rates post-2011. The second column is the average annual improvement in the log age-standardised mortality rate if the mortality rate trends for each age group are restored to pre-2011 levels for the given cause of death $k$, $w^{[k]}$, with the associated 95$\%$ confidence intervals for $w^{[Obs]}$ in the third column. The 95$\%$ confidence intervals are obtained using 5000 bootstrap iterations. The fourth column is the additional rate of improvement due to restored cause-of-death rates compared to the rate of improvement for the Unadjusted rates scenario, $w^{[k]} - w^{[Obs]}$.

\begin{table}[h!]
	\centering
	\caption{Effects of changes in the cause-specific trends on log age-standardised mortality rates for males in England and Wales [Unadjusted rates post-2011 improvement $w^{[Obs]}$ = 0.00457 (95$\%$ CI: 0.00316, 0.00596), All-causes-adjusted post-2011 improvement $w^{[\cdot]}$ = 0.02128 (0.01985, 0.02276)]}
	\begin{tabular}{|c|c|c|c|c|}
		\hline
		\textbf{\begin{tabular}{@{}c@{}}{\ul Cause}\\ {\ul of}\\ {\ul Death}\end{tabular}} & \textbf{\begin{tabular}{@{}c@{}}{\ul Cause-reverted}\\ {\ul average annual}\\ {\ul improvement in} \\{\ul log ASMR}\\ {\ul post 2011} \\{\ul ($w^{[k]}$)}\end{tabular}} & {\ul \textbf{95\% CI}} & \textbf{\begin{tabular}{@{}c@{}}{\ul Additional}\\ {\ul potential annual}\\ {\ul improvement in}\\{\ul log ASMR} \\ {\ul post 2011}\\{\ul ($w^{[k]} - w^{[Obs]}$)}\end{tabular} } \\ \hline
		Cancer                  & 0.00468                              & (0.00313, 0.00621)                    & 0.00011                                                       \\ \hline
		Circulatory             & 0.01078                              & (0.00937, 0.01219)                    & 0.00621                                                      \\ \hline
		Digestive               & 0.00451                              & (0.00306, 0.00585)                    & -0.00005                                                       \\ \hline
		Endocrine               & 0.00546                              & (0.00400, 0.00699)                    & 0.00090                                                       \\ \hline
		External                & 0.00598                              & (0.00452, 0.00753)                    & 0.00141                                                       \\ \hline
		Genitourinary           & 0.00365                              & (0.00217, 0.00504)                    & -0.00092                                                      \\ \hline
		Infectious              & 0.00424                              & (0.00277, 0.00572)                    & -0.00032                                                      \\ \hline
		Mental                  & 0.00746                              & (0.00603, 0.00891)                    & 0.00289                                                       \\ \hline
		Musculoskeletal         & 0.00472                              & (0.00292, 0.00608)                    & -0.00005                                                       \\ \hline
		Nervous                 & 0.00726                             & (0.00572, 0.00871)                    & 0.00316                                                       \\ \hline
		Respiratory             & 0.00669                              & (0.00516, 0.00817)                    & 0.00213                                                        \\ \hline
		Other                   & 0.00554                              & (0.00401, 0.00693)                    & 0.00098                                                       \\ \hline
	\end{tabular}
	\label{MaleNBASMR}
\end{table}

\clearpage
\newpage

\begin{table}[h!]
	\centering
	\caption{Effects of changes in the cause-specific trends on log age-standardised mortality rates for females in England and Wales [Unadjusted rates post-2011 improvement $w^{[Obs]}$ = 0.00107 (95$\%$ CI: $-$0.00100, 0.00325), All-causes-adjusted post-2011 improvement $w^{[\cdot]}$ = 0.01420 (0.01219, 0.01639)]}
	\begin{tabular}{|c|c|c|c|c|}
		\hline
		\textbf{\begin{tabular}{@{}c@{}}{\ul Cause}\\ {\ul of}\\ {\ul Death}\end{tabular}} & \textbf{\begin{tabular}{@{}c@{}}{\ul Cause-reverted}\\ {\ul average annual}\\ {\ul change in} \\{\ul log ASMR}\\{\ul post 2011}\\{\ul ($w^{[k]}$)}\end{tabular}} & {\ul \textbf{95\% CI}} & \textbf{\begin{tabular}{@{}c@{}}{\ul Additional}\\ {\ul annual}\\ {\ul improvement in}\\{\ul log ASMR} \\ {\ul post 2011}\\{\ul ($w^{[k]} - w^{[Obs]}$)}\end{tabular}} \\ \hline
		Cancer                  & 0.00049                            & (-0.00160, 0.00252)                    & -0.00058                                                      \\ \hline
		Circulatory             & 0.00485                               & (0.00284, 0.00707)                    & 0.00378                                                       \\ \hline
		Digestive               & 0.00081                             & (-0.00118, 0.00290)                    & -0.00026                                                       \\ \hline
		Endocrine               & 0.00167                              & (-0.00035, 0.00383)                    & 0.00059                                                       \\ \hline
		External                & 0.00211                              & (0.00006, 0.00416)                    & 0.00104                                                        \\ \hline
		Genitourinary           & -0.00076                             & (-0.00279, 0.00128)                   & -0.00184                                                      \\ \hline
		Infectious              & 0.00035                              & (-0.00020, 0.00249)                    & -0.00072                                                      \\ \hline
		Mental                  & 0.00402                              & (0.00191, 0.00608)                    & 0.00295                                                       \\ \hline
		Musculoskeletal         & 0.00103                              & (-0.00117, 0.00306)                    & -0.00004                                                       \\ \hline
		Nervous                 & 0.00511                              & (0.00298, 0.00713)                    & 0.00403                                                       \\ \hline
		Respiratory             & 0.00290                              & (0.00079, 0.00494)                    & 0.00182                                                       \\ \hline
		Other                   & 0.00303                              & (0.00085, 0.00520)                    & 0.00196                                                        \\ \hline
	\end{tabular}
	\label{FemaleNBASMR}
\end{table}

Tables \ref{NBLE0Male} through \ref{NBLE65Female} display the effects of the reversion of cause-specific trends on the life expectancies and expected remaining lifetimes at birth and age 65 respectively, given in months per year.

\begin{table}[h!]
	\centering
	\caption{Effect of changes in cause-specific mortality trends on male life expectancies at birth in England and Wales [Unadjusted rates post-2011 improvement $v^{[Obs]}$= 0.656 (95$\%$ CI: 0.544, 0.813), All-causes-adjusted post-2011 improvement $v^{[\cdot]}$ = 2.986 (2.885, 3.088)]}
	\begin{tabular}{|c|c|c|c|c|}
		\hline
		\textbf{\begin{tabular}{@{}c@{}}{\ul Cause}\\ {\ul of}\\ {\ul Death}\end{tabular}} & \textbf{\begin{tabular}{@{}c@{}}{\ul Cause-reverted} \\ {\ul average annual}\\ {\ul improvement in}\\{\ul life expectancy}\\{\ul at birth}\\ {\ul post 2011}\\{\ul (Months per year)}\\{\ul ($v^{[k]}$)}\end{tabular}} & {\ul \textbf{95\% CI}}       & \textbf{\begin{tabular}{@{}c@{}}{\ul Potential}\\ {\ul improvement}\\ {\ul lost}\\ {\ul annually}\\{\ul due to changes}\\ {\ul in cause-specific}\\{\ul mortality trends}\\{\ul post 2011 (months)}\\{\ul ($v^{[k]} - v^{[Obs]}$)}\end{tabular}} \\ \hline
		Cancer          & 0.725                   & (0.568, 0.881) & 0.069                             \\ \hline
		Circulatory     & 1.567                    & (1.413, 1.733)   & 0.911                                \\ \hline
		Digestive       & 0.617                  & (0.457, 0.773) & -0.039                           \\ \hline
		Endocrine       & 0.794                & (0.621, 0.942) & 0.138                           \\ \hline
		External        & 1.069                & (0.922, 1.235)   & 0.413                         \\ \hline
		Genitourinary   & 0.571                & (0.413, 0.724) & -0.085                          \\ \hline
		Infectious      & 0.616                & (0.465, 0.773) & -0.039                           \\ \hline
		Mental          & 0.927                 & (0.771, 1.101) & 0.271                            \\ \hline
		Musculoskeletal & 0.647              & (0.487, 0.815) & -0.009                             \\ \hline
		Nervous         & 0.914             & (0.760, 1.075) & 0.258                           \\ \hline
		Respiratory     & 0.942                 & (0.768, 1.104) & 0.286                            \\ \hline
		Other           & 0.700                 & (0.528, 0.869) & 0.044                              \\ \hline
	\end{tabular}
	\label{NBLE0Male}
\end{table}

\clearpage
\newpage

\begin{table}[h!]
	\centering
	\caption{Effect of changes in cause-specific mortality trends on female life expectancies at birth in England and Wales [Unadjusted rates post-2011 improvement $v^{[Obs]}$ = 0.234 (95$\%$ CI: 0.133, 0.387), All-causes-adjusted post-2011 improvement $v^{[\cdot]}$ = 1.992 (1.896, 2.008)]}
	\begin{tabular}{|c|c|c|c|c|}
		\hline
		\textbf{\begin{tabular}{@{}c@{}}{\ul Cause}\\ {\ul of}\\ {\ul Death}\end{tabular}} & \textbf{\begin{tabular}{@{}c@{}}{\ul Cause-reverted} \\ {\ul average annual}\\ {\ul improvement in}\\{\ul life expectancy}\\{\ul at birth}\\ {\ul post 2011}\\{\ul (Months per year)}\\{\ul ($v^{[k]}$)}\end{tabular}} & {\ul \textbf{95\% CI}}       & \textbf{\begin{tabular}{@{}c@{}}{\ul Potential}\\ {\ul improvement}\\ {\ul lost}\\ {\ul annually}\\{\ul due to changes}\\ {\ul in cause-specific}\\{\ul mortality trends}\\{\ul post 2011 (months)}\\{\ul ($v^{[k]} - v^{[Obs]}$)}\end{tabular}} \\ \hline
		Cancer          & 0.191                & (-0.049, 0.434)    & -0.043                           \\ \hline
		Circulatory     & 0.767               & (0.515, 0.993)    & 0.533                              \\ \hline
		Digestive       & 0.201                & (-0.033, 0.431)    & -0.033                          \\ \hline
		Endocrine       & 0.319                 & (0.077, 0.541)    & 0.085                         \\\hline
		External        & 0.457                 & (0.221, 0.678)    & 0.223                     \\ \hline
		Genitourinary   & 0.044                 & (-0.195, 0.244) & -0.190                          \\ \hline
		Infectious      & 0.153              & (0.083, 0.383)  & -0.081                          \\ \hline
		Mental          & 0.550          & (0.319, 0.780)    & 0.316                             \\ \hline
		Musculoskeletal & 0.229                 & (-0.015, 0.452)    & -0.005                          \\ \hline
		Nervous         & 0.662               & (0.404, 0.906)    & 0.421                           \\ \hline
		Respiratory     & 0.475              & (0.251, 0.706)    & 0.241                          \\\hline
		Other           & 0.456                   & (0.210, 0.701)    & 0.222                              \\ \hline
	\end{tabular}
	\label{NBLE0Female}
\end{table}

\begin{table}[h!]
	\centering
	\caption{Effect of changes in cause-specific mortality trends on male expected remaining lifetimes at age 65 in England and Wales [Unadjusted rates post-2011 improvement $v^{[Obs]}$ = 0.591 (95$\%$ CI: 0.451, 0.735), All-causes-adjusted post-2011 improvement $v^{[\cdot]}$ = 2.405 (2.292, 2.511)]}
	\begin{tabular}{|c|c|c|c|c|}
		\hline
		\textbf{\begin{tabular}{@{}c@{}}{\ul Cause}\\ {\ul of}\\ {\ul Death}\end{tabular}} & \textbf{\begin{tabular}{@{}c@{}}{\ul Cause-reverted} \\ {\ul average annual}\\ {\ul improvement in}\\{\ul expected remaining}\\ {\ul lifetime}\\{\ul at age 65}\\ {\ul post 2011}\\{\ul (Months per year)}\\{\ul ($v^{[k]}$)}\end{tabular}} & {\ul \textbf{95\% CI} }      & \textbf{\begin{tabular}{@{}c@{}}{\ul Potential}\\ {\ul improvement}\\ {\ul lost}\\ {\ul annually}\\{\ul due to changes}\\ {\ul in cause-specific}\\{\ul mortality trends}\\{\ul post 2011 (months)}\\{\ul ($v^{[k]} - v^{[Obs]}$)}\end{tabular}} \\ \hline
		Cancer          & 0.579                & (0.430, 0.727) & -0.013                           \\ \hline
		Circulatory     & 1.257                 & (1.100, 1.414)   & 0.666                            \\ \hline
		Digestive       & 0.614                & (0.455, 0.778) & 0.023                           \\ \hline
		Endocrine       & 0.678                  & (0.522, 0.841) & 0.086                         \\ \hline
		External        & 0.672                & (0.514, 0.821) & 0.080                            \\ \hline
		Genitourinary   & 0.499               & (0.348, 0.658) & -0.093                           \\ \hline
		Infectious      & 0.564                & (0.404, 0.727) & -0.028                            \\ \hline
		Mental          & 0.910               & (0.761, 1.066) & 0.318                             \\ \hline
		Musculoskeletal & 0.590                 & (0.440, 0.737)  & -0.001                            \\ \hline
		Nervous         & 0.898               & (0.732, 1.060) & 0.307                       \\ \hline
		Respiratory     & 0.841                  & (0.672, 0.991) & 0.250                             \\ \hline
		Other           & 0.705                   & (0.545, 0.863) & 0.114                      \\ \hline
	\end{tabular}
	\label{NBLE65Male}
\end{table}

\clearpage
\newpage

\begin{table}[h!]
	\centering
	\caption{Effect of changes in cause-specific mortality trends on female expected remaining lifetimes at age 65 in England and Wales [Unadjusted rates post-2011 improvement $v^{[Obs]}$ = 0.102 (95$\%$ CI: $-$0.138, 0.337), All-causes-adjusted post-2011 improvement $v^{[\cdot]}$ = 1.559 (1.417, 1.713)]}
	\begin{tabular}{|c|c|c|c|c|}
		\hline
		\textbf{\begin{tabular}{@{}c@{}}{\ul Cause}\\ {\ul of}\\ {\ul Death}\end{tabular}} & \textbf{\begin{tabular}{@{}c@{}}{\ul Cause-adjusted} \\ {\ul average annual}\\ {\ul improvement in}\\{\ul expected remaining}\\ {\ul lifetime}\\{\ul at age 65}\\ {\ul post 2011}\\{\ul (Months per year)}\\{\ul ($v^{[k]}$)}\end{tabular}} & {\ul \textbf{95\% CI}}           & \textbf{\begin{tabular}{@{}c@{}}{\ul Potential}\\ {\ul improvement}\\ {\ul lost}\\ {\ul annually}\\{\ul due to changes}\\ {\ul in cause-specific}\\{\ul mortality trends}\\{\ul post 2011} (months)\\{\ul ($v^{[k]} - v^{[Obs]}$)}\end{tabular}} \\ \hline
		Cancer          & 0.023               & (-0.231, 0.279)  & -0.079                         \\ \hline
		Circulatory     & 0.487                 & (0.240, 0.740)     & 0.385                           \\ \hline
		Digestive       & 0.080                & (-0.181, 0.311)   & -0.022                            \\ \hline
		Endocrine       & 0.162                    & (-0.088, 0.411)     & 0.060                         \\ \hline
		External        & 0.174                   & (-0.064, 0.423)     & 0.072                            \\ \hline
		Genitourinary   & -0.111                  & (-0.342, 0.141) & -0.213                        \\ \hline
		Infectious      & 0.021                   & (-0.225, 0.273)  & -0.081                         \\ \hline
		Mental          & 0.458               & (0.210, 0.690)     & 0.355                            \\ \hline
		Musculoskeletal & 0.098                    & (-0.139, 0.360)   & -0.004                        \\ \hline
		Nervous         & 0.590                  & (0.328, 0.833)     & 0.488                           \\ \hline
		Respiratory     & 0.308                  & (0.067, 0.573)     & 0.205                          \\ \hline
		Other           & 0.332                   & (0.056, 0.581)       & 0.229                            \\ \hline
	\end{tabular}
	\label{NBLE65Female}
\end{table}

Tables \ref{WHOMalesB} to \ref{WHOFemalesB} display WHO projected trends for the remainder of the cause-of-death groups for males and females.

\begin{table}[h!]
	\centering
	\caption{WHO projected annual change in log mortality rates by cause - Male - Part 2}
	\begin{tabular}{|c|c|c|c|c|c|c|}
		\hline
		{\ul \textbf{Age Group}} & {\ul \textbf{INF}} & {\ul \textbf{MEN}} & {\ul \textbf{MUS}} & {\ul \textbf{NER}} & {\ul \textbf{RES}} & {\ul \textbf{OTH}} \\ \hline
		\textbf{\textless{}1}    & -0.04799           & 0.00579            & -0.01704           & -0.02063           & -0.0554            & -0.03035           \\ \hline
		\textbf{1-4}             & -0.04799           & 0.00579            & -0.01704           & -0.02063           & -0.0554            & -0.03035           \\ \hline
		\textbf{5-9}             & -0.04661           & -0.01263           & -0.02457           & -0.01079           & -0.01782           & -0.01585           \\ \hline
		\textbf{10-14}           & -0.04661           & -0.01263           & -0.02457           & -0.01079           & -0.01782           & -0.01585           \\ \hline
		\textbf{15-19}           & -0.03899           & 0.01394            & -0.00923           & -0.00328           & -0.01161           & -0.0175            \\ \hline
		\textbf{20-24}           & -0.03899           & 0.01394            & -0.00923           & -0.00328           & -0.01161           & -0.0175            \\ \hline
		\textbf{25-29}           & -0.03899           & 0.01394            & -0.00923           & -0.00328           & -0.01161           & -0.0175            \\ \hline
		\textbf{30-34}           & -0.04567           & 0.00808            & -0.01258           & -0.00142           & -0.0158            & -0.00696           \\ \hline
		\textbf{35-39}           & -0.04567           & 0.00808            & -0.01258           & -0.00142           & -0.0158            & -0.00696           \\ \hline
		\textbf{40-44}           & -0.04567           & 0.00808            & -0.01258           & -0.00142           & -0.0158            & -0.00696           \\ \hline
		\textbf{45-49}           & -0.04567           & 0.00808            & -0.01258           & -0.00142           & -0.0158            & -0.00696           \\ \hline
		\textbf{50-54}           & -0.02462           & -0.01697           & -0.0081            & -0.00071           & -0.0193            & -0.00891           \\ \hline
		\textbf{55-59}           & -0.02462           & -0.01697           & -0.0081            & -0.00071           & -0.0193            & -0.00891           \\ \hline
		\textbf{60-64}           & -0.02462           & -0.01697           & -0.0081            & -0.00071           & -0.0193            & -0.00891           \\ \hline
		\textbf{65-69}           & -0.02462           & -0.01697           & -0.0081            & -0.00071           & -0.0193            & -0.00891           \\ \hline
		\textbf{70-74}           & 0.00664            & -0.00696           & -0.0005            & 0.01887            & -0.00682           & -0.01295           \\ \hline
		\textbf{75-79}           & 0.00664            & -0.00696           & -0.0005            & 0.01887            & -0.00682           & -0.01295           \\ \hline
		\textbf{80-84}           & 0.00664            & -0.00696           & -0.0005            & 0.01887            & -0.00682           & -0.01295           \\ \hline
		\textbf{85+}             & 0.00664            & -0.00696           & -0.0005            & 0.01887            & -0.00682           & -0.01295           \\ \hline
	\end{tabular}
	\label{WHOMalesB}
\end{table}

\clearpage
\newpage

\begin{table}[h!]
	\centering
	\caption{WHO projected annual change in log mortality rates by cause - Female - Part 1}
	\begin{tabular}{|c|c|c|c|c|c|c|}
		\hline
		{\ul \textbf{Age Group}} & {\ul \textbf{CAN}} & {\ul \textbf{CIR}} & {\ul \textbf{DIG}} & {\ul \textbf{END}} & {\ul \textbf{EXT}} & {\ul \textbf{GEN}} \\ \hline
		\textbf{\textless{}1}    & -0.02216           & -0.01305           & -0.01644           & -0.01810           & 0.00570            & -0.01651           \\ \hline
		\textbf{1-4}             & -0.02216           & -0.01305           & -0.01644           & -0.01810           & 0.00570            & -0.01651           \\ \hline
		\textbf{5-9}             & -0.02226           & -0.01757           & -0.03126           & -0.01772           & 0.00325            & -0.03114           \\ \hline
		\textbf{10-14}           & -0.02226           & -0.01757           & -0.03126           & -0.01772           & 0.00325            & -0.03114           \\ \hline
		\textbf{15-19}           & -0.01478           & -0.01277           & -0.00856           & -0.01151           & -0.00323           & -0.01651           \\ \hline
		\textbf{20-24}           & -0.01478           & -0.01277           & -0.00856           & -0.01151           & -0.00323           & -0.01651           \\ \hline
		\textbf{25-29}           & -0.01478           & -0.01277           & -0.00856           & -0.01151           & -0.00323           & -0.01651           \\ \hline
		\textbf{30-34}           & -0.01101           & -0.01507           & -0.00311           & 0.00402            & 0.00000            & -0.00391           \\ \hline
		\textbf{35-39}           & -0.01101           & -0.01507           & -0.00311           & 0.00402            & 0.00000            & -0.00391           \\ \hline
		\textbf{40-44}           & -0.01101           & -0.01507           & -0.00311           & 0.00402            & 0.00000            & -0.00391           \\ \hline
		\textbf{45-49}           & -0.01101           & -0.01507           & -0.00311           & 0.00402            & 0.00000            & -0.00391           \\ \hline
		\textbf{50-54}           & -0.00588           & -0.02129           & -0.00684           & -0.00591           & -0.00568           & -0.00419           \\ \hline
		\textbf{55-59}           & -0.00588           & -0.02129           & -0.00684           & -0.00591           & -0.00568           & -0.00419           \\ \hline
		\textbf{60-64}           & -0.00588           & -0.02129           & -0.00684           & -0.00591           & -0.00568           & -0.00419           \\ \hline
		\textbf{65-69}           & -0.00588           & -0.02129           & -0.00684           & -0.00591           & -0.00568           & -0.00419           \\ \hline
		\textbf{70-74}           & -0.00519           & -0.02610           & -0.01347           & -0.00780           & -0.00899           & -0.00494           \\ \hline
		\textbf{75-79}           & -0.00519           & -0.02610           & -0.01347           & -0.00780           & -0.00899           & -0.00494           \\ \hline
		\textbf{80-84}           & -0.00519           & -0.02610           & -0.01347           & -0.00780           & -0.00899           & -0.00494           \\ \hline
		\textbf{85+}             & -0.00519           & -0.02610           & -0.01347           & -0.00780           & -0.00899           & -0.00494           \\ \hline
	\end{tabular}
	\label{WHOFemalesA}
\end{table}

\begin{table}[h!]
	\centering
	\caption{WHO projected annual change in log mortality rates by cause - Female - Part 2}
	\begin{tabular}{|c|c|c|c|c|c|c|}
		\hline
		{\ul \textbf{Age Group}} & {\ul \textbf{INF}} & {\ul \textbf{MEN}} & {\ul \textbf{MUS}} & {\ul \textbf{NER}} & {\ul \textbf{RES}} & {\ul \textbf{OTH}} \\ \hline
		\textbf{\textless{}1}    & -0.04971           & 0.00791            & -0.01608           & -0.01990           & -0.05641           & -0.02969           \\ \hline
		\textbf{1-4}             & -0.04971           & 0.00791            & -0.01608           & -0.01990           & -0.05641           & -0.02969           \\ \hline
		\textbf{5-9}             & -0.05429           & -0.01970           & -0.01111           & -0.01779           & -0.01914           & -0.01962           \\ \hline
		\textbf{10-14}           & -0.05429           & -0.01970           & -0.01111           & -0.01779           & -0.01914           & -0.01962           \\ \hline
		\textbf{15-19}           & -0.03966           & 0.00800            & -0.03053           & -0.00008           & -0.01561           & -0.01450           \\ \hline
		\textbf{20-24}           & -0.03966           & 0.00800            & -0.03053           & -0.00008           & -0.01561           & -0.01450           \\ \hline
		\textbf{25-29}           & -0.03966           & 0.00800            & -0.03053           & -0.00008           & -0.01561           & -0.01450           \\ \hline
		\textbf{30-34}           & -0.02763           & 0.00801            & -0.00925           & 0.00393            & -0.00461           & -0.00633           \\ \hline
		\textbf{35-39}           & -0.02763           & 0.00801            & -0.00925           & 0.00393            & -0.00461           & -0.00633           \\ \hline
		\textbf{40-44}           & -0.02763           & 0.00801            & -0.00925           & 0.00393            & -0.00461           & -0.00633           \\ \hline
		\textbf{45-49}           & -0.02763           & 0.00801            & -0.00925           & 0.00393            & -0.00461           & -0.00633           \\ \hline
		\textbf{50-54}           & -0.01110           & -0.01119           & -0.01136           & 0.00583            & -0.00690           & -0.00661           \\ \hline
		\textbf{55-59}           & -0.01110           & -0.01119           & -0.01136           & 0.00583            & -0.00690           & -0.00661           \\ \hline
		\textbf{60-64}           & -0.01110           & -0.01119           & -0.01136           & 0.00583            & -0.00690           & -0.00661           \\ \hline
		\textbf{65-69}           & -0.01110           & -0.01119           & -0.01136           & 0.00583            & -0.00690           & -0.00661           \\ \hline
		\textbf{70-74}           & 0.00051            & -0.00439           & -0.00909           & 0.01030            & -0.00695           & -0.01269           \\ \hline
		\textbf{75-79}           & 0.00051            & -0.00439           & -0.00909           & 0.01030            & -0.00695           & -0.01269           \\ \hline
		\textbf{80-84}           & 0.00051            & -0.00439           & -0.00909           & 0.01030            & -0.00695           & -0.01269           \\ \hline
		\textbf{85+}             & 0.00051            & -0.00439           & -0.00909           & 0.01030            & -0.00695           & -0.01269           \\ \hline
	\end{tabular}
	\label{WHOFemalesB}
\end{table}

\end{document}